\documentclass[runningheads,fleqn]{svmult}

\usepackage{makeidx}   
\usepackage{graphicx}  
\usepackage{subeqnar}  
\usepackage{multicol}  
\usepackage{physmult}  
\makeindex             

\def\bk{{\bf k}}

\def\bkom{(\bk,\omega)}

\def\em{{\cal E}_m}
\def\en{{\cal E}_n}

\def\tc{T_c}
\def\fullint{\int_{-\infty}^{+\infty}d\omega}
\def\akw{A(\bk,\omega)}

\def\y_124{YBa$_2$Cu$_4$0$_8$}
\def\etal{{\it et al.}}
\def\br{{\bf r}}
\def\bR{{\bf R}}
\def\om{\omega}
\begin{document}
\title*{Photoemission in the High $T_{c}$ Superconductors}
\titlerunning{Photoemission in the High $T_{c}$ Superconductors}
\toctitle{Photoemission in the High $T_{c}$ Superconductors}
\author{J.C. Campuzano$^{1,2}$, M.R. Norman$^{2}$, and M. Randeria$^{3}$}
\authorrunning{J.C. Campuzano, M.R. Norman, and M. Randeria}
\tocauthor{J.C. Campuzano, M.R. Norman, and M. Randeria}
\institute{$^{1}$Department of Physics, University of Illinois at Chicago, 
Chicago, IL  60607 \\
$^{2}$Materials Science Division, Argonne National Laboratory, Argonne, IL  
60439 \\
$^{3}$Tata Institute of Fundamental Research, Mumbai 400005, India}
\tableofcontents
\maketitle         
\abstract{
We review angle resolved photoemission spectroscopy (ARPES) results on the 
high $T_c$ superconductors, focusing primarily on results obtained on the 
quasi-two dimensional cuprate Bi$_2$Sr$_2$CaCu$_2$O$_8$ and its single layer 
counterpart Bi$_2$Sr$_2$CuO$_6$.  The topics treated include the basics of 
photoemission and methodologies for analyzing spectra, normal state 
electronic structure including the Fermi surface, the superconducting energy
gap, the normal state pseudogap, and the electron self-energy as determined
from photoemission lineshapes.}

\section{Introduction}

Angle resolved photoemission spectroscopy (ARPES) has played a major 
role in the elucidation of the electronic excitations in the high 
temperature cuprate superconductors. Several reasons have contributed 
to this development. First, the great improvement in experimental 
resolution, both in energy and momentum, aided by the large energy 
scales present in the cuprates, allows one to see features on the 
scale of the superconducting gap. More recently the resolution has 
improved to such an extent, that now features in traditional 
superconductors like Nb and Pb, with energy scales of a meV, can be 
observed by ARPES \cite{Yokoya}.
 
Second, most studies have focused on $Bi_2Sr_2CaCu_2O_8$ (Bi2212) and
its single layer counterpart, $Bi_2Sr_2CuO_6$ (Bi2201).  These materials
are characterized by weakly coupled BiO layers, with the longest 
interplanar separation in the cuprates. This results in a natural 
cleavage plane, with minimal charge transfer. This is crucial for 
ARPES, since it is a surface sensitive technique; for the photon 
energies typically used, the escape depth of the outgoing electrons 
is only of the order of $\sim 10 \AA$.  For this reason, we elect to
concentrate on these materials in the current article, since the
data are known to be reproducible among the various groups.

The third reason is the quasi-two dimensionality of the electronic 
structure of the cuprates, which permits one to unambiguously determine 
the momentum of the initial state from the measured final state momentum, 
since the component parallel to the surface is conserved in photoemission.
Moreover, in two dimensions, ARPES directly probes the the single particle 
spectral function, and therefore offers a complete picture of the many 
body interactions inherent in these strongly correlated systems.

There are other reasons to point out as well.  For example, the very 
incoherent nature of the excitations near the Fermi energy in the
normal state does not allow the application of traditional techniques, 
such as the de Haas van Alphen effect. The electrons simply do not live long 
enough to complete a cyclotron orbit.  Another important reason why ARPES 
has played a major role is the highly anisotropic nature of the electronic 
excitations in the cuprates, which means that momentum resolved probes are 
desirable when attempting to understand these materials.

It is satisfying to compare the earlier ARPES papers to more recent ones, 
and observe the remarkable progress the field has undergone. But this 
satisfaction must be tempered by the knowledge that the current literature 
is still full of unresolved issues. Undoubtedly, some of what is described 
here will be viewed in a new light in the future, and we will try to point 
out the contentious points, but the reader must keep in mind that the 
writers, by necessity, approach this task with their own set of biases. 
For a complementary view, the reader is refered to the recent review by 
Damascelli, Shen and Hussain \cite{RMP}. An earlier review of ARPES 
studies of high $T_c$ superconductors can be found in the book by Lynch 
and Olson \cite{LYNCH}.

\section{Basics of Angle-Resolved Photoemission}

The basics of angle-resolved photoemission have been 
described in detail in the literature\cite{HUFNER}.
We will limit ourselves to a brief review of
some salient points, emphasizing those aspects of the technique which 
will be useful in understanding ARPES studies of high $\tc$ superconductors. 
We start by looking at a simple independent particle picture, and 
subsequently include the effects of strong interactions.

\begin{figure}
\includegraphics[width=1\textwidth]{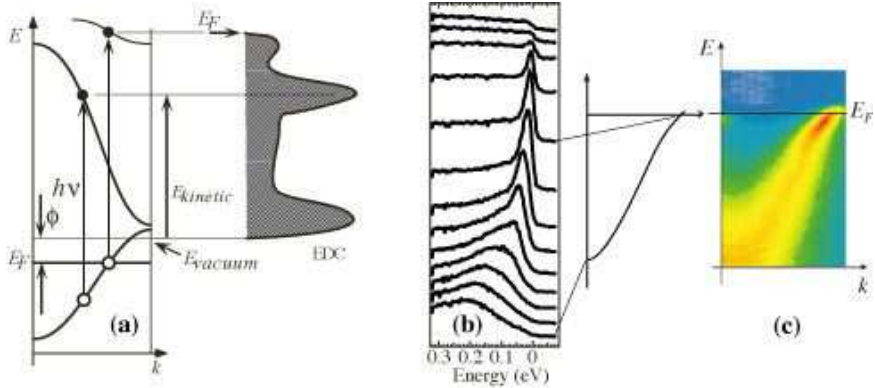}
\caption[]{a: Independent particle approximation view of ARPES; To the right 
is the resulting energy distribution curve (EDC); b: Experimental EDCs along
a path in momentum space in Bi2212; c: Intensity map from these data.}
\label{arpes.eps}
\end{figure}

The simplest model of ARPES is the three step model \cite{3STEP}, 
which separates the process into photon absorption, electron 
transport through the sample, and emission through the surface. In 
the first step, the incident photon with energy $h\nu$
is absorbed by an electron in an 
occupied initial state, causing it to be promoted to an unoccupied 
final state, as shown in Fig.~\ref{arpes.eps}a. There is conservation of 
energy, such that 
\begin{equation}
    h\nu =BE+\Phi +E_{\rm kin}
\label{econservation}
\end{equation}
where $\Phi$ is the work function, and $BE$ and $E_{\rm kin}$ are the binding 
and kinetic energies of the electron, respectively. We will not 
discuss the subsequent two steps of the three step model, 
photoelectron transport to the surface and transmission through the 
surface into the vacuum, as they only affect the number of 
emitted electrons, and thus the absolute intensity \cite{HUFNER}.

The kinetic energy of the electrons is measured by an electron energy 
analyzer. 
If the number of emitted electrons is then plotted as a 
function of their kinetic energy, as shown in Fig.~\ref{arpes.eps}, 
peaks are found whenever an allowed transition takes place. Eq. (1) 
then yields the kinetic energy of the electron if the work function 
is known. Measuring $\Phi$ accurately is not a trivial task, but 
fortunately, in the case of a metallic sample one does not need to know $\Phi$. 
By placing the sample in electrical contact with a good
metal (e.g., polycrystalline gold) one
can measure the binding energy in the sample with respect to the
chemical potential (Fermi level $E_f$) of Au.
The photoemission signal from Au will simply be a Fermi function 
convolved with the experimental energy resolution, and from the mid point
of its leading edge one estimates $E_f$.


The existence of the sample surface breaks (discrete) translational
invariance in the direction normal to the surface. However, two-dimensional
translational invariance in the the directions parallel to the surface is 
still preserved, and thus ${\bf k}_{\parallel}$, 
the component of the electron momentum parallel to the surface,
is conserved in the emission process. 
This allows us to obtain the in-plane momentum of the
initial state by identifying it with the parallel momentum of the 
emitted electron
\begin{equation}
k_{\parallel}=\sqrt {{{2mE_{\rm kin}} \over \hbar }}\sin \theta \cos \phi
\label{kparallel}
\end{equation}
of the outgoing electron emitted along the $(\theta,\phi)$ direction
with a kinetic energy $E_{\rm kin}$, as shown in Fig.~\ref{selectionrules}a.

The momentum perpendicular to the sample surface $k_z$ is not conserved,
and thus knowledge of the final state $k_z^f$ does not permit one to say
anything useful about the initial state $k_z^i$, except in the case 
where ${\bf k}_{\parallel}=0$, in which case $k_z^i=k_z^f$. We should 
emphasize that a given outgoing electron corresponds to a fixed, but 
apriori unknown, $k_z^i$. In materials with three-dimensional electronic 
dispersion it is, of course, essential to fully characterize the initial 
state, and many techniques have been developed to estimate $k_z^i$;
see Ref.~\cite{HUFNER}. Aside from making ${\bf k}_{\parallel}=0$ as 
described above, one can vary the incident photon energy $h\nu$ thereby 
changing the kinetic energy of the outgoing electron. For fixed 
${\bf k}_{\parallel}$ this amounts to changing the initial state $k_z$. 
(One also has to take into account the changes in intensity brought about 
by the photon energy dependence of the matrix elements).

We will not be much concerned about methods of $k_z$ determination here, since
the high $T_c$ cuprates are quasi-two-dimensional (2D) materials with, 
in many cases (e.g., Bi2212), no observable $k_z$-dispersion. 
In fact, this makes ARPES data from 2D materials much easier to interpret, 
and is one of the reasons for the great success of the ARPES technique for the
cuprates. In the remainder of this article, we will use the symbol
${\bf k}$ to simply denote the two-dimensional momentum parallel to
the sample surface, unless explicitly stated otherwise.

%
%
%
%

In the independent particle approximation, the ARPES intensity as a function
of momentum ${\bf k}$ (in the 2D Brillouin zone) and energy $\omega$
(measured with respect to the chemical potential) 
is given by Fermi's Golden Rule as
\begin{equation}
I(k,\omega )\propto \left| {\left\langle {\psi _f} \right|{\bf A\cdot 
p}\left| {\psi _i} \right\rangle } \right|^2 f(\omega)
\delta (\omega -\varepsilon _k)
\label{MATRIX}
\end{equation}
where $\psi_{i}$ and $\psi_{f}$ are the initial and final states, $\bf p$ 
is the momentum operator, and $\bf A$ the vector potential of the incident
photon \cite{SURFACEPE}. Here $f(\omega) = 1/[\exp(\omega/T) + 1]$ is
the Fermi function at a temperature $T$ in units where $\hbar = k_B = 1$.
It ensures the physically obvious constraint that photoemission only probes
{\it occupied} electronic states.

For noninteracting electrons, the emission at a given ${\bf k}$ is 
at a sharp energy $\epsilon_{\bf k}$ corresponding to
the initial state dispersion. As we will discuss below, the effect of
interactions (either electron-electron or electron-phonon) 
is to replace the delta function by the 
one-particle spectral function, which is a non-trivial function
of both energy and momentum.  Although now the peaks will in general
shift in energy and acquire a width, the overall intensity is still 
governed by the same matrix element as in the non-interacting case.
Thus the lessons learned from studying general properties of the
matrix elements are equally applicable in the fully interacting case.

\subsection{Matrix Elements and Selection Rules}

The dipole matrix element 
$M_{fi} = \left\langle{\psi_f}\right|{\bf A}\cdot{\bf p}
\left|{\psi_i}\right\rangle$ is in general a function of the
the momentum ${\bf k}$, of the incident photon energy $h\nu$,
and of the polarization ${\bf A}$ of the incident light.
Here we discuss dipole selection rules which arise from very
general symmetry constraints imposed on $M_{fi}$, and which
are very useful in interpreting ARPES data.
Our approach is a simplified version of Hermanson's analysis
\cite{HERMANSON}. 

\begin{figure}
\includegraphics[width=1\textwidth]{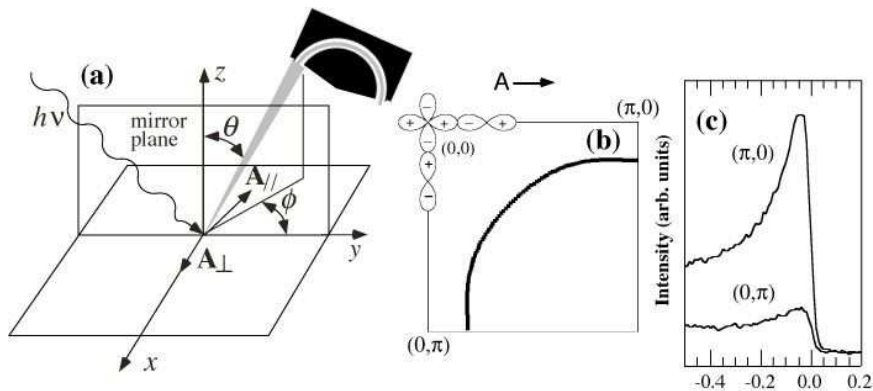}
\caption[]{
(a) Arrangement of the photon beam and detector in
order to make use of the photoemission selection rules.
(b) Symmetry of the Cu$3d_{x^{2}-y^{2}}$ orbitals hybridized with the
O$2p$ orbitals. (c) EDCs showing the symmetry of the
orbitals in (b) obtained at $\hbar\nu=$22 eV.
}
\label{selectionrules}
\end{figure}

Let the photon beam be incident along a plane of
mirror symmetry of the sample (${\cal M}$), with
detector placed in the same mirror plane; see Fig.~\ref{selectionrules}(a). 
The final state $\psi_f$ must then be even with respect to reflection
in ${\cal M}$, because if it were odd the wave function would vanish at
the detector.  
(For this discussion it is simpler to imagine changing the photon
polarization, keeping the detector fixed.  In the actual experiment, however,
it is the polarization which is fixed and
the detector is moved when checking dipole selection rules.)

The dipole transition is allowed if the entire matrix element
has an overall even symmetry. Thus two possibilities arise.
(1) For an initial state $\psi_i$ which
is even with respect to ${\cal M}$, the light polarization ${\bf A}$
must also be even, i.e. parallel to ${\cal M}$. 
(2) For an initial state odd with respect to ${\cal M}$, 
${\bf A}$ must also be odd, i.e.  perpendicular to ${\cal M}$. 
This can be summarized as:
\begin{equation}
\left\langle{\psi_f}\right|{\bf A}\cdot\hat{\bf
p}\left|{\psi_i}\right\rangle
\left\{{\matrix{{\psi_i\,\,even\,\,\left\langle + \right|+\left| +
\right\rangle
\,\,\Rightarrow \,\,{\bf A}\,\,even}\cr
{\psi_i\,\,odd\,\,\left\langle + \right|-\left| - \right\rangle 
\,\,\Rightarrow \,\,{\bf A}\,\,odd}\cr
}}\right.
\end{equation}

Consider hybridized $Cu3d-O2p$ initial states, as shown in 
Fig.~\ref{selectionrules}(b), 
which have a $d_{x^2 - y^2}$ symmetry about a Cu site.
These states are even under reflection in $(0,0)-(\pi,0)$ 
(i.e. the plane defined by this symmetry axis and the z-axis)
and odd with respect to $(0,0) - (\pi,\pi)$.
Therefore, measurement along the $(0,0)-(\pi,0)$ direction will be 
dipole-allowed (forbidden) if the polarization vector $\textbf{A}$ is 
parallel (perpendicular) to this axis. Fig.~\ref{selectionrules}(c)
shows that, consistent with an initial state which is even about 
$(0,0)-(\pi,0)$, the signal is maximized when ${\bf A}$ lies in the 
mirror plane and minimized when ${\bf A}$ is perpendicular to this plane.
(The reasons for small non-zero intensity in the dipole forbidden
geometry are the small, but finite, ${\bf k}$-window of the experiment
and the possibility of a small misalignment of the sample).
Similarly, we have checked experimentally (for Bi2212 in the
Y-quadrant where there are no superlattice complications) 
that the initial state is
consistent with odd symmetry about $(0,0) - (\pi,\pi)$ (see also
Ref.~\cite{SHEN_REVIEW}).

While the dipole matrix elements are strongly photon energy dependent,
the selection rules are, of course, independent of photon energy.
This has been checked by measurements at 22 eV and 34 eV.\cite{mesot01}
All of these results are consistent with the fact that we are probing
$Cu3d-O2p$ initial one-electron states with $d_{x^2 - y^2}$ symmetry.
We will see below that these selection rules are extremely useful in
disentangling the main CuO$_2$ ``band'' from its umklapp
images due to the superlattice in Bi2212.

\subsection{The One-Step Model}
The three-step model of photoemission gives a very useful 
zeroth order description of the photoemission process, but it needs
to be put on a firmer footing, both as regards the calculation
of photoemission intensities from ab-initio electronic structure 
calculations, and also for developing a deeper understanding
of how many-body effects influence
ARPES spectra. We now briefly review the so called ``one-step''
model of photoemission with these goals in mind.

Ever since Hertz and Hallwachs discovered photoemission\cite{HERTZ}, 
it is known that the photo-electron current at the detector is 
proportional to the incident photon flux, i.e., to the {\it square} 
of the vector potential. Thus photoemission measures a {\it nonlinear} 
response function, and the relevant correlation function is a three 
current correlator, as first emphasized by Schaich and Ashcroft \cite{ASHCROFT}.
It is instructive to briefly review their argument. 
As in standard response function calculations,
let's look at an expansion of the current at the detector (the response)
in powers of the vector potential of incident photons (the perturbation).
Let $\bR$ be the location of the detector in vacuum and $\br$ 
denote points inside the sample. 
The zeroth order contribution $\langle 0|j_\alpha (\bR,t)|0 \rangle$ vanishes
as there are no currents flowing in the ground state 
$|0\rangle$ of the unperturbed system.
The linear response also vanishes, with
$\langle 0|j_\alpha (\bR,t)j_\beta(\br,t')|0 \rangle = 0$ and 
$\langle 0|j_\alpha (\br,t')j_\beta(\bR,t)|0 \rangle = 0$,
since there are no particles at the detector, 
in absence of the electromagnetic field, and $j_\beta (\bR,t)|0 \rangle = 0$.
Thus the leading term which survives is
\begin{eqnarray}\nonumber\langle j_\gamma (\bR ,t) \rangle \propto 
\int d\br' dt' d\br'' dt'' A_\alpha (\br',t')A_\beta (\br'' ,t'') \\
\langle 0|j_\alpha (\br',t')j_\gamma (\bR ,t)j_\beta (\br'', t'')|0 \rangle
\label{threecurrent}
\end{eqnarray}
where only current operators
{\it inside} the sample act on the unperturbed ground state on either side 
and the current at the detector is sandwiched in between. 

\begin{figure}
\sidecaption
\includegraphics[width=.5\textwidth]{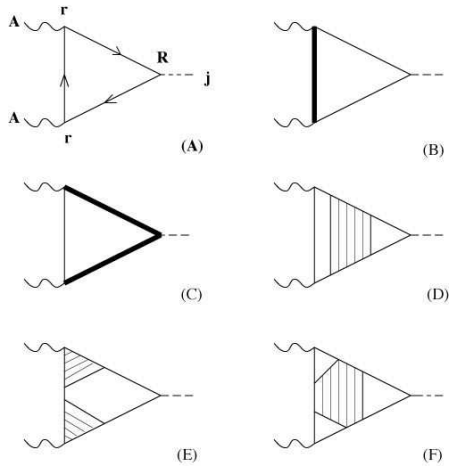}
\caption[]{Diagrams contributing to the three-current correlator
of ARPES. The physical processes that each diagram represents is
discussed in the text.}
\label{feynman2}
\end{figure}

The three current correlation function can be represented by 
the triangle diagram shown in Fig.~\ref{feynman2}(A)
\cite{CAROLI}, where the line between the two external photon 
vertices is the Greens function of the ``initial state'' or 
``photo-hole'' and the two lines connecting the photon vertex 
to the current at the detector represent the ``photo-electron'' 
which is emitted from the solid. 
There is a large literature on the evaluation of the
(bare) triangle diagram incorporating the results of ab-initio electronic 
structure calculations of a semi-infinite solid, including the effects
of realistic surface termination and multiple scattering effects 
in the photo-electron final states.
Detailed calculations using this formalism were first carried out
by Pendry \cite{PENDRY} for a system with one atom 
per unit cell, and later generalized to more complex crystals
\cite{LARSON,LINDBERG,HOPKINSON}. 

This approach is very reliable for calculating photoemission
intensities, and gives important information about ``matrix element
effects'', i.e., the dependence of the ARPES intensity on ${\bf k}$, and
on the incident photon energy and polarization. (It does not,
however, shed light on the many-body aspects of the lineshape).
Such ab-initio methods have been extensively used by 
Bansil, Lindroos and coworkers \cite{BANSIL} for the high $\tc$ cuprates.

\begin{figure}
\sidecaption
\includegraphics[width=1\textwidth]{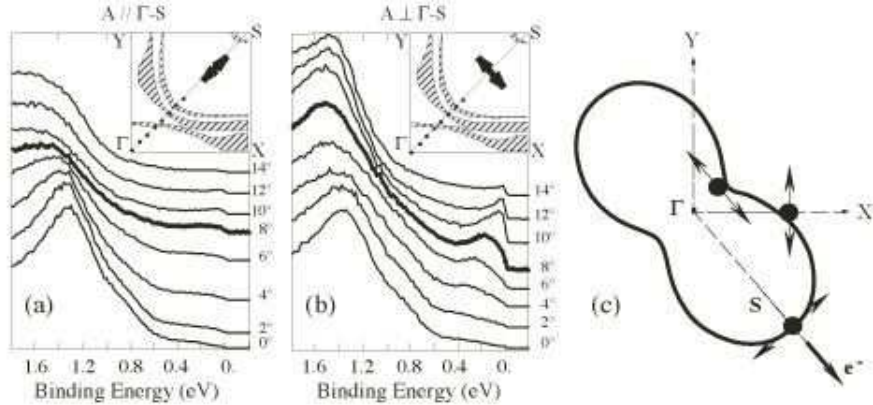}
\caption[]{Comparison of the photoemission intensity of states along the
zone diagonal
in YBCO with (a) even polarization and (b) odd polarization, showing that
this state is odd with respect to reflection about the diagonal to the
Cu-O bond direction. (c) Comparison of the photoemission intensity to the
observed one as a function of polarization. The detailed polarization
comparison gives confidence to the interpretation of the observed peak
as arising from the Cu-O plane states.}
\label{onestepcomp.eps}
\end{figure}

As an example of the usefulness of ab-initio methods, we show the comparison
between the observed and calculated ARPES intensities for YBCO
in Fig.~\ref{onestepcomp.eps}.
Since the calculated intensity is a sensitive function of the
termination plane, such a comparison suggests that 
the crystal breaks at the chains, a conclusion later confirmed
by STM measurements \cite{DELOZANNE}. In fact, it was the complexity of 
the termination in YBCO, plus the fact that none of the possible cleavage 
planes of YBCO are charge neutral, and thus involve 
significant charge transfer, that convinced us to focus primarily on the
Bi-based cuprates (Bi2212 and Bi2201). These materials have two adjacent
BiO planes which are van der Waals bonded, and thus cleavage leads to
neutral surfaces which are not plagued by uncontrolled surface effects.
The highly two-dimensional Bi-based cuprates are thus ideal from
the point of view of ARPES studies.

\subsection{Single-Particle Spectral Function}

Although the one step model gives a reasonable interpretation of the 
overall {\it intensity} in the photoemission process, 
much of what we really want to learn from the experiment relates 
to the spectral {\it lineshape}, which, as we will show, is strongly influenced 
by correlations. 
%
%
We must then ask ourselves the important question of how the
initial state lineshape enters the ARPES intensity and to 
what extent this is revealed in the observed lineshape.
Although a fully rigorous justification of a simple interpretation
of ARPES spectra is not available at the present time, a reasonable
case can be made for analyzing the data in terms of the single particle
spectral function of the initial state.

We begin by considering the various many body renormalizations
of the bare triangle diagram which are shown
in Fig.~\ref{feynman2}. These renormalizations can arise, in principle,
from either electron-phonon or electron-electron interactions.
These self energy effects and vertex corrections are
easy to draw, but impossible to evaluate in any controlled
calculation. Nevertheless, they are useful in obtaining a
qualitative understanding of the the various processes and estimating 
their relative importance. Diagram (B) represents the 
self-energy corrections to the occupied initial state that 
we are actually interested in studying; (C) and (D)
represent final state line-width broadening and inelastic scattering;
(E) is a vertex correction that describes the interaction of
the escaping photo-electron with the photo-hole in the solid;
(F) is a vertex correction which combines features of (D) and (E).
(An additional issue in a 
quantitative theory of photoemission is related to the modification 
of the external vector potential inside the medium, i.e., 
renormalizations of the photon line. These are considered in detail 
in Ref. \cite{BANSIL}).

If the sudden approximation is valid,
we can neglect the vertex corrections:
the outgoing photo-electron is moving so fast that
it has no time to interact with the photo-hole.
Let us make simple time scale estimates for the cuprates
with 15 - 30 eV (ultraviolet) incident photons. 
The time $t$ spent by the escaping photo-electron in the vicinity of 
the photo-hole is the time available for their interaction. 
A photoelectron with a kinetic energy of (say) 20 eV has a velocity 
$v = 3 \times 10^8$ cm/s. The relevant length scale, which is the smaller 
of the screening radius (of the photo-hole) and the escape depth, 
is $\sim 10 \AA$. Thus $t = 3 \times 10^{-16}$ s, 
which should be compared with the time scale for 
electron-electron interactions (which are the dominant source of 
interactions at the high energies of interest): 
$t_{ee} = (2 \pi) / \omega_p = 4 \times 10^{-15}$ s, using a plasma 
frequency $\omega_p = 1$ eV for the cuprates (this would be even slower 
if c-axis plasmons are involved). {\it If} $t \ll t_{ee}$, then we can 
safely ignore vertex corrections. 
From our very crude estimate $t/t_{ee} = 0.1$, so that 
the situation with regard to the validity 
of the impulse approximation is not hopeless, but clearly, experimental 
checks are needed, and we present these in the next subsection.

Very similar estimates can be made for renormalizations
of the outgoing photo-electron due to its interaction
with the medium; again electron-electron interactions dominate at the
energies of interest. The relevant length scale here is the
escape depth, which leads to a process of self-selection:
those electrons that actually make it to the detector with
an appreciable kinetic energy have suffered no collisions in the medium.
Such estimates indicate that the ``inelastic background'' 
must be small and we will show how to experimentally obtain its precise 
dependence on $\bk$ and $\om$ later.

Finally, one expects that final state linewidth corrections are small
for quasi-2D materials based on the estimates made by 
Smith {\it et al.} \cite{SMITH}
In fact, this is yet another reason for the ease of interpreting 
ARPES data in quasi-2D materials.
A clear experimental proof that these effects are negligible for Bi2212
will be presented later, where it will be seen that deep in
the superconducting state, a resolution limited leading edge is 
obtained for the quasiparticle peak. 

\subsection{Spectral Functions and Sum Rules}

Based on the arguments presented in the preceding subsection,
we assume the validity of the sudden approximation and ignore both 
the final-state linewidth broadening and the additive extrinsic 
background. Then (B) is the only term that survives from all the terms
described in Fig.~\ref{feynman2} and the ARPES intensity is 
given by \cite{hedin,NK}
\begin{equation}\label{intensity}
I\bkom = I_0(\bk) f(\omega) A\bkom
\end{equation}
where $\bk$ is the initial state momentum in the 2D Brillouin zone and $\om$ 
the energy relative to the chemical potential. The prefactor
$I_0(\bk)$ includes all the  kinematical factors and the 
square of the dipole matrix element (shown in Eq.~\ref{MATRIX}),
$f(\omega)$ is the Fermi function, and $A\bkom$ is the one-particle spectral
function which will be described in detail below. 

We first describe some general consequences
of Eq.~\ref{intensity} based on sum rules 
and their experimental checks. The success of 
this strategy employed by Randeria {\it et al.} \cite{NK}
greatly strengthens the case for a simple $A\bkom$ interpretation
of ARPES data.

The one-particle spectral function represents the probability of 
adding or removing a particle from the interacting many-body system,
and is defined as
$\akw = - (1/\pi) {\rm Im}G(\bk,\omega + i0^+)$
in terms of the Green's function. 
It can be written as the sum of two pieces
$\akw = A_{-}\bkom + A_{+}\bkom$, where 
the spectral weight to add an electron to the system is given by
$A_{+}\bkom = Z^{-1}\sum_{m,n} e^{-\beta\em}
|\langle n|c^\dagger_\bk |m \rangle|^2 \delta(\omega + \em - \en)$,
and that to extract an electron is
$A_{-}\bkom = Z^{-1}\sum_{m,n} e^{-\beta\em}
|\langle n|c_\bk |m \rangle|^2 \delta(\omega + \en - \em)$.
Here $|m\rangle$ is an exact eigenstate of the many-body system
with energy $\em$, $Z$ is the partition function and $\beta = 1/T$.
It follows from these definitions that $A_{-}\bkom = f(\omega) \akw$
and $A_{+}\bkom = \left[1 - f(\omega)\right] \akw$, where
$f(\omega) = 1/[\exp(\beta\omega) + 1]$ is the Fermi function.
Since an ARPES experiment involves removing an electron from the 
system, the simple golden rule Eq.~\ref{MATRIX} can be generalized
to yield an intensity proportional to $A_{-}\bkom$

We now discuss various sum rules for $A\bkom$ and their possible
relevance to ARPES intensity $I\bkom$. While the
prefactor $I_0$ depends on $\bk$ and also on 
the incident photon energy and polarization, 
it does {\it not} have any significant $\omega$ or $T$ dependence.
Thus the energy dependence of the ARPES lineshape
and its $T$ dependence are completely characterized by 
the spectral function and the Fermi factor.
The simplest sum rule $\fullint \akw = 1$ is not useful for ARPES since
it involves both occupied (through $A_{-}$)
{\it and unoccupied} states ($A_{+}$). 
Next, the density of states (DOS) sum rule
$\sum_\bk \akw = N(\omega)$ is also not directly useful
since the prefactor $I_0$ has very strong $\bk$-dependence. 
However it may be useful to $\bk$-sum ARPES data
$N_p(\omega) = \sum_\bk I_0(\bk)\akw$ in an attempt to simulate the
angle-integrated photoemission intensity.

The important sum rule for ARPES is
\begin{equation}\label{nofk}
\fullint f(\omega) \akw = n(\bk),
\end{equation}
which directly relates the energy-integrated ARPES intensity
to the momentum distribution
$n(\bk) = \langle c^\dagger_\bk c_\bk \rangle$.
(The sum over spins is omitted for simplicity).
Somewhat surprisingly, the usefulness of this sum rule 
has been overlooked in the ARPES literature prior to Ref.~\cite{NK}.

We first focus on the Fermi surface $\bk = \bk_F$.
One of the major issues, that we will return to several times
in the remainder of this article, will be the question of how to
define ``$\bk_F$'' at finite temperatures in a strongly correlated
system which may not even have well-defined quasiparticle excitations,
and how to determine it experimentally. For now, we simply
define the Fermi surface to be the locus of gapless excitations in
$\bk$-space in the normal state, so that $A(\bk_F,\omega)$ has a
peak at $\omega = 0$.

To make further progress with Eq.~(\ref{nofk}), we need to
make a weak particle-hole symmetry assumption:
$A(\bk_F,-\omega) = A(\bk_F,\omega)$ for ``small'' $\omega$,
where ``small'' means those frequencies for which
there is significant $T$-dependence in the spectral function.
It then follows that \cite{NK}
$\partial n(\bk_F)/\partial T = 0$, i.e.,
{\it the integrated area under the EDC at} $\bk_F$
{\it is independent of temperature}.
To see this, rewrite Eq.~(\ref{nofk}) as
$n(\bk_F) = {1\over 2} - 
{1 \over 2}\int_0^\infty d\omega\tanh\left(\omega/2T\right)
\left[A(\bk_F,\omega) - A(\bk_F,-\omega) \right]$, and
take its $T$-derivative. It should be emphasized that
we cannot say anything about the {\it value} of $n(\bk_F)$,
only that it is $T$-independent. (A much stronger assumption,
$A(\bk_F,-\omega) = A(\bk_F,\omega)$ for {\it all} $\omega$,
is sufficient to give $n(\bk_F)=1/2$ independent of $T$).
We emphasize the approximate nature of the $\bk_F$-sum-rule
since there is no exact symmetry that enforces it.

We note that we did not make use of any properties of
the spectral function other than the weak particle-hole symmetry assumption,
and to the extent that this is also valid in the superconducting state, our
conclusion $\partial n(\bk_F)/\partial T = 0$ holds equally
well below $T_c$. There is the subtle issue of the meaning of
``$\bk_F$'' in the superconducting state. In analogy with the Fermi surface
 as the
``locus of gapless excitations'' above $T_c$, we can define
the ``minimum gap locus''  below $T_c$. We will describe this
in great detail in Sect. 5.1 below; it suffices to note here that
``$\bk_F$'' is independent of temperature, within experimental
errors, in both the normal and superconducting states of the systems studied
thus far \cite{DING95}.

\begin{figure}
\sidecaption
\includegraphics[width=.5\textwidth]{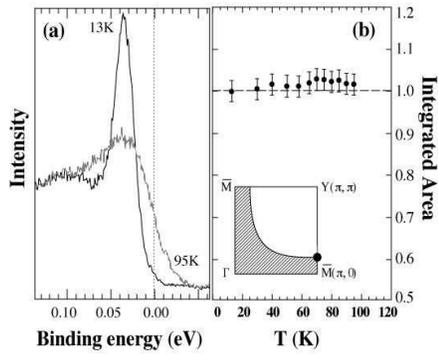}
\caption[]{
(a) ARPES spectra for Bi2212 at the $\bk_F$ point shown in the zone inset
in the normal and
superconducting states. (b) Integrated intensity as a function of 
temperature at the same $k$ point. 
}
\label{edcvstatef}
\end{figure}

In Fig.~\ref{edcvstatef}(a) we show ARPES spectra for near-optimal Bi2212
($T_c=87$ K) at $\bk_F$ along $(\pi,0)$ to $(\pi,\pi)$ at
two temperatures: $T=13$ K, which is well below $T_c$,
and $T=95$ K, which is in the normal state. The two data sets
were normalized in the positive energy region,
which after normalization was chosen to be the common zero baseline.
(The essentially $\omega$-independent emission at positive energies,
which is not cut-off by the Fermi function, is due to higher harmonics of
the incident photon beam, called ``second order light'').
For details, see Ref.~\cite{NK}.

In Fig.~\ref{edcvstatef}(b) we plot the integrated intensity
as a function of $T$ and find that, in spite of the remarkable
changes in the lineshape from 95K to 13K, the integrated intensity 
at $\bk_F$ is very weakly $T$-dependent, verifying the sum rule
$\partial n(\bk_F)/\partial T = 0$. 
The error bars come from the 
normalization due to the low count rate in the $\omega >0$ background.

Let us discuss several other potential complications in testing the
$T$-independence of the integrated intensity at $\bk_F$.
Note that matrix elements have no effect on this result, 
since they are $T$-independent. 
The same is true for a $T$-independent additive extrinsic background.
In an actual experiment the observed intensity 
will involve convolution of Eq.~\ref{intensity} with 
the energy resolution and a sum over the $\bk$-values within the
momentum window. While energy resolution is irrelevant 
to an integrated quantity, sharp $\bk$-resolution is of the essence.

The $T$-independence of the integrated intensity
is insensitive to the choice of the integration cutoff
at negative $\omega$, provided it is chosen beyond the dip feature.
It has been observed experimentally that
the normalized EDCs at $\bk_F$ have identical intensities for
all $\omega < -100$meV. This is quite reasonable,
since we expect the spectral functions to
be the same for energies much larger than the scale associated with
superconductivity. The fact that one has to go to 100meV (much
larger than $T_c \sim 10 meV$)
in order to satisfy the sum rule suggests that electron-electron
interactions are involved in superconductivity.

\subsection{Analysis of ARPES Spectra: EDCs and MDCs}

In the preceding subsections we have presented evidence in favor
of a simple spectral function interpretation of ARPES data on
the quasi 2D high $\tc$ cuprates. In the process, we saw that the ARPES
intensity in Fig.~\ref{edcvstatef} has a nontrivial lineshape which
has a significant temperature dependence. We now introduce some of the
basic ideas which will be used throughout the rest of the article
to analyze and understand the ARPES lineshape.

The one-electron Green's function can be generally written as 
$G^{-1}\bkom = G_0^{-1}\bkom - \Sigma\bkom$ where
$G_0\bkom = 1/[\omega - \varepsilon_\bk]$ is the Green's
function of the noninteracting system, $\varepsilon_\bk$ is the
``bare'' dispersion, and the (complex) self-energy 
$\Sigma\bkom = \Sigma^{\prime}\bkom + i\Sigma^{\prime\prime}\bkom$
encapsulates the effects of all the many-body interactions.
Then using its definition in terms of ${\rm Im}G$, we obtain
the general result
\begin{equation}
\akw={1 \over \pi }{{\Sigma^{\prime\prime}\bkom} \over {\left[
{\omega -\varepsilon_\bk-\Sigma^{\prime}\bkom} \right]^2+
\left[ {\Sigma^{\prime\prime}\bkom} \right]^2}}
\label{akw}
\end{equation}
We emphasize that this expression is entirely general,
and does not make any assumptions about the validity
of perturbation theory or of Fermi liquid theory.

New electron energy analyzers, which measure the photoemitted 
intensity as a function of energy and momentum simultaneously, allow 
the direct visualization of the spectral function, as shown in 
Fig.~\ref{spectralfn}, and have also suggested new ways
of plotting and analyzing ARPES data.\hfill\break
(a) In the traditional energy distribution curves (EDCs), the measured
intensity $I\bkom$ is plotted as a function of $\omega$ (binding energy) 
for a fixed value of $\bk$; and \hfill\break
(b) In the new \cite{Valla99} momentum distribution curves (MDCs),
$I\bkom$ is plotted at fixed $\omega$ as a function of $\bk$.

\begin{figure}
\sidecaption
\includegraphics[width=.5\textwidth]{spectralfn.epsf}
\caption[]{
(a) The ARPES intensity as a function of ${\bf k}$ and $\omega$ at
$h\nu$=22eV and T=40K. main is the main band, and SL a superlattice image.
(b) A constant $\omega$ cut (MDC) from (a).
(c) A constant ${\bf k}$ cut (EDC) from (a).
The diagonal line in the zone inset shows the location of the ${\bf k}$
cut; the curved line is the Fermi surface.
}
\label{spectralfn}
\end{figure}

Until a few years ago, the only data available were in the form of
EDCs, and even today this is the most useful way to analyze
data corresponding to gapped states (superconducting and pseudogap
phases). These analyses will be discussed in great detail in subsequent 
sections. We note here some of the issues in analyzing EDCs and then contrast 
them with the MDCs. First, note that the EDC lineshape is non-Lorentzian
as a function of $\omega$ for two reasons. The trivial reason is the 
asymmetry introduced by the Fermi function $f(\omega)$ which chops off
the positive $\omega$ part of the spectral function. (We will discuss
later on ways of eliminating the effect of the Fermi function).
The more significant reason is that the self energy has non-trivial
$\omega$ dependence and this makes even the full $A\bkom$ non-Lorentzian
in $\omega$ as seen from Eq.~\ref{akw}. Thus one is usually forced to
model the self energy and make fits to the EDCs. At this point one is
further hampered by the lack of detailed knowledge of the additive
extrinsic background which itself has $\omega$-dependence. (Although,
as we shall see, the MDC analysis give a new way of determining 
this background).

The MDCs obtained from the new analyzers have certain
advantages in studying gapless excitations near the Fermi surface
\cite{Valla99,Kink,Adam01}.
In an MDC the intensity is plotted as a function of $k$ varying normal 
to the Fermi surface in the vicinity of a fixed $\bk_F(\theta)$,
where $\theta$ is the angle parametrizing the Fermi surface. 
For $k$ near $k_F$ we may linearize the bare dispersion
$\varepsilon_\bk \simeq v_F^0(k - k_F)$, where $v_F^0(\theta)$ is
the bare Fermi velocity. We will not explicitly show the $\theta$
dependences of $k_F$, $v_F^0$, or other quantities considered
below, in order to simplify the notation.

Next we make certain simplifying assumptions about the remaining
$\bk$-dependences in the intensity $I\bkom$. We assume that: 
(i) the self-energy $\Sigma$ is essentially independent of 
$k$ normal to the Fermi surface, but can have arbitrary dependence
on $\theta$ along the Fermi surface; and (ii) the prefactor $I_0(\bk)$ does
not have significant $k$ dependence over the range of interest.
It is then easy to see from Eqs.~\ref{intensity} and \ref{akw}
that $I\bkom$ plotted as function $k$ (with fixed 
$\omega$ and $\theta$) has the following lineshape.
The MDC is a Lorentzian: \hfill\break
(a) centered at $k = k_F + [\omega - \Sigma^\prime(\omega)]/v_F^0$; with
\hfill\break
(b) width (HWHM) $W_M = |\Sigma^{\prime\prime}(\omega)|/v_F^0$.

Thus the MDC has a very simple lineshape, and its peak position
gives the renormalized dispersion, while its width is proportional
to the imaginary self energy. The consistency of the assumptions made 
in reaching this conclusion may be tested by
simply checking whether the MDC lineshape is fit by a Lorentzian or not.
Experimentally, excellent Lorentzian fits are invariably obtained
(except when one is very near the bottom of the ``band'' or in a 
gapped state\cite{MDC01}).

Finally, note that the external background in the case of MDCs is
also very simple. One can fit the MDC (at each $\omega$)
to a Lorentzian plus a constant (at worst Lorentzian plus linear in $k$) 
background. From this one obtains the value of the external background
including its $\omega$ dependence. Now this $\omega$-dependent 
background can be subtracted off from the EDC also, if one wishes
to. Note that estimating this background
was not possible from an analysis of the EDCs alone.

\section{The Valence Band}

The basic unit common to all cuprates is the copper-oxide plane, $CuO_{2}$. 
Some compounds have a tetragonal cell, $a = b$, such as the $Tl$ compounds, 
but most have an orthorhombic cell, with 
$a$ and $b$ differing by as much as 3\% in YBCO. 
There are two notations used in the literature for the reciprocal cell.
The one used here, appropriate for Bi2212 and Bi2201, has $\Gamma-M$ along 
the $Cu-O$ bond direction, with $M \equiv (\pi,0)$, and $\Gamma-X(Y)$ along 
the diagonal, with $Y \equiv (\pi,\pi)$.  The other notation, appropriate to 
YBCO, has $\Gamma-X(Y)$ along the $Cu-O$ bond direction and 
$\Gamma-S$ along the diagonal.  This difference occurs 
because the orthorhombic distortion in one 
compound is rotated $45^{\circ}$ with respect to the other.
The main effect of the orthorhombicity in Bi2212 and Bi2201 is
the superlattice modulation along the $b$ axis, with $Q_{SL}$ parallel to
$\Gamma-Y$.  Except when refering to this modulation, we will assume tetragonal
symmetry in our discussions.  For a complete review of the electronic structure
of the cuprates, see Ref.~\cite{PICKETT}.
 
The Cu ions are four fold coordinated to planar oxygens.  Apical (out
of plane oxygens) exist in some structures (LSCO), but not in others.  Either
way, the apical bond distance is considerably longer than the planar one,
so in all cases, the cubic point group symmetry of the Cu ions is lowered, 
leading to the highest energy Cu state having $d_{x^2-y^2}$
symmetry.  As the atomic $3d$ and $2p$ states are nearly degenerate,
a characteristic which distinguishes cuprates from other 3d transition
metal oxides, the net result is a strong bonding-antibonding splitting
of the Cu $3d_{x^2-y^2}$ and O $2p$ $\sigma$ states, with all other states
lying in between.  In the stochiometric (undoped) material, Cu is in a $d^{9}$
configuration, leading to the upper (antibonding) state being half filled.
According to band theory, the system should be a metal.  But in the
undoped case, integer occupation of atomic orbitals is possible, and
correlations due to the strong on-site Coulomb repulsion on the Cu sites
leads to an insulating state.  That is, the antibonding band
``Mott-Hubbardizes'' and splits into two, one completely filled (lower Hubbard
band), the other completely empty (upper Hubbard band) \cite{PHIL}.

\begin{figure}
\includegraphics[width=1\textwidth]{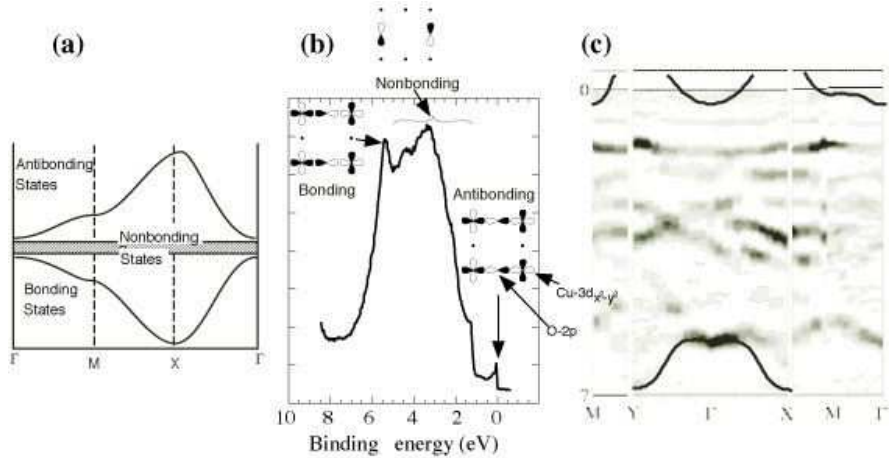}
\caption[]{(a) Simple three band estimate
of the electronic structure 
of the Cu-O plane states; (b) EDC showing the whole valence band at 
the ($\pi,0$) point; c) Intensity map of the whole valence band obtained 
by taking the second derivative of spectra such as the one in (b).  The
orbitals in (b) are based on the three band model,
where black and white lobes corresponding to positive and negative
wavefunctions.}
\label{valenceband}
\end{figure}

On the other hand, for dopings characteristic of the superconducting state,
a large Fermi surface is observed by ARPES (as discussed in the next
section).  Thus, to a first approximation, the basic electronic structure in 
this doping range can be understood from simple band theory considerations.
The simplest approximation is to consider the single Cu $3d_{x^2-y^2}$ 
and two O $2p$ (x,y) orbitals.  The resulting secular equation is \cite{OKA}
\begin{eqnarray}
\left(
\begin{array}{ccc}
                \epsilon_{d} & 2t\sin(k_{x}a) & -2t\sin(k_{y}a) \\
                2t\sin(k_{x}a) & \epsilon_{p} & 0 \\
                -2t\sin(k_{y}a) & 0 & \epsilon_{p}
\end{array}
\right)
\label{secular}
\end{eqnarray}
where $\epsilon_{d}$ is the atomic $3d$ orbital energy, $\epsilon_{p}$ is
the atomic $2p$ orbital energy, and $t$ is the Cu-O hopping integral.
Diagonalization of this equation leads to a non-bonding eigenvalue
$E_{nb}=\epsilon_{p}$, and
\begin{equation}
E_{\pm} = \frac{\epsilon_{p}+\epsilon_{d}}{2} \pm
\sqrt{(\frac{\epsilon_{p}-\epsilon_{d}}{2})^{2} + 4t^{2}(\sin^{2}(k_{x}a)
+ \sin^{2}(k_{y}a))}
\label{eigen}
\end{equation}
where + refers to antibonding, - to bonding.  This dispersion is shown in
Fig.~\ref{valenceband}(a)

In Fig.~\ref{valenceband}(b), we show an ARPES spectrum obtained at the 
$(\pi,0)$ point of the Brillouin zone for Bi2212. Three distinct features 
can be observed: the bonding state at roughly -6eV, the antibonding 
state near the Fermi energy, and the rest of the states in between.  This
rest consists of the non-bonding state mentioned above, as well as the
remainder of the Cu $3d$ and O $2p$ orbitals, plus states originating
from the other (non Cu-O) planes.
It is difficult to identify all of these ``non-bonding'' states, as their 
close proximity and broadness causes them to overlap in energy.  Perhaps 
surprisingly, the overall picture of the electronic structure of the valence 
band has the structure that one would predict by the simple chemical arguments
given above, as shown in Fig.~\ref{valenceband}(c).
The most important conclusion that 
one can derive from Fig~\ref{valenceband} is the early
prediction by Anderson \cite{RVB0}, namely that there is a 
single state relevant to transport and superconducting properties. 
This state, the antibonding state in Fig~\ref{valenceband}, is well 
separated from the rest of the states, and therefore any reasonable 
theoretical description of the physical properties of these novel 
materials should arise from this single state.

Despite these simple considerations, correlation effects do play a major 
role, even in the doped state.  The observed antibonding band width is about 
a factor of
2-3 narrower than that predicted by band theory \cite{OLSON}.  In fact,
the correlation effects are the ones we are most interested in, as they
give rise to the many unusual properties of the cuprates, including the
superconducting state.

As mentioned above, in the magnetic insulating regime, the picture is quite 
different.  The antibonding band splits into two, with the chemical potential 
lying in the gap \cite{WELLS}.  Upon doping with electrons or holes, the 
chemical potential 
would move up or down.  One would then expect to observe 
small Fermi surfaces, hole pockets centered at $(\pi/2,\pi/2)$ with a volume
equal to the hole doping, $x$.  These can be generated at the mean field level
by folding Fig.~\ref{valenceband}(a) back into the magnetic zone
with Q=$(\pi,\pi)$, and turning
on an interaction, $U_{eff}$, between the two folded bands.
This can be contrasted with the large Fermi surface of the unfolded case,
typically centered at $(\pi,\pi)$ with a volume $1+x$.
It is still an open question whether there is a continuous evolution between
these two limits, or whether there is a discontinuous change at 
a metal-insulator transition point. In any case, a 
proper description of the electronic structure must take the 
strong electron-electron correlations into account, even in the 
superconducting regime.

\section{Normal State Dispersion and the Fermi Surface}
\label{NormalState}

The Fermi surface is one of the central concepts in the theory of metals,
with electronic excitations near the Fermi
surface dominating all the low energy properties of the system.
In this Section we describe the use of ARPES to
elucidate the electronic structure and the Fermi surface of the high $T_c$
superconductors.

It is important to discuss these results in detail because ARPES is the
only experimental probe which has yielded useful information about
the electronic structure and the Fermi surface of the planar Cu-O
states which are important for high $T_c$ superconductivity.
Traditional tools for studying the Fermi surface such as
the deHaas-van Alphen effect have not yielded useful information
about the cuprates, because of the need for very high magnetic fields, and
possibly because of the lack of well defined quasiparticles. 
Other Fermi surface probes like positron annihilation are hampered by the 
fact that the positrons appear to preferentially probe spatial regions 
other than the Cu-O planes.

The first issue facing us is: what do we mean by a Fermi surface in a
system at high temperatures where there are no well-defined quasiparticles?
(Recall that quasiparticles, if they exist, manifest themselves as 
sharp peaks in the one-electron spectral function whose width is less 
than their energy, and lead to a jump discontinuity in the momentum 
distribution at $T=0$.)
Clearly, the traditional $T=0$ definition of a Fermi surface defined by the
jump discontinuity in $n(\bk)$ is not useful for the cuprates.
First, the systems of interest are superconducting at low temperatures.
But even samples which have low $\tc$'s have
normal state peak widths at $E_f$ which are an order of
magnitude broader than the temperature\cite{Adam00,Adam01}. 
If, as indicated both
by ARPES and transport, sharp quasiparticle excitations do not exist above
$T_c$, there is no possibility of observing a thermally-smeared,
resolution-broadened, discontinuity in $n(\bk)$.

It is an experimental fact that in the cuprates
ARPES sees broad peaks which disperse as a function of momentum and
go through the chemical potential at a reasonably well-defined momentum.
We can thus adopt a practical definition of the ``Fermi surface'' in
these materials as ``the locus of gapless excitations''. 

Historically, the first attempts to determine the Fermi surface in cuprates
were made on YBCO\cite{CAMPUZANO_90}, however, surface effects as well
as the presence of chains appear to complicate the picture, so we will 
focus principally on Bi2212 and Bi2201, which have been studied the most
intensively. Other cuprates which have also been studied by
ARPES, include the electron-doped material NCCO \cite{NCCO}
and, more recently, LSCO as a function of hole doping \cite{LSCO}.

We discuss below various methods used for the determination of the
spectral function peaks in the vicinity of $E_f$.
In addition, we supplement these methods with momentum distribution studies,
taking due care of matrix element complications.
We will then discuss three topics: the extended saddle-point
in the dispersion, the search for bilayer splitting in Bi2212, and
(in Section 6.4) the doping dependence of the Fermi surface. 

\subsection{Normal State Dispersion in Bi2212: A First Look}

We begin with the results obtained by using the
traditional method of deducing the dispersion and Fermi surface by
studying the EDC peaks as a function of momentum.
This method was used for the cuprates
by Campuzano \etal \cite{CAMPUZANO_90}, Olson \etal \cite{OLSON}, and 
Shen and Dessau \cite{SHEN_REVIEW}, culminating in the  
very detailed study of Ding \etal \cite{DING_96}.
The use of EDC peak dispersion has some limitations
which we discuss below.
Nevertheless, it has led to very considerable understanding
of the overall electronic structure, Fermi surface, and of 
superlattice effects in Bi2212, and therefore it is worthwhile to
review its results first, before turning to more refined methods. 

The main results of Ding \etal \cite{DING_96} on the
electronic dispersion and the Fermi surface
in the normal state ($T = 95$K) of near-optimal OD Bi2212 ($T_c = 87$K)
using incident photon energies of 19 and 22 eV 
are summarized in Fig.~\ref{ns.1}. The peak positions of the EDC's
as a function of $\bk$ are marked with various symbols in 
Fig.~\ref{ns.1}(b). The filled circles are for odd initial states
(relative to the corresponding mirror plane), open circles for even initial
states, and triangles for data taken in a mixed geometry (i.e. the photon
polarization was at $45^\circ$ to the mirror plane).
The Fermi surface crossings corresponding to these dispersing states
are estimated from the ${\bf k}$-point at which the EDC peak positions 
go through the chemical potential when extrapolated from the occupied side. 
The $k_F$ estimates are plotted as open symbols in Fig.~\ref{ns.1}(a).

We use the following square lattice notation
for the 2D Brillouin zone of Bi2212:
$\Gamma {\bar M}$ is along the CuO bond direction, with
$\Gamma = (0,0)$, ${\bar M} = (\pi,0)$, $X=(\pi,-\pi)$ and $Y=(\pi,\pi)$
in units of $1/a^*$, where $a^* = 3.83\AA$ is the separation
between near-neighbor planar Cu ions.
(The orthorhombic a axis is along $X$ and b axis along $Y$).

\begin{figure}
\sidecaption
\includegraphics[width=.5\textwidth]{ns1.epsf}
\caption[]{
Fermi surface (a) and dispersion (b) obtained from normal state
measurements.  The thick lines are obtained by a tight binding fit to the
dispersion data of the main band with the thin lines $(0.21\pi,0.21\pi)$
umklapps and the dashed lines $(\pi,\pi)$ umklapps of the main band.  Open
circles in (a) are the data.  In (b), filled circles are for odd initial states
(relative to the corresponding mirror plane), open circles for even initial
states, and triangles for data taken in a mixed geometry.  The
inset of (b) is a blowup of $\Gamma X$.
}
\label{ns.1}
\end{figure}

In addition to the symbols obtained from data in Fig.~\ref{ns.1}, there 
are also several curves which clarify the significance of all of the 
observed features. The thick curve in Fig.~\ref{ns.1}(b) is $\epsilon(\bk)$, 
a six-parameter tight-binding fit \cite{NORMAN_95a} to the dispersion
data in the $Y$-quadrant; this represents the main CuO$_2$ ``band''.
It cannot be overemphasized that, although this dispersion looks very much
like band theory (except for an overall renormalization of the bandwidth
by a factor of 2 to 3), the actual normal state lineshape is highly
anomalous. As discussed in Section 7 below, there are no well defined
quasiparticles in the normal state.

The thin curves in  Fig.~\ref{ns.1}(b) are $\epsilon(\bk \pm {\bf Q})$, 
obtained by shifting the main band fit by $\pm{\bf Q}$ respectively, where
${\bf Q} = (0.21\pi, 0.21\pi)$ is the superlattice (SL) vector known from 
structural studies \cite{SUPERLATTICE}.
We also have a few data points lying on a dashed curve
$\epsilon(\bk + {\bf K}_\pi)$ with ${\bf K}_\pi = (\pi,\pi)$;
this is the ``shadow band'' discussed below.

The thick curve in Fig.~\ref{ns.1}(a) is the Fermi surface contour
obtained from the main band fit, while the Fermi surfaces 
corresponding to the SL bands are the thin lines and that for
the shadow band is dashed.
It is very important to note that the shifted dispersion curves
and Fermi surfaces provide an excellent description
of the data points that do not lie on main band. 
We note that the main Fermi surface is a large hole-like barrel centered
about the $(\pi,\pi)$ point whose enclosed area
corresponds to approximately 0.17 holes per planar Cu.
One of the key questions is why {\it only one} CuO main band is found in
Bi2212 which is a bilayer material with two CuO planes per unit cell. 
We postpone discussion of this important issue to
end of this Section. 

The ``shadow bands'' seen above, were first observed by Aebi et al. \cite{AEBI}
in ARPES experiments done in a mode similar to the MDCs by
measuring as a function of $\bk$ the intensity 
$\int_{\delta\om} d\om f(\om)\akw$ integrated over a small range 
$\delta\om$ near $\om = 0$. 
The physical origin of these ``shadow bands'' is not certain at the
present time. They were predicted early on
to arise from short ranged antiferromagnetic correlations \cite{KAMPF}.
In this case the effect should become stronger with
underdoping toward the AFM insulator, for which there is little
experimental evidence \cite{DING_97}. An alternative explanation is that
the shadow bands are of structural origin: Bi2212 has a face-centered 
orthorhombic cell with two inequivalent Cu sites per plane, which by itself
could generate a $(\pi,\pi)$ foldback.
Interestingly, it has been recently observed that the shadow band intensity 
is maximal at optimal doping \cite{KORDYUK}.

We now turn to the effect of the superlattice (SL) on the ARPES
spectra.  This is very important, since a lack of understanding of
these effects has led to much confusion regarding such basic
issues as the Fermi surface topology (see below), and 
the anisotropy of the SC gap (see Section 5).
The data strongly suggest \cite{BIO_SL} that these ``SL bands''
arise due to diffraction of the outgoing photoelectron 
off the structural superlattice distortion (which lives
primarily) on the Bi-O layer, thus leading to 
``ghost'' images of the electronic structure at $\epsilon_{\bf k \pm Q}$.

We emphasize the use of polarization selection rules 
(discussed in Section 2.1) for ascertaining various important
points. First, we use them to carefully check the absence of a Fermi
crossing for the main band along $(0,0) - (\pi,0)$, i.e. $\Gamma\bar{M}$. 
Note that the Fermi crossing that we do see near $(\pi,0)$ along
$\Gamma\bar{M}$ in Fig.~\ref{ns.1}(a) is clearly associated with a
superlattice umklapp band, as seen both from the dispersion data
in Fig.~\ref{ns.1}(b) and its polarization analysis.
This Fermi crossing is only seen in the $\Gamma\bar{M}\perp$ (odd)
geometry both in our data and in earlier work \cite{DESSAU_93}.
Emission from the main $d_{x^2 - y^2}$ band,
which is even about $\Gamma\bar{M}$, is dipole forbidden, and one
only observes a weak superlattice signal crossing $E_f$.
(We will return below to newer data at different incident photon energies
where the possibility of a Fermi crossing along $\Gamma\bar{M}$ is
raised again).

Second, we use polarization selection rules 
to disentangle the main and SL bands in the $X$-quadrant
where the main and umklapp Fermi surfaces are very close together; 
see Fig.~\ref{ns.1}(a).
The point is that $\Gamma X$ (together with the $z$-axis) and, similarly
$\Gamma Y$, are mirror planes, and an initial state
arising from an orbital which has $d_{x^2-y^2}$ symmetry about a
planar Cu-site is odd under reflection in these mirror planes.
With the detector placed in the mirror plane the final state
is even, and one expects a dipole-allowed transition when the photon
polarization ${\bf A}$ is perpendicular to (odd about)
the mirror plane, but no emission when the polarization is parallel to
(even about) the mirror plane.
While this selection rule is obeyed along $\Gamma Y$ it is violated
along $\Gamma X$. In fact this apparent violation of selection rules
in the X quadrant was a puzzling feature of all previous
studies \cite{SHEN_REVIEW} of Bi2212.
It was first pointed out in Ref.~\cite{NORMAN_95b}, and
then experimentally verified in Ref.~\cite{DING_96},
that this ``forbidden'' $\Gamma X \vert\vert$ emission
originates from the SL umklapps.
We will come back to the $\Gamma X \vert\vert$ emission in the
superconducting state below.

\subsection{Improved Methods for Fermi Surface Determination}

We now discuss more recently developed methodologies for Fermi surface 
determination. The need for these improvements arises in part
from the practical difficulty of determining precisely where 
in ${\bf k}$-space a state goes through $E_f$.
This problem is particularly severe in the vicinity of the $(\pi,0)$ point
of the zone where one is beset by the following complications in both
Bi2201 and 2212:
 \hfill\break
(1) The normal state ARPES peaks are very broad.
This has important implications about the (non-Fermi-liquid) nature of the
the normal state, as discussed later (section 7).
\hfill\break
(2) The electronic dispersion near $(\pi,0)$
is anomalously flat (``extended saddle-point'').  
\hfill\break
(3) In Bi2212 the superlattice structure complicates the interpretation of 
the data as described above. Fortunately this complication is greatly reduced
in Pb-doped Bi2201 and 2212.
\hfill\break
(4) The final complication comes from the strong variation
of the $\bk$-dependent ARPES matrix elements with incident photon energy.
This makes the use of changes in absolute intensity as a function of $\bk$
to estimate Fermi surface crossings highly questionable.

Note that the first two points are intrinsic problems intimately related to the
physics of high $T_c$ superconductivity, the third is a material-specific 
problem, while the last is specific to the technique of ARPES.
Nevertheless, all of these issues must be dealt with adequately before
ARPES data on Bi2212 can be used to yield definitive results on the
Fermi surface.

{\bf Eliminating the Fermi function:} 
Recall that the peak of the EDC as a function of $\om$ corresponds to that 
of $f(\omega)A\bkom$, which in general does {\it not} coincide with 
the peak of the spectral function $A({\bf k},\omega)$. 
If one has a broad spectral function, which at $k_F$ is centered about 
$\omega = 0$, then the peak of the EDC will be at $\omega < 0$
(positive binding energy), produced by the Fermi function $f(\om)$
chopping off the peak of $A$, in addition to resolution effects.

Since the goal is to study the dispersive peaks in $A({\bf k},\omega)$,
rather than in the EDC, one must effectively  eliminate
the Fermi function from the observed intensity. 
We present two ways of achieving this goal, and illustrate it with
data on Bi2201 where it permits us to study the broad and weakly
dispersive spectral peaks (points (1) and (2) above) near $(\pi,0)$
without the additional complication of the superlattice.

One approach is simply to divide the data by the Fermi function; more
accurately one divides the measured intensity $I(\textbf{k}_F,\omega)$
by $f(\om) \ast R(\om)$,
the convolution of the Fermi function with the energy resolution. 
Although this does not rigorously give the
spectral function (because of the convolution), this is a good
approximation in situations where the energy resolution is very sharp.
An excellent example of such an approach can be seen in the work
of Sato \etal \cite{sato01} on Pb-doped Bi2201
(Bi$_{1.80}$Pb$_{0.38}$Sr$_{2.01}$CuO$_{6-\delta}$)
which is overdoped with a $T_c < 4K$.
In Fig.~\ref{ns.2} we show the raw data in the vicinity of $M$ 
at various temperatures in the left panel and the corresponding
data ``divided by the Fermi function'' in the right panel. The
``divided'' data permits one to clearly observe the dispersion 
well above $E_f$ particularly at high temperatures, and thereby 
identify the Fermi crossings with a great degree of confidence.

\begin{figure}
\includegraphics[width=.8\textwidth]{ns2.epsf}
\sidecaption
\caption{
Left panel: temperature dependence of ARPES intensity of OD Bi2201
($T_c < 4K$) along the $(\pi,0) - (\pi,\pi)$ cut. Right panel:
same divided by the Fermi function at each temperature convoluted with
a Gaussian of width 11 meV (energy resolution). Dotted lines show the
energy above $E_f$ at which the Fermi function takes a value of 0.03.
The solid line in the 140 K data indicates the peak positions obtained
by fitting with a Lorentzian.
}
\label{ns.2}
\end{figure}

An alternative approach \cite{mesot01} to eliminating the Fermi function is to
symmetrize the data. For each ${\bf k}$ define the symmetrized intensity
by $I_{\rm sym}(\textbf{k}_F,\omega)
= I(\textbf{k}_F,\omega) + I(\textbf{k}_F,-\omega)
= I_0 \, A_{sym}(\textbf{k}_F,\omega)$. It is easy to show that
$I_{\rm sym}(\om)$ will exhibit a local minimum, or dip, at $\om = 0$
for an occupied ${\bf k}$ state, while it will show a local maximum
at $\om = 0$ for an unoccupied ${\bf k}$ state.
In practice, then, the Fermi crossing $\bk_F$ is determined as follows:
All EDCs along a cut are symmetrized and $\bk_F$
is identified as the boundary in $\bk$-space between points where
$I_{\rm sym}$ has a local maximum versus a local minimum at $\omega = 0$.
As shown in Ref.~\cite{mesot01}, these arguments work even in the
presence of finite resolution effects. 
We note that this method, and the one presented above, for eliminating the
Fermi function require a very accurate determination of the
chemical potential (zero of binding energy).

\begin{figure}
\includegraphics[width=.8\textwidth]{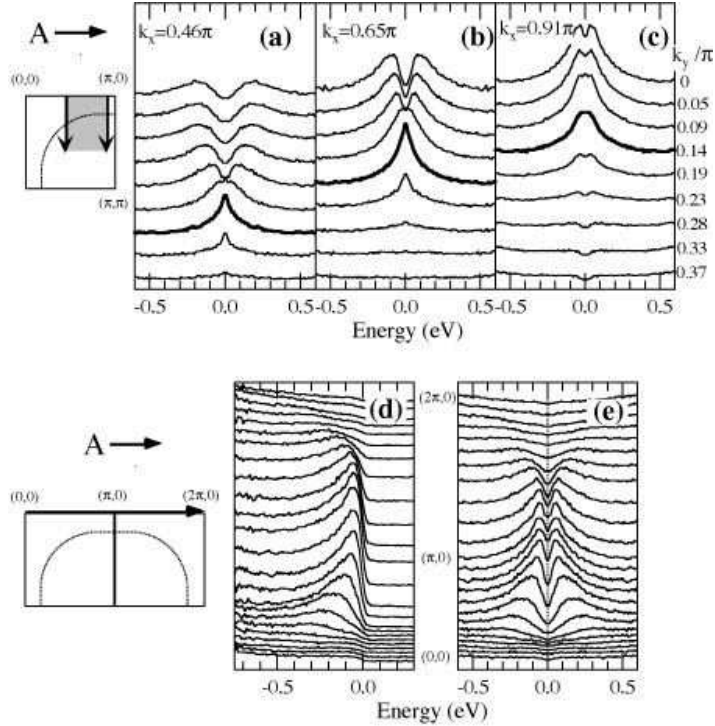}
\caption[]{
(a,b,c) Symmetrized EDCs along cuts parallel to $(\pi,0)$ to 
$(\pi,\pi)$ for OD 23K Bi2201 at $T=25 K$ using $h\nu = 22 eV$
photons. The curve corresponding to $\bk_F$ along each cut is determined 
as explained in the text and is shown by a thick line. The lower panel
(d) shows EDCs along the $(0,0) \to (2\pi,0)$ direction; panel (e) shows
symmetrized curves corresponding to the data in (d) and shows the
absence of a Fermi surface crossing along $(0,0)$ to $(\pi,0)$.
}
\label{ns.3}
\end{figure}

In Fig.~\ref{ns.3} we show the results \cite{mesot01} of a symmetrization
analysis for an OD 23K Bi2201 (Bi$_{1.6}$Pb$_{0.4}$Sr$_{2}$CuO$_{6-\delta}$)
sample. 
From the raw data along cuts parallel to $(\pi,0)$ to $(\pi,\pi)$
(see Fig.~7 of Ref.~\cite{mesot01}) and along $(0,0) \to (2\pi,0)$ 
(shown in Fig.~\ref{ns.3}(d)) one sees broad peaks whose dispersion
is very flat near $(\pi,0)$, thus making it hard to determine
${\bf k}_F$ from EDC dispersion alone.
Nevertheless, the symmetrized data provide completely unambiguous results: 
in the top panels (a,b, and c) of Fig.~\ref{ns.3} 
we illustrate the use of symmetrized data to determine
${\bf k}_F$ using the criterion described above.
Two other features about this analysis are worth noting. 
First, on approaching $\bk_F$ from
the occupied side, resolution effects are expected to 
lead to a flat topped symmetrized spectrum. Second, one expects an
intensity drop in the symmetrized spectrum upon crossing $\bk_F$, assuming
that matrix elements are not strongly $\bk$-dependent. Both of these
effects are indeed seen in the data and further help in deducing $\bk_F$. 

It is equally important to be able to ascertain the {\it absence} of a
Fermi crossing along a cut. 
In this respect, the raw data along $(0,0)-(\pi,0)-(2\pi,0)$ 
in Fig.~\ref{ns.3}(d)is difficult to interpret: the ``flat band'' 
remains extremely close to $E_f$ but does it cross $E_f$? 
It is simple to see from the symmetrized data in Fig.~\ref{ns.3}(e)
the answer is ``no''. The symmetrized data do not show a peak centered 
at $\om = 0$ for any $\bk$, thus establishing
the absence of a Fermi crossing along this cut.

{\bf Intensity Plots:}
A different method for determining the Fermi surface is to make a 
2D plot as a function of $\bk$ of $\int_{\delta\omega} d\omega I(\bk,\omega)$,
the observed intensity integrated over a suitably chosen narrow
energy interval $\delta\omega$ about the Fermi energy.
At first sight this seems to be a very direct way to find out
the $\bk$-space locus on which the low energy excitations live.
However, as we discuss below, one has to be very careful in interpreting
such plots since one is now focusing on the {\it absolute} intensity
of the ARPES signal, which can be strongly affected by the 
$\bk$-dependence of the photoemission matrix elements.

\begin{figure}
\includegraphics[width=.5\textwidth]{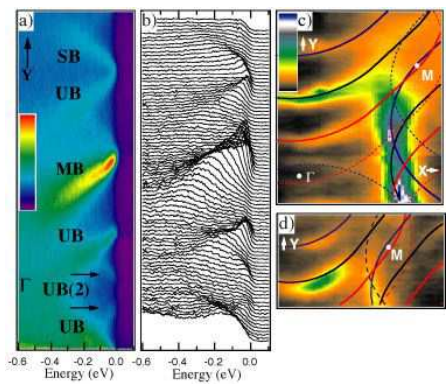}
\caption[]{
(a) Intensity $I({\bf k},\omega)$ and (b) EDCs along $\Gamma$Y
measured on an optimally doped sample (T$_{c}$=90K) at T=40K with 33 eV
photons polarized along $\Gamma$X.
(c) Integrated intensity (-100 to +100 meV) covering the X
and Y quadrants of the Brillouin zone. Data were collected on a regular
lattice of ${\bf k}$ points (spacing $1^{\circ}$ along $\Gamma$X and
$0.26^{\circ}$ along $\Gamma$Y).
(d) Integrated intensity ($\pm$ 40 meV) as in (c), but in the normal
state (T = 150K).
Overlaid on (c) is the main band (black), $\pm$ umklapps (blue/red),
and $\pm$ 2nd order umklapps (dashed blue/red lines)
Fermi surfaces from a tight binding fit \protect\cite{NORMAN_95a}.
}
\label{ns.4}
\end{figure}

This approach was pioneered in the cuprates by Aebi and coworkers
\cite{AEBI}, and in recent times with the availability of Scienta
detectors with dense $\bk$-sampling it has been used by several 
groups \cite{saini97,comment99,chuang99,fretwell00,borisenko00}. 
We show as an example in Fig.~\ref{ns.4}
results from our group \cite{fretwell00} 
on an optimally doped $T_c = 90$K Bi2212 sample \cite{caveat}
using $33$eV incident photons.

Before commenting on the controversies about the Fermi surface crossing 
near the $(\pi,0)$ point, we first examine EDCs along the
$\Gamma$Y direction (middle panel of Fig.~\ref{ns.4},
where the left panel shows a two dimensional plot of the energy and
momentum dependent intensity of photoelectrons along the $\Gamma$Y cut).
All of the features -- main band (MB), superlattice umklapp bands
(UB) and the shadow band (SB) -- seen before \cite{DING_96} and discussed above
are confirmed. We also see weaker, second order 
umklapps from the superlattice (corresponding to $\pm(0.42,0.42)$, twice
the superlattice wavevector), which confirms the diffraction origin of the
superlattice bands. 

We now turn to panels (c,d) of Fig.~\ref{ns.4}
where we plot the integrated intensity within a $\pm$100 meV window
about the chemical potential. We note the very rapid suppression of
intensity beyond $\sim0.8\Gamma$M \cite{saini97}, which does not occur in
data taken with 22eV incident photons. This has led some authors
\cite{chuang99,feng99} to suggest the existence of an electron-like Fermi 
surface with a crossing at this point.
However, Fretwell \etal \cite{fretwell00} and independently
Borisenko \etal \cite{borisenko00} have argued that
this Fermi crossing along $(0,0)$ to $(\pi,0)$ is actually due to 
one of the umklapp bands, and the near optimally
doped Fermi surface is indeed hole like as earlier shown by
Ding \etal \cite{DING_96}.
\begin{figure}
\includegraphics[width=.5\textwidth]{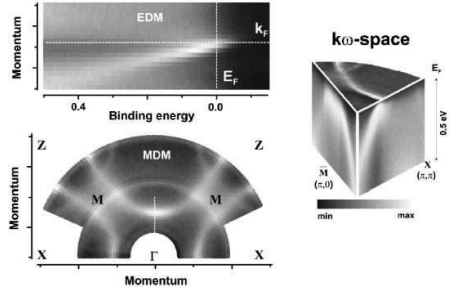}
\caption[]{
Upper panel: Energy distribution map from the $\Gamma-(\pi,\pi)$
direction in the Brillouin zone of Pb-doped BSCCO recorded at room 
temperature. Lower panel: momentum distribution map of Pb-doped BSCCO, 
recorded at
room temperature (raw data). White horizontal
dashed line represents a ${\bf k}_{F}$-EDC, vertical ones
correspond to an $\it{E}_{F}$-MDC. In both cases the gray
scale represents the photoemission intensity as indicated.
The inset shows the three dimensional
$(\it{k}_{x}, \it{k}_{y}, \omega)$ space, which is probed in ARPES of
quasi-2D systems (from Ref.\cite{borisenko00}).
}
\label{borisenko}
\end{figure}

This can be seen most clearly in Pb-doped samples, where the
umklapp bands are not visible.  In 
Fig.~\ref{borisenko}, we show the Fermi energy intensity map of
Borisenko {\it et al.} \cite{borisenko00}, where the hole surface centered 
around $(\pi,\pi)$ and its shadow band partner are quite apparent.

One of the main reasons for the controversy surrounding the
topology of the optimal doped Fermi surface is the fact that data
taken at different incident photon energies $h\nu$ lead to different
intensity patterns. Our assertion, based on Refs.~\cite{fretwell00,mesot01},
is that the superlattice umklapp band is more noticeable at $h\nu =33$eV 
compared with 22 eV since matrix element effects suppress 
the main band intensity at 33 eV. For a detailed discussion about how
to discriminate between a main band and superlattice Fermi crossing,
we refer the reader to the cited papers, and also
to Refs.~\cite{DING_96,mesot01} for the use of polarization selection 
rules for this purpose.

{\bf Matrix Element Effects:}
There is an important general lesson to be learned from the above
discussion which is equally relevant for the $n(\bk)$ methods to be discussed
below. Changes in the ARPES intensities (either integrated over a small
energy window or over a large energy range) as a function of $\bk$ can
be strongly affected by matrix element effects.
This is, of course, obvious from the expression for the
the ARPES intensity:
$I(\textbf{k},\omega) = I_0({\bf k};\nu;\hat{\bf A}) f(\om) A({\bf k},\omega)$.
The key question is: after integration over the appropriate range in $\omega$,
how do we differentiate between the $\bk$ dependence coming from the
matrix elements $I_0$ (which we are not interested in per-se) from the 
$\bk$ dependence coming from the spectral function?

One possibility is to have {\it a priori} information about the
matrix elements from electronic structure calculations \cite{BANSIL}.
But as we now show, even in the absence of such information, one
can experimentally separate the effects of a strong ${\bf k}$-variation
of the matrix element from a true Fermi surface crossing.
The basic idea is to exploit the fact that changing the incident
photon energy one only changes the ARPES matrix elements and not
the spectral function (or the resulting momentum distribution) 
of the initial states.

\begin{figure}
\includegraphics[width=.8\textwidth]{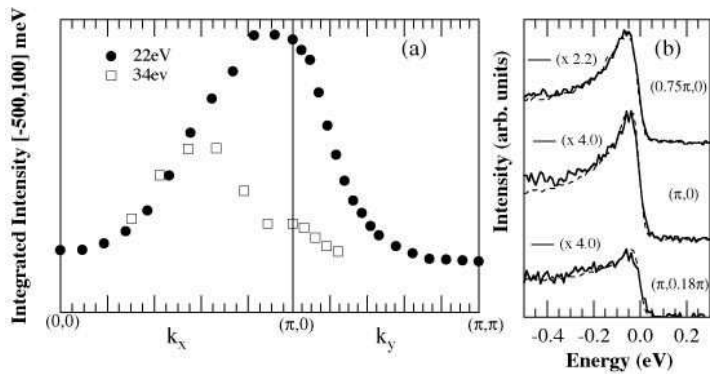}
\caption[]{
Bi2201-OD23K, (a) Integrated intensity 
(over the range is $-500$ to $100$ meV)
along the $(0,0) \to (\pi,0) \to (\pi,\pi)$ directions
for two incident photon energies $h\nu=$22 and 34 eV.
(b) Comparison of the ARPES lineshape measured at 22 (dashed lines) and 34 eV
(solid lines) at three different ${\bf k}$ points. One curve has been scaled 
by the multiplicative constant indicated to make it lie on top of the other.
}
\label{ns.5}
\end{figure}

We will use Bi2201 to illustrate our point since it has
all the complications (points (1),(2) and (4)) listed above without
the superlattice (point (3)). Fig.~\ref{ns.5} (a), from Ref.~\cite{mesot01},
shows the intensity (integrated over a large energy range) as a function
of $\bk$ for OD 23K Bi2201, and highlights the differences between 
data obtained at 22 eV and 34 eV incident photon energies. 
At 22 eV the maximum intensity occurs
close to $(\pi,0)$ and decreases both toward $(0,0)$ and $(\pi,\pi)$,
while at 34 eV there is a strong depression of intensity on
approaching $(\pi,0)$, resulting in a shift of the intensity maximum
away from $(\pi,0)$. 

From the discussion of Section 2.4, we can write the integrated intensity
as: $I(\textbf{k})=\int_{-\infty}^{+\infty}d\omega I(\textbf{k},\omega)
= I_0({\bf k};\nu;{\bf A}) n({\bf k})$.
We attribute this loss in intensity around $(\pi,0)$
at 34 eV seen in Fig.~\ref{ns.5}(a) to
strong ${\bf k}$-dependence of $I_0$ rather than $n(\bk)$.
Experimentally we prove this by showing that
the EDCs at the same point in the Brillouin zone obtained at
the two different photon energies show exactly the
same lineshape, i.e. one can be rescaled onto the other
as shown in Fig.~\ref{ns.5}(b).
As an independent check of this, we have also shown that the
symmetrization analysis leads to the same conclusion that
there is no Fermi crossing along $(0,0) \to (\pi,0)$; see
Fig.~9 of Ref.~\cite{mesot01}.

For completeness, we note that another possible source of incident 
$h\nu$-dependence in ARPES is $k_z$-dispersion. If there was c-axis 
dispersion, different photon energies would probe initial states with
different $k_z$ values consistent with energy conservation.
However the scaling shown in Fig.~\ref{ns.5}(b) proves that it is the
{\it same} two-dimensional ($k_z$-independent) initial state which
is being probed in the data shown here, and the $h\nu$-dependence
arises entirely from the different final states that the matrix element
couples to.

{\bf Methods based on the Momentum Distribution}
Finally we turn to the use of the momentum distribution
sum rule \cite{NK} (discussed in Section 2.4) in determining the Fermi surface.
In principle, locating the rapid variation of $n({\bf k})$ offers a 
very direct probe of the Fermi surface which we emphasize is not 
restricted to Fermi liquids. (The $T=0$ momentum distribution
for known non-Fermi liquid systems, such as Luttinger liquids in
one dimension, do show a inflection point singularity at ${\bf k}_F$.)
However in practice, one needs to be very careful about the
$\bk$-dependence of the matrix elements, 
as clearly recognized in the original proposal \cite{NK}.

Here we discuss two approaches using the $n({\bf k})$ method to
obtain information about the Fermi surface.
The first method is to study the $\bf k$-space gradient of the
logarithm of the integrated intensity. The second method is to
study the temperature-dependence of the integrated intensity and use the
approximate sum rule \cite{NK} $\partial n(\bk_F)/\partial T = 0$
discussed in Section 2.4.

The gradient method was used in our early work 
\cite{DING_96,CAMPUZANO_96} where ${\bf k}_F$ was estimated from the
location of $\max \left| {\nabla _{\bf k}n({\bf k})} \right|$.
The same method has also been successfully used later by other authors
\cite{HUEFNER2,SCHABEL}.
In the presence of strong matrix element effects, it is even more useful
to plot the magnitude of the logarithmic gradient: 
$|\nabla_{\bf k}I({\bf k})|/I({\bf k})$
which emphasizes the rapid changes in the integrated intensity.
The logarithmic gradient filters out the less abrupt changes in
the matrix elements and helps to focus on the intrinsic variations
in $n({\bf k})$.

\begin{figure}
\includegraphics[width=.7\textwidth]{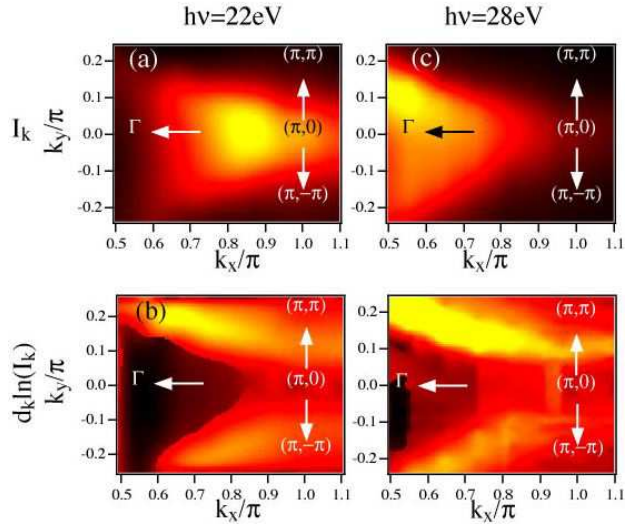}
\caption{
Bi2201-OD0K: (a and c) Integrated intensity (over -350 to +50 meV)
$I({\bf k})$
measured at $\hbar\nu=$22 and 28 eV around the $(\pi,0)$ point. Notice
that the intensity maximum depends strongly upon the photon energy
$\hbar\nu$.(b and d) Corresponding gradient of the logarithm,
$|\nabla_{\bf k}I({\bf k})|/I({\bf k})$, the maxima which
correspond to Fermi crossings and clearly show that, \textit{independent
of the photon energy}, the Fermi surface consists of a hole barrel centered
around $(\pi,\pi)$.
}
\label{ns.6}
\end{figure}

As an example we show in Fig.~\ref{ns.6} the results \cite{mesot01}
of such an analysis for an OD 0K Bi2201 sample. 
In the top panels (a) and (c) we show the integrated intensity 
$I({\bf k})$ around the $(\pi,0)$ point
obtained at two different photon energies: 22 eV and 28 eV respectively.
In the lower panels 11(b) and (d), we plot 
$|\nabla_{\bf k}I({\bf k})|/I({\bf k})$.
Note that there are large differences between the
two top panels, due to different matrix elements at 22 eV and 28 eV.
However, as explained above, the logarithmic gradients in the bottom
panels, which are more influenced by the intrinsic $n(\bk)$,
are much more similar.
The Fermi surface can be clearly seen as two high intensity arcs
curving away from the $(\pi,0)$ point. Once matrix element effects
are taken care of, the Fermi surface results obtained at the 
two different photon energies are quite similar, and 
in good agreement with the results obtained from independent
methods like symmetrization on the same data set \cite{mesot01}.

\begin{figure}
\sidecaption
\includegraphics[width=.5\textwidth]{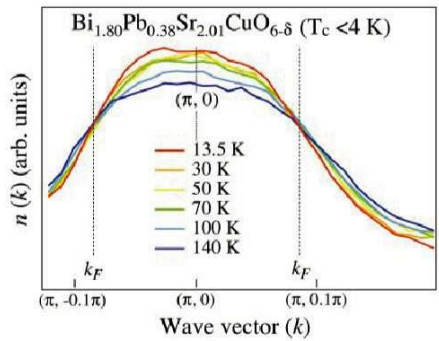}
\caption[]{
Temperature dependence of the integrated ARPES intensity $I(\bk)$
along the $(\pi,0)-(\pi,\pi)$ cut, obtained by integrating the
ARPES intensity from $- 100$ to $100$ meV. (An angle-integrated
temperature-independent background is subtracted before integration).
}
\label{ns.7}
\end{figure}

Our final method for the determination of a Fermi crossing goes back to
the sum rule \cite{NK} that we had introduced earlier,
$\partial n(\bk_F)/\partial T = 0$. Assuming that the matrix elements
are $T$-independent on the temperature scales of interest, this immediately
implies the integrated intensity at (and only at) $\bk_F$ is $T$-independent.
In Section 2.4,  we had used this sum rule to get confidence in the
validity of the single-particle spectral function interpretation of ARPES
by verifying at $\bk_F$ assuming that $\bk_F$ was known (by some other means).
Now we can invert the logic: we can look at the $T$ dependence of the
integrated intensity, and identify $\bk_F$ as that point in $\bk$-space
where the integrated intensity is $T$-independent.
This is illustrated in Fig.~\ref{ns.7} from the work of
Sato et al. \cite{sato01}, who determine $\bk_F$ along the $(\pi,0) - (\pi,\pi)$
cut for a highly OD Bi2201 sample. The result is completely
consistent with that obtained by other methods, such as
``division by the Fermi function'' on the same sample (see Fig.~\ref{ns.2}).

{\bf EDCs vs MDCs:}
To conclude this discussion, let us note a very recent development
for determining $k_F$ and the near-$E_f$ dispersion based
on the MDCs, which are plots of the ARPES intensity as a function of
$k$ (in this case normal to the expected Fermi surface), at various
fixed values of $\omega$. As shown in Section 2.5, the MDC peak position
in the vicinity of the Fermi surface, i.e, near $(\bk = \bk_F,\omega =0)$
is given by: $k = k_F + [\omega - \Sigma^\prime(\omega)]/v_F^0$.

Thus $k_F$ is determined by the peak location of the MDC
at $\omega = 0$. The fully renormalized Fermi velocity
$v_F = v_F^0/[1 - \partial\Sigma^\prime/\partial\om]$ is given by
the slope of the MDC peak dispersion. We note that the 
factor arising from the $k$-dependence of the self-energy is
already included in $v_f^0$, so that
$v_f^0 = v_f^{\rm bare} [1 + \partial\Sigma^\prime/\partial\varepsilon_\bk]$.
(To see this, note that the analysis of Section 2.5
can be easily generalized to retain the first order term
$\partial\Sigma^\prime/\partial\varepsilon_\bk$ without spoiling the
Lorentzian lineshape of the MDC provided this $k$-dependence does not
enter $\Sigma^{\prime\prime}$).

As discussed earlier, the above results derive from the Lorentzian lineshape
of the MDC which arises when three conditions are satisfied: 
the matrix elements do not depend on $k$, the self energy
does not depend on $k$ (except for the $(k-k_F)$ variation of
$\Sigma^\prime$ noted above) and dispersion can be 
linearized near the Fermi surface. The validity of these assumptions
can be checked self-consistently by the Lorentzian MDC lineshape and
the dispersion deduced from the data.

The significance of this approach is that, as emphasized by
Kaminski \etal \cite{Adam01}, the dispersions of the EDC and MDC 
peak positions are actually different in the cuprates;
see Fig.~\ref{fig8.17}(a) in Section 7.7. 
This difference arises due to
the non-Fermi liquid nature of the normal state, so that the
EDC peak dispersion is {\it not} given by the condition 
$\omega - v_F^0(k-k_F) -\Sigma^\prime =0$ but also involves
in general $\Sigma^{\prime\prime}$. In contrast the MDC peak
dispersion is rigorously described by the expression described above,
and is much simpler to interpret.

We expect that the MDC method for determining the dispersion
and $k_F$ which has thus far been used mainly along the zone
diagonal, will eventually be the method of choice, except when
one is very close to the bottom of the band where linearization
fails.


\subsection{Summary of Results on the Optimally Doped Fermi Surface:}

We have discussed a large number of methods for determination of the
Fermi surface in Bi2201 and Bi2212 in the previous subsection. 
These include: (a) dispersion of EDC peaks through $E_f$,
(b) dispersion of peaks after division by the Fermi function,
(c) symmetrization, (d) maps of intensities at $E_f$, (e) gradient
of $n(\bk)$, (f) $T$-dependence of $n(\bk)$, and (g) MDC dispersion.
In addition we also discussed using the $h\nu$-dependence of the data and 
polarization selection rules to eliminate matrix element effects 
and to identify superlattice Fermi crossings.

The reader might well ask: why so many different methods? The reason
is that the development of all of these methods has taken place to deal 
with the complications of accurately identifying the Fermi surface 
in the presence of the four problems listed at the beginning of the
preceding subsection. Each method has its pros and cons, so that some, 
like (b) and (c) require very accurate $E_f$ determination, which is
not the case in (e) and (f) which use energy-integrated intensities.
Most methods require dense sampling in $\bf k$ space, while method (f)
requires in addition data at several temperatures.

Given the complications of the problem at hand it is important to
look for crosschecks and consistency between various ways of 
determining the Fermi surface. We believe that for optimally doped
Bi2201 and 2212 there is unambiguous evidence for a single hole barrel
centered about the $(\pi,\pi)$ point enclosing a Luttinger volume
of $(1+x)$ holes where $x$ is the hole doping.
We discuss further below the issues of the doping-dependence of the
Fermi surface and of bilayer splitting in Bi2212.


\subsection{Extended Saddle Point Singularity}

The very flat dispersion near the $(\pi,0)$ point 
observed in all of the data is striking.
Specifically, along $(0,0)$ to $(\pi,0)$ there is an
intense spectral peak corresponding to the main band, which disperses
toward $E_F$  but stays just below it.
This is often called the ``flat band'' or ``extended saddle point'', 
and appears to exist in all cuprates, 
though at different binding energies in different materials 
\cite{GOFRON,DESSAU_93,SHEN_REVIEW}.

In our opinion this flat band is not a consequence of the
bare electronic structure, but rather a many-body effect,
because a tight-binding description of
such a dispersion requires fine-tuning (of the ratio of the next-near
neighbour hopping to the near-neighbour hopping) which would be
unnatural even in one material, let alone many.

An important related issue is whether this flat band leads to a singular
density of states.  It is very important to recognize that,
while Fig.~\ref{ns.1}(b) {\it looks like} a conventional band structure,
the dispersing states whose ``peak positions'' are plotted
are extremely broad, with a width comparable to binding energy,
and these simply cannot be thought of as quasiparticles.
This general point is true at all $\bk$'s, but specifically for
the flat band region it has the effect of spreading out the spectral weight
over such a broad energy range that any singularity in the DOS
would be washed out. This is entirely consistent with the fact that
other probes (tunneling, optics, etc.) do not find any evidence for a
singular density of states either.

\subsection{Bilayer Splitting?}

On very general grounds, one expects that the two CuO$_2$ layers in
a unit cell of Bi2212 should hybridize to produce two electronic
states which are even and odd under reflection in a mirror plane
mid-way between the layers. Where are these two states? Why then did we
find only one main ``band'' and only one Fermi surface in Bi2212? 

Let us first recall the predictions of electronic structure
calculations \cite{BAND_THEORY}.
In systems like Bi2212, the intra-bilayer hopping as a function 
of the in-plane momentum $\bk$ is of the form \cite{OKA,ILT}
$t_\perp(\bk) = - t_z (\cos k_x - \cos k_y)^2$.
Thus the two bilayer states are necessarily degenerate along the
zone diagonal. However they should have a maximum splitting
at $\bar M = (\pi,0)$ of order 0.25 eV, which may be somewhat
reduced by many-body interactions.

Depending on the exact doping levels and on the presence of Bi-O Fermi
surface pockets, 
which are neither treated accurately in the theory nor
observed in the ARPES data, we must obtain one of the two following 
situations: (1) the bilayer antibonding (A) state is unoccupied
while the bonding (B) state is occupied at $(\pi,0)$. This would lead to
an A Fermi crossing along $(0,0) - (\pi,0)$ and a B Fermi crossing
along $(\pi,0) - (\pi,\pi)$. As described at great length above we
did not find evidence for a main band Fermi crossing along
$(0,0) - (\pi,0)$ at least for the near optimal doped sample, therefore
this possibility is ruled out.

(2) The second possibility is that both the A and B bilayer states are
occupied at the $(\pi,0)$. In this case, there should be two (in principle,
distinct) Fermi crossings along $(\pi,0) - (\pi,\pi)$, although they might
be difficult to resolve in practice. Nevertheless, one would definitely 
expect to see two distinct occupied states at the $(\pi,0)$ point.
Unfortunately, the normal state spectrum at $(\pi,0)$ is so broad
in the optimally doped and underdoped materials that it is hard to make a 
clear case for bilayer splitting. Thus an effort was made to search for
this effect in the superconducting state at $T \ll T_c$, 
when a sharp feature (quasiparticle peak) is seen (see
Fig.~\ref{ns.8}) and one might hope that the bilayer splitting should 
be readily observable. 

\begin{figure}
\sidecaption
\includegraphics[width=.5\textwidth]{ns8.epsf}
\caption{
Low temperature (T=13K) EDC's of near optimal
$T_c = 87$K Bi2212 at $(\pi,0)$ for various
incident photon angles.  The solid (dashed) line is $18^\circ$ ($85^\circ$) from the normal.  The inset shows the height of the sharp peak for data
normalized to the broad bump, at different incident angles.
}
\label{ns.8}
\end{figure}

The issue then is how to interpret the peak-dip-hump structure
seen in the ARPES lineshape at $(\pi,0)$ in Fig.~\ref{ns.8}. 
The peak-dip-hump structure will be discussed at length in
Section 7 below. Nevertheless, here we will briefly
address the question of whether:
(I) the peak and the hump are the two bilayer split states,
which are resolved below $T_c$ once the peak becomes sharp?
Or (II) is the non-trivial line shape due to many-body effects 
in a single spectral function $\akw$?  

Three pieces of evidence will be offered in favor of hypothesis
(II) as opposed to (I), so that no bilayer spliting is observable
even in the superconducting state of near optimal doped Bi2212.
The first evidence comes from studying the polarization
dependence of the ARPES matrix elements.
For case (I) there are two independent matrix elements which,
in general, should vary differently with photon polarization
${\bf A}$, and thus the intensities of the two features should 
vary independently as ${\bf A}$ is varied. On the other hand,
for case (II), the intensities of the two features are
governed by a single matrix element. As shown in Fig.~\ref{ns.8}
it was found in Ref.~\cite{DING_96} that by varying the
z-component of ${\bf A}$, the peak and hump intensities scale 
together, and thus the peak-dip-hump are all part of a 
single spectral function for Bi2212. 

A second piece of evidence comes from a comparison \cite{Mike97}
of the normal and superconducting state dispersions near the $(\pi,0)$
point, which will be discussed in detail in connection with
Figs.~\ref{fig8.8} and \ref{fig8.9} of Section 7.4. From these 
data, we argue that there is no evidence for a feature above $T_c$, which
would correspond to the dispersionless quasiparticle peak below $T_c$.
Thus the dispersionless peak must be of many-body origin. 

The third and final piece of evidence comes from both ARPES and
SIS tunneling. In the ARPES data \cite{JC99} shown in Fig.~\ref{fig8.15}
of Section 7.6 one sees the striking fact that while
the energy scales of both the quasiparticle peak and the $(\pi,0)$ hump
increase with underdoping, their ratio is essentialy doping independent.
Since the location of the peak is the (maximum) superconducting gap
-- as discussed in detail in Section 5 -- the $(\pi,0)$ hump 
energy scales with the gap. SIS tunneling data also finds the same 
correlation on a very wide range of materials (including some which 
have a single CuO plane) whose gap energies vary by a factor of 
30 \cite{JohnZ96}. This provides very strong evidence that both the peak 
and hump are related to many-body effects and not manifestations of 
bilayer spliting.

We note that Anderson \cite{ANDERSON} had predicted that
many-body effects within a single layer could
destroy both the quasiparticles and the coherent bilayer splitting
{\it in the normal state}.
But why the splitting should not be visible in the superconducting
state, where sharp quasiparticles do exist, is not clear from a theoretical  
point of view.

Recently, the above picture has been challenged by a number of
authors \cite{BILAY1,BILAY2,BILAY3}.  What has become clear is that bilayer 
splitting is indeed present for heavily overdoped Bi2212 samples, and has been 
seen now by several groups, including our own.  In Fig.~\ref{feng1}, we
show (a) the bilayer split Fermi surfaces and (b) the bilayer split EDCs
observed by the Stanford group for a heavily overdoped ($T_{c}$=65K) Bi2212 
sample.  Note that the bilayer splitting can even be seen in the umklapp
bands.
The resulting dispersion is reproduced in Fig.~\ref{feng2}, where one sees
that the momentum dependence of the splitting follows that expected from 
electronic structure considerations \cite{OKA,ILT}.

\begin{figure}
\sidecaption
\includegraphics[width=.5\textwidth]{feng1.epsf}
\caption{
(a) Bilayer-split Fermi surfaces of heavily overdoped
OD65; the two weaker features are their superstructure counter
parts. Solid and dashed lines represent the bonding and antibonding
Fermi surfaces, respectively. (b) Normal state photoemission spectra
of Bi2212 taken at $(\pi,0)$ for three different doping levels. Data
were taken with $h\nu$ = 22.7 eV photon. Bars indicate identified feature
positions, and triangles indicate possible feature positions. 
(from Ref.~\cite{BILAY1}).
}
\label{feng1}
\end{figure}

\begin{figure}
\sidecaption
\includegraphics[width=.5\textwidth]{feng2.epsf}
\caption{
(a) Dispersion extracted from heavily overdoped OD65. (b) Energy splitting
along the antibonding Fermi surface, which is obtained from data
shown in Fig.\ref{feng1}. It is simply the binding energy of the bonding
band, since the binding energy of the antibonding band is zero at its Fermi 
surface. The curve is 
$\frac{1}{2} t_{\bot ,\exp} [\cos (k_xa)-\cos (k_ya)]^2$, 
where $t_{\bot ,\exp }$ = 44 $\pm$5 meV.
Error bars are due to the uncertainties in determining the energy
position (from Ref.\cite{BILAY1}).
}
\label{feng2}
\end{figure}

How this effect
evolves as a function of doping, though, is still controversial.  In
particular, if it is present for optimal doping, it is difficult to resolve.
Moreover, some authors who advocate bilayer splitting still argue that
the peak/dip/hump structure in the superconducting state is largely a
many-body effect \cite{BILAY1} as we adovcate here.
This is supported by the fact that the same lineshape is seen in trilayer
Bi2223 \cite{TRILAYER}, where one would expect three features if layer 
splitting were causing these effects.  And similar lineshapes have been
seen by tunneling in single layer systems, in support of a many-body
interpretation.
The reader is refered to Ref.~\cite{RMP}, where the issue of bilayer
splitting is discussed in greater detail than here.

\section{Superconducting Energy Gap}

In this Section, we will first establish how the superconducting 
(SC) gap manifests 
itself in ARPES spectra, and then directly map out its variation with 
$\bk$ along the Fermi surface. This is the only available technique for 
measuring the momentum dependence of the energy gap, and complements 
phase-sensitive tests of the order parameter symmetry \cite{OP_SYMMETRY}.
Thus ARPES has played an important role \cite{SHEN_93}, \cite{DING_GAP} 
in establishing the $d$-wave order 
parameter in the high $T_c$ superconductors \cite{OP_SYMMETRY}. 
At the end of the Section, we will discuss the doping dependence of the
SC gap and its anisotropy, and the implications of this study
for various low temperature experiments like thermal conductivity and
penetration depth.

\subsection{Particle-Hole Mixing}

To set the stage for the experimental results it may be useful to recall 
particle-hole (p-h) mixing in the BCS framework (even though, as we shall 
see in Section 7, there are aspects of the data which are dominated 
by many body effects beyond weak coupling BCS theory). 
The $BCS$ spectral function is given by 
\begin{equation}
A({\bf k},{\omega})=
u_{\bf k}^2 {\Gamma}/{\pi}(({\omega}-E_{\bf k})^2+{\Gamma}^2)
+ v^2_{\bf k}{\Gamma}/{\pi}(({\omega}+
E_{\bf k})^2+{\Gamma}^2)
\label{bcsakw}
\end{equation}
where the coherence factors are 
$v^2_{\bf k}=1-u_{\bf k}^2={1\over2}(1-{\epsilon}_{\bf k}/E_{\bf k})$
and $\Gamma$ is a phenomenological linewidth. The normal state energy 
${\epsilon}_{\bf k}$ is measured from  $E_f$ and the Bogoliubov 
quasiparticle energy is 
$E_{\bf k}=\sqrt{\epsilon_{\bf k}^2+\vert\Delta({\bf k})\vert^2}$, 
where $\Delta({\bf k})$ is the gap function. Note that only the second 
term in Eq.~\ref{bcsakw}, with the $v_{\bf k}$-coefficient, would be 
expected to make a significant contribution to the EDCs at low 
temperatures.  

\begin{figure}
\sidecaption
\includegraphics[width=.3\textwidth]{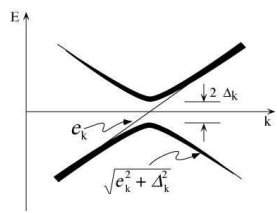}
\caption[]{
Schematic dispersion in the normal (thin line) and 
superconducting (thick lines) states following BCS theory.
The thickness of the superconducting state lines indicate 
the spectral weight given by the BCS coherence factors $u$ and $v$.
}
\label{bcsmodel}
\end{figure}

In the normal state above $T_c$, the peak of $A({\bf k},{\omega})$ is 
at $\omega = \epsilon_{\bf k}$ as can be seen by setting $\Delta = 0$ in 
Eq.~\ref{bcsakw}. We would thus expect to see in ARPES a spectral peak 
which disperses through zero binding energy as ${\bf k}$ goes through 
${\bf k}_F$ (the Fermi surface). In the superconducting state, the 
spectrum changes from $\epsilon_{\bf k}$ to $E_{\bf k}$; see 
Fig.~\ref{bcsmodel}. As ${\bf k}$ approaches the Fermi surface the 
spectral peak shifts towards lower binding energy, but no longer crosses 
$E_f$. Precisely at ${\bf k}_F$ the peak is at 
$\omega = |\Delta({\bf k}_F)|$, which is the closest it gets to $E_f$. 
This is the manifestation of the gap in ARPES.
Further, as ${\bf k}$ goes beyond ${\bf k}_F$,
in the region of states which were unoccupied above $T_c$,
the spectral peak {\it disperses back}, receding away from $E_f$, 
although with a decreasing intensity (see Eq.~\ref{bcsakw}). This is
the signature of p-h mixing. 

\begin{figure}
\sidecaption
\includegraphics[width=.5\textwidth]{particlehole.epsf}
\caption[]{
(a) Superconducting state and (b) normal state EDC's for a near 
optimal $T_c = 87$K Bi2212 sample for a set of $\bk$ values
(in units of $1/a$) shown in the Brillouin zone at the top.
Note the different binding energy scales in panels (a) and (b).
}
\label{particlehole}
\end{figure}

\begin{figure}
\includegraphics[width=.6\textwidth]{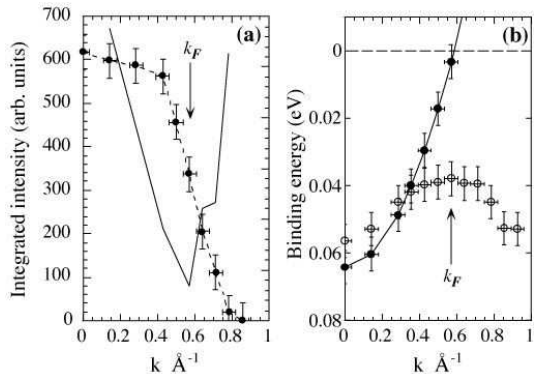}
\caption[]{
(a) Integrated intensity versus ${\bk}$  from the normal
state data of Fig.~\ref{particlehole} (b) shown by black dots
gives information about the momentum distribution $n(\bk)$.
The derivative of the integrated intensity is shown by the
black curve (arbitrary scale). The Fermi surface crossing 
$\bk = \bk_F$  is identified by the minimum in the derivative.
(b) Normal state dispersion (closed circles) and SC state 
dispersion (open circles) obtained from EDC's of Fig.~\ref{particlehole}.
Note the back-bending of the SC state dispersion for $\bk$ beyond
$\bk_F$ which is a clear indication of particle-hole mixing.
The SC state EDC peak position at $\bk_F$ is an estimate of the 
SC gap at that point on the Fermi surface.
}
\label{sc_dispersion}
\end{figure}

Experimental evidence for particle-hole mixing in
the SC state was first given in Ref.~\cite{CAMPUZANO_96}.
In Fig.~\ref{particlehole} we show normal
and SC state spectra for Bi2212 for $\bk$'s along
the cut shown in the inset. In the normal state data in panel (b) we
see the electronic state dispersing through $E_f$: the $\bk$'s go from
occupied (top of panel) to unoccupied states (bottom of panel).
The normal state dispersion is plotted as black dots in
Fig.~\ref{sc_dispersion} (b).
The $\bk_F$ obtained from this dispersion is in agreement with that
estimated from the $|\nabla_{\bk} n(\bk)|$ analysis of the normal state
data shown in Fig.~\ref{sc_dispersion} (a).

We see from Fig.~\ref{particlehole} (a) that the
SC state spectral peaks do not disperse
through the chemical potential, rather they first approach $\om =0$
and then recede away from it.
The difference between the normal and SC state dispersions
is clearly shown in Fig.~\ref{sc_dispersion} (b).

There are three important conclusions to be drawn from 
Fig.~\ref{sc_dispersion} (b). First, the bending back of the SC state 
spectrum for $\bk$ beyond $\bk_F$ is direct evidence for p-h 
mixing in the SC state.  Second, the energy of 
closest approach to $\om=0$ is related to the SC gap that has opened up 
at the FS, and a quantitative estimate of this gap will be described below. 
Third, the location of closest approach to $\om=0$ (``minimum gap'') 
coincides, within experimental uncertainties, with the $\bk_F$ obtained 
from analysis of normal state data. 

In fact by taking cuts in $\bk$-space
which which are perpendicular to the normal state Fermi surface one can
map out the ``minimum gap locus'' in the SC state, or for that matter in 
any gapped state (e.g., the pseudogap regime to be discussed in the following
Section). We emphasize that particle-hole mixing leads to the appearance of
the ``minimum gap locus'' and this locus in the gapped state
gives information about the underlying Fermi surface. 
(By this we mean the Fermi surface on which the SC state gap
appears below $T_c$).
In fact, the observation of p-h mixing in the ARPES spectra
is a clear way of asserting that 
the gap seen by ARPES is due to superconductivity rather than of some 
other origin, e.g., charge- or spin-density wave formation. 

\subsection{Quantitative Gap Estimates}

The first photoemission measurements of the SC gap in the cuprates was 
by Imer \etal \cite{IMER89} using angle-integrated photoemission,
and by Olson and coworkers \cite{olson2} using angle-resolved
photoemission. The first identification of
a large gap anisotropy consistent with d-wave pairing was made by
Shen and coworkers \cite{SHEN_93}. Ding \etal \cite{DING95,DING_GAP}
subsequently made quantitative fits to the SC state spectral function
to study the gap anisotropy in detail.

We now discuss the quantitative extraction of the gap 
at low temperatures ($T \ll T_c$) following
Ding \etal \cite{DING_GAP}. In Fig.~\ref{dwavegap}, 
we show the $T= 13$K EDCs for an 87K $T_c$ sample for 
various points on the main band FS in the $Y$-quadrant. Each spectrum 
shown corresponds to the minimum observable gap along a set of ${\bf k}$ 
points normal to the FS, obtained from a dense sampling of ${\bf k}$-space 
\cite{DENSE}. We used 22 eV photons in a $\Gamma Y\perp$ polarization,
with a 17 meV (FWHM) energy resolution, and a $\bk$-window of radius 
0.045$\pi/a^*$.

\begin{figure}
\sidecaption
\includegraphics[width=.5\textwidth]{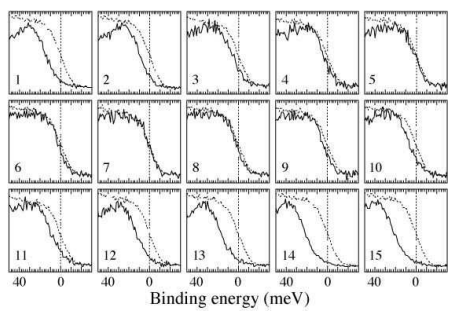}
\caption[]{
Bi2212 spectra (solid lines) for an 87K $T_c$ sample at 13K
and Pt spectra (dashed lines)
versus binding energy (meV) along the Fermi surface in the $Y$ quadrant.
The photon polarization and BZ locations of the data points
are shown in inset to Fig.~\protect\ref{dwavegap}
}
\label{gaparoundfs}
\end{figure}

The simplest gap estimate is
obtained from the mid-point shift of the leading edge of Bi2212
relative to a good metal (here Pt) in electrical contact with the sample.  
This has no obvious quantitative validity,
since the Bi2212 EDC is a spectral function while
the polycrystalline Pt spectrum (dashed curve in Fig.~\ref{gaparoundfs})
is a weighted density of states whose leading edge is an 
energy-resolution limited Fermi function.
We see that the shifts (open circles in Fig.~\ref{dwavegap}) indicate
a highly anisotropic gap which vanishes in the nodal directions,
confirming earlier results by Shen {\it et al.} \cite{SHEN_93}.
These results are qualitatively similar to one obtained
from the fits described below.

Next we turn to modeling \cite{DING95,DING_GAP}
the SC state data in terms of spectral functions.
It is important to ask how can we model the non-trivial line shape
(with the dip-hump structure at high $\om$) in the absence of a 
detailed theory, and, second, how do we deal with the extrinsic background?
We argue as follows:
in the large gap region near $(\pi,0)$, we see a linewidth collapse
for frequencies smaller than $\sim 3\Delta$ upon cooling well below $T_c$.
Thus for estimating the SC gap at the low temperature,
it is sufficient to look at small frequencies, and to focus on
the coherent piece of the spectral function with a resolution-limited
leading edge. 
(Note this argument fails at higher temperatures, e.g., just below
$T_c$).  This coherent piece is modeled by the BCS spectral function, 
Eq.~\ref{bcsakw}.

The effects of experimental resolution are taken into account via
\begin{equation}
{\tilde I}({\bf k},\omega)= I_0 \int_{\delta\bk} d\bk'
\int_{-\infty}^{+\infty} d\omega' R(\omega - \omega') 
f(\omega') A(\bk',\omega')
\end{equation}
where $R(\omega)$, the energy resolution, is a normalized Gaussian 
and $\delta\bk$ is the $\bk$-window of the analyzer.
In so far as the fitting procedure is concerned,
all of the incoherent part of the spectral function
is lumped together with the experimental background into one
function which is added to the ${\tilde I}$ above. 
Since the gap is determined by fitting the
resolution-limited leading edge of the EDC,
its value is insensitive to this drastic simplification.
To check this, we have made an independent set of fits to the 
small gap data where we do not use any background fitting function,
and only try to match the leading edges, not the full spectrum.
The two gap estimates are consistent within a meV.
Once the insensitivity of the gap to the assumed background is
established, there are only two free parameters
in the fit at each $\bk$: the overall intensity $I_0$ and the
gap $|\Delta|$; the dispersion ${\epsilon}_{\bf k}$ is known
from the normal state study, the small linewidth $\Gamma$ is
dominated by the resolution.

The other important question is the justification for using
a coherent spectral function to model the rather broad EDC along 
and near the zone diagonal. As far as the early data
being discussed here is concerned, such a description
is self-consistent \cite{DING95,DING_GAP},
though perhaps not unique,
with the entire width of the EDC accounted for by the
large dispersion (of about 60 meV within our ${\bf k}$-window) 
along the zone diagonal.  More recent data taken along $(0,0)$ to $(\pi,\pi)$
with a momentum resolution of $\delta{\bf k} \simeq 0.01 \pi/a^*$
fully justifies this assumption by resolving coherent nodal quasiparticles
in the SC state; see Section 7.7. 

\begin{figure}
\sidecaption
\includegraphics[width=.5\textwidth]{dwavegap.epsf}
\caption[]{
$Y$ quadrant gap in meV versus angle on the Fermi surface
(filled circles) from fits to the data of Fig.~\protect\ref{gaparoundfs}. 
Open symbols
show leading edge shift with respect to Pt reference.
The solid curve is a d-wave fit to the filled symbols.
}
\label{dwavegap}
\end{figure}

The gaps extracted from fits to the spectra of Fig.~\ref{gaparoundfs}
are shown as filled symbols in Fig.~\ref{dwavegap}.
For a detailed discussion of the the error bars (both on the 
the gap value and on the Fermi surface angle), and also of sample-to-sample
variations in the gap estimates, we refer the reader to 
Ref.~\cite{DING_GAP}.
The angular variation of the gap obtained from the fits
is in excellent agreement with $|\cos(k_x) - \cos(k_y)|$ form.
The ARPES experiment cannot
of course measure the phase of the order parameter, but this result
is in strong support of $d_{x^2-y^2}$ pairing \cite{OP_SYMMETRY}.
Such an order parameter arises naturally in theories with
strong correlations and/or antiferromagnetic spin fluctuations
\cite{DWAVE}.  Moreover, the functional form of the
anisotropy we find is consistent with electrons in the Cooper pair
residing on neighboring Cu sites.  That is, ARPES gives information
on the spatial range of the pair interaction which is difficult to
obtain from other techniques.

Finally we comment on the temperature dependence of the gap.
Unfortunately with increasing temperature the linewidth grows
and a simple BCS-type modeling (valid for $T\ll T_c$)
of just the coherent part of the spectral function is not possible.
While $T$-dependence of the leading edge can certainly be used as
a rough guide, this estimate is affected by both the gap and the
linewidth (the diagonal self-energy). We will discuss methods for
modeling the SC state data in Section 7. 

For completeness, we add a remark clarifying the earlier
observation of two nodes in the $X$-quadrant \cite{DING95},
and the related non-zero gap along $\Gamma X$ in the $\Gamma X \vert\vert$
geometry \cite{DING95,KELLY}.
It was realized soon afterwards that these observations were
related to gaps on the superlattice bands \cite{NORMAN_95b},
and not on the main band. To prove this experimentally, the
$X$-quadrant gap has been studied in the $\Gamma X \perp$ geometry
\cite{DING_GAP} and found to be consistent 
with $Y$-quadrant $d_{x^2-y^2}$ result described above.

\subsection{Doping Dependence of the SC Gap}

There are two important issues to be addressed about the doping
dependence: How does the maximum gap change with doping?
and how does the gap anisotropy evolve with doping?

ARPES results on underdoped samples show that the maximum gap 
{\it increases} \cite{HARRIS_96,SNS97} with decreasing hole
concentration. For a recent compilation of gap versus doping
results \cite{JC99} in Bi2212, see Fig.~\ref{fig8.15} of Section 
7.6.
Identical results have also been obtained by
tunneling spectroscopy \cite{STM_GAP}. 
This was at first quite unexpected since $T_c$ decreases as one underdopes 
from optimality. The fact that $2\Delta/k_B T_c$
is not constant with doping and can become an order of magnitude larger
than its BCS weak-coupling value is very clear evidence that the
SC phase transition on the underdoping side is qualitatively different
from the BCS transition, a point that we will return to in the next 
Section on pseudogaps.

We next turn to the question of the SC gap anisotropy as a function of doping.
In this case it is the behavior of the gap in the vicinity of the node
which is the most important since all low temperature $T \ll T_c$
thermodynamic and transport properties are controlled by thermal 
excitation of quasiparticles in the vicinity of the nodes
(where the SC gap vanishes on the Fermi surface).
These low energy excitations have a Dirac-like spectrum
$E(\bk) = \sqrt{v_F^2 k_{\perp}^2 + v_2^2 k_{\parallel}^2}$ where
$k_{\perp}$ ($k_{\parallel}$) are the components of the $\bk$ perpendicular
(tangential) to the Fermi surface, and measured from the nodal point.
Here $v_F$ is the nodal Fermi velocity controlling the dispersion
perpendicular to the Fermi surface, while 
$v_2 = \partial\Delta_\bk/\partial k_{\parallel}$ is the
the gap slope at the nodal point.

The density of states for low energy excitations is then given by
$N(\om) = 2 \omega/(\pi v_F v_2)$ for $\omega \ll \Delta$
(in units where $\hbar = 1$). This leads to characteristic temperature
dependences for various low temperature properties with 
coefficients essentially determined by $v_F$ and $v_2$.
Specifically for $T \ll T_c$, the specific heat goes like
$C(T) = c_1 T /(v_F v_2)$; the thermal conductivity goes like
$\kappa(T)/T = c_2\left(v_F/v_2 + v_2/v_F\right)$ and the superfluid
density goes like $\rho_s(T) = \rho_s(0) - c_3(v_F/v_2)T$.
Here $c_1$ and $c_2$ are known constants while $c_3$ contains
(apriori unknown) multiplicative Fermi liquid parameters
\cite{MILLIS,DURST}.

Clearly ARPES has a unique ability to independently 
measure both $v_F$, using the dispersion of the nodal quasiparticle, 
and $v_2$, using the slope of the SC gap at the node. One can then
make detailed comparison with various bulk measurements, and
thereby test the validity of the description of the low temperature 
properties of the SC state in terms of weakly interacting quasiparticles
\cite{MILLIS,DURST,CHIAO}. 
This is particularly important given the lack of sharply defined 
quasiparticles in the normal state of the cuprates.

With these motivations in mind, 
a detailed measurement of the shape of the superconducting gap in Bi2212
as a function of doping was carried out by Mesot \etal \cite{MESOT}. 
Although these measurements were carried out at a time when
the energy and momentum resolutions were lower than those
currently available, they still give useful information because the 
gap over the full range of angles over the irreducible zone was 
measured. 
Using the simple BCS spectral function fits (described above) and
taking into account the measured dispersion and the known energy
and momentum resolutions, Mesot \etal found the results 
shown in Fig.~\ref{gapvsdoping}. The simple d-wave gap
$\Delta =\Delta _0\cos (2\phi )$ (Fig.~\ref{dwavegap}) is modified by 
the addition of the first harmonic
$\Delta_{\bf k}=\Delta_{\rm max}[B\cos(2\phi)+(1-B)\cos(6\phi)]$, 
with $0\leq B \leq 1$. 
Note that the $\cos(6\phi)$ term in the Fermi surface harmonics
can be shown to be closely related to the tight binding function
$\cos(2k_x) - \cos(2k_y)$, which represents next nearest neighbours
interaction, just as $\cos(2\phi)$ is closely related to the near neighbor
pairing function $\cos(k_x)-\cos(k_y)$.
From Fig.~\ref{gapvsdoping} we find that while the overdoped
data are consistent with $B \simeq 1$, the parameter $B$ decreases
as a function of underdoping.

\begin{figure}
\sidecaption
\includegraphics[width=.5\textwidth]{gapvsdoping.epsf}
\caption[]{
Values of the superconducting gap as a function of the Fermi surface
angle $\phi$ obtained for a series of Bi2212 samples with varying doping.
Note two different UD75K samples were measured, and the UD83K sample has 
a larger doping due to sample aging\protect\cite{DING_97}. The solid lines 
represent the best fit using the gap function:
$\Delta_{k}=\Delta_{\rm max}[B\cos(2\phi)+(1-B)\cos(6\phi)]$
as explained in the text. The dashed line in the panel of an UD75K sample 
represents the gap function with B=1.
}
\label{gapvsdoping}
\end{figure}

Before discussing the doping dependence of the results in detail,
let us first look at the comparisons between ARPES and other 
probes near optimality. Low temperature specific heat data on Bi2212 
is not available, but the thermal conductivity has been measured
by Taillefer and coworkers \cite{CHIAO} on an optimal $T_c = 89$K sample. 
Now one can use the $\kappa/T$ formula given above with the coefficient
$c_2 = k_B^2 n / 3\hbar d$, where $n/d$ is the stacking density of 
CuO$_2$ planes. It is important to note that $c_2$ is {\it not}
renormalized by either vertex corrections or Fermi liquid parameters
\cite{DURST}. Thus one infers $v_F/v_2 = 19$ from the thermal transport
data \cite{CHIAO}, which is in remarkable agreement with the ARPES estimate
\cite{MESOT} of $v_F/v_2 \simeq 20$ coming from the measured
values of $v_F = 2.5 \times 10^7$ cm/s and $v_2 = 1.2 \times 10^6$ cm/s 
for the near optimal $T_c = 87$K sample of Fig.~\ref{gapvsdoping}.
(We note that in Ref.~\cite{MESOT} 
we used the notation $v_\Delta$ for one-half the gap slope at the
node, which is related to $v_2$ defined above by
$v_2 = 2 v_\Delta/\hbar k_F$.)

The comparison of ARPES results with the slope of the superfluid density 
$\rho_s \sim 1/\lambda^2$ obtained from penetration depth measurements
is more complicated, as discussed in more detail in Ref.~\cite{MESOT},
for two reasons. Experimentally, there seem to be discrepancies between
the $d\lambda^{-2}/dT$ results of various groups, and theoretically the
slope of $\rho_s$ is renormalized by doping-dependent
Fermi liquid parameters \cite{MILLIS,DURST} which are not known apriori.
These parameters characterize the residual interactions between
the nodal quasiparticles in the superconducting state.
Thus, e.g., even at optimality the slope $d\lambda^{-2}/dT$ obtained
using the ARPES estimate of $v_F/v_2$ and ignoring Fermi liquid 
renormalization is almost three times as large as the experimentally
measured value of $d\lambda^{-2}/dT$ in near optimal Bi2212
\cite{LAMBDA}. This indicates the importance of Fermi liquid renormalizations
in order to make a quantitative comparison. For more details
on the doping dependence of these renormalizations, see 
Ref.~\cite{MILLIS,MESOT,M2S}.

Finally, let us return to the doping dependence of the results of
Fig.~\ref{gapvsdoping}. In contrast to the maximum gap (at the 
$(\pi,0)-(\pi,\pi)$ Fermi surface crossing) increasing as a function of 
underdoping, noted earlier, we see the gap slope at the node,
which determines $v_2$, decreasing with underdoping. This is a result
of the doping dependence of the $B$ parameter introduced above.

This surprising observation raises several questions.
First, could the flattening at the node be, in fact,
evidence for a ``Fermi arc'', a line of gapless excitations,
in the underdoped materials,
especially since such arcs are seen above $T_{c}$ 
(See Sect. 6.3 for further discussion of Fermi arcs in underdoped 
materials). Given the error bars on gap estimates in 
Fig.~\ref{gapvsdoping}, it is impossible to
rule out arcs in all the samples. Nevertheless, it is clear that
there are samples (especially OD87K, UD80K and UD75K) where there
is clear evidence in favor of a point node rather than an arc
at low temperatures. Furthermore, it is very important to note
that a linear $T$ dependence of $\rho_s(T)$ at low
temperature, for all doping levels, in clean samples gives independent
evidence for point nodes, at least in YBCO \cite{HARDY}.

Second, is the change in gap anisotropy intrinsic, or related to impurity
scattering\cite{HAAS}?  We can eliminate the latter explanation on two grounds.
The maximum gap {\it increases} as the doping is reduced, 
opposite to what would be expected from pair breaking due to
impurities.  Also, impurity scattering is expected to lead to a
characteristic ``tail" to the leading edge \cite{FN}, for which there
is no evidence in the observed spectra.

Thus the flattening near the nodes with underdoping does appear
to be an intrinsic feature which may be related to the increased
importance of longer range pairing interactions as one approaches
the insulator. It would be of great interest to study the details
of the doping dependence of the gap anisotropy with the new Scienta
detectors which have greatly improved energy and momentum resolution.

\section{Pseudogap}

In this Section we describe one of the most fascinating developments 
in the study of high $T_c$ superconductors: the appearance of a 
pseudogap above $T_c$ which is seen most prominently on 
the underdoped side of the cuprate phase diagram. 
Briefly the ``pseudogap'' phenomenon is the loss of low energy
spectral weight in a window of temperatures $T_c < T < T^*$;
see Fig.~\ref{phase_diagram}.
The pseudogap regime has been probed by many techniques like 
NMR, optics, transport, tunneling, $\mu$SR and specific heat; 
for reviews and references, see Refs.~\cite{TIMUSK,MR_VARENNA}.
ARPES, with its unique momentum-resolved capabilities,
has played a central role in elucidating the pseudogap phenomenon
\cite{MARSHALL,LOESSER,DING_NATURE,DING_97,Nature98}.

We will discuss in this Section ARPES results 
on the anisotropy of the pseudogap, its $T$-dependence,
its doping dependence, and its effect on the normal state Fermi surface.
We emphasize that for the most part we will focus on the
``low energy'' or leading edge pseudogap, and only mention
ARPES evidence for the ``high energy pseudogap'' toward the end.
We will conclude the Section with a summary of the constraints
put by the ARPES data on various theoretical descriptions of the 
pseudogap.

\begin{figure}
\sidecaption
\includegraphics[width=.5\textwidth]{phase_diagram.epsf}
\caption[]{
$T^*$ (triangles for determined values and squares for
lower bounds) and $T_c$ (dashed line) as a function of hole
doping $x$. The $x$ values for a measured $T_c$ were obtained by
using the empirical relation $T_{c}/T_{c}^{max}=1-82.6(x-0.16)^{2}$
\cite{PRES} with $T_{c}^{max}=$95 K. Also shown is the low temperature
(maximum) superconducting gap $\Delta(0)$ (circles). Note the similar
doping trends of $\Delta(0)$ and $T^*$.
}
\label{phase_diagram}
\end{figure}

\subsection{Pseudogap near $(\pi,0)$}

In the underdoped materials, $T_c$ is suppressed by lowering
the carrier (hole) concentration as shown in Fig.~\ref{phase_diagram}.
In the samples used by our group 
\cite{DING_NATURE,DING_97,Nature98} underdoping was achieved by 
adjusting the oxygen partial pressure 
during annealing the float-zone grown crystals. These crystals also have
structural coherence lengths of at least 1,250$\AA$ as seen from
x-ray diffraction, and optically flat surfaces upon cleaving,
similar to the slightly overdoped $T_c$ samples discussed above.
We denote the underdoped (UD)
samples by their onset $T_c$: the 83K sample has a transition width of 2K
and the highly underdoped 15K and 10K have transition widths $> 5$K.
Other groups have also studied samples where underdoping was
achieved by cation substitution \cite{MARSHALL,LOESSER}.

\begin{figure}
\includegraphics[width=.8\textwidth]{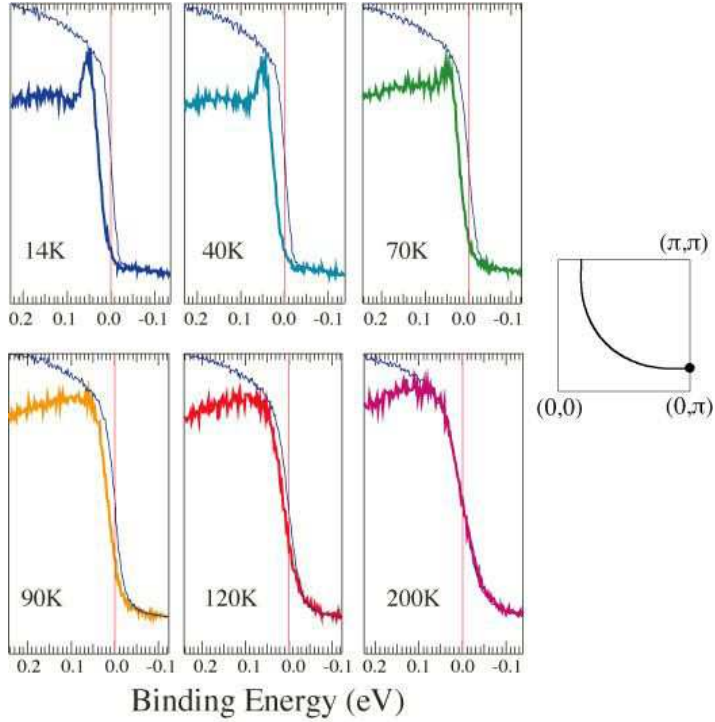}
\caption{
ARPES spectra along the $(\pi,0) \to (\pi,\pi)$ direction for an 83 K
underdoped sample at various temperatures (solid curves). The thin
curves in each panel are reference spectra from polycrystalline Pt used to
accurately determine the zero of binding energy at each temperature.
}
\label{SixPanels}
\end{figure}

We now contrast the remarkable properties of the underdoped samples
with the near-optimal Bi2212 samples which we have been mainly
focusing on thus far. We will first focus on the behavior
near the $(\pi,0)$ point where the most dramatic effects occur,
and come back to the very interesting $\bk$-dependence later.
In Fig.~\ref{SixPanels} \cite{SNS97} we show the $T$-evolution of the
ARPES spectrum at the $(\pi,0) \to (\pi,\pi)$ Fermi crossing for
an UD 83K sample. At sufficiently high temperature, the leading edge of the 
UD spectrum at $\bk_F$ and the reference Pt spectrum coincide, but below 
a crossover temperature $T^*\simeq 180$K the leading edge midpoint
of the spectrum shifts below the chemical potential. In 
Fig.~\ref{SixPanels} one can clearly see a loss of low energy spectral 
weight at 120K and 90K.  It must be emphasized that this 
gap-like feature is seen in the normal (i.e., non-superconducting) state
for $T_c = 83K < T < T^* = 180K$.

The doping dependence of the temperature $T^*$, below which a 
leading-edge pseudogap appears near $(\pi,0)$, 
is shown in Fig.~\ref{phase_diagram}. Remarkably $T^*$  increases
with underdoping, in sharp contrast with $T_c$, but very similar to
the low temperature SC gap, a point we will return to at the end
of the Section. The region of the phase diagram
between $T_c$ and $T^*$ is called the pseudogap region.

It is important to emphasize that our understanding of the lightly
UD samples (e.g., the UD 83K sample) is the best among the UD materials.
In such samples all three regimes -- the SC state below $T_c$,
the pseudogap regime between $T_c$ and $T^*$ and the gapless
``normal'' regime above $T^*$  -- can be studied in detail.
In contrast, in the heavily UD samples (e.g. the UD 10K and UD 15K samples),
not only is the SC transition broad, 
one also has such low $T_c$'s and such high $T^*$'s 
that only the pseudogap regime is experimentally
accessible. Nevertheless, the results on the heavily underdoped samples
appear to be a natural continuation of the weakly underdoped materials
and the results (the trends of gap and $T^*$)
on the low $T_c$ samples are in qualitative agreement with  those
obtained from other probes (see Ref.~\cite{TIMUSK,MR_VARENNA}).

The $T$-dependence of the leading-edge midpoint shift
appears to be completely smooth through the SC transition
$T_c$. In other words, the normal state pseudogap evolves
smoothly into the SC gap below $T_c$.
Nevertheless, there {\it is} a characteristic change in the lineshape
in passing through $T_c$ associated with the appearance of
of a sharp feature below $T_c$ in Fig.~\ref{SixPanels}.
This can be identified as the coherent quasiparticle 
peak for $T \ll T_c$. The existence of a SC state quasiparticle peak
is quite remarkable given that the normal state spectra of UD materials 
are even broader than at optimality, and
in fact become progressively broader with underdoping.
In fact, the low temperature SC state spectra near $(\pi,0)$ 
in the UD systems (see Fig.~\ref{fig8.15} of Section 7.6)
are in many ways quite similar to those at optimal doping, with the one
crucial difference that the spectral weight in the coherent 
quasiparticle peak diminishes rapidly with underdoping 
\cite{FENG_00,DING_00}.

\subsection{Anisotropy of the Pseudogap}

We have already indicated that the pseudogap above $T_c$ near the $(\pi,0)$
point of the zone evolves smoothly through $T_c$ into the large SC gap 
below $T_c$, and thus the two also have the same magnitude. Since the SC
gap has the $d$-wave anisotropy (discussed in detail in the preceding Section),
it is natural to ask: what is the $\bk$-dependence of the pseudogap 
above $T_c$?

The first ARPES studies \cite{MARSHALL,LOESSER,DING_NATURE}
showed that the pseudogap is also highly anisotropic and has a 
$\bk$-dependence which is very similar to that of the SC gap below
$T_c$. Later work \cite{Nature98} further clarified the situation by
showing that the anisotropy has a very interesting temperature dependence.
We now describe these developments in turn.

\begin{figure}
\includegraphics[width=.5\textwidth]{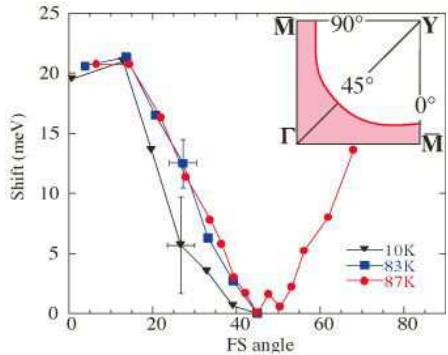}
\caption{
Momentum dependence of the gap
estimated from the leading-edge shift in samples with $T_{c}$'s
of 87K (slightly overdoped), 83K (UD) and 10K (UD), measured at 14K. 
For the sake of comparison between samples we made
vertical offsets so that the shift at $45^\circ$ is zero;
the offsets are $-3$meV for the 83K and $+2$meV for the 10K sample.
The inset shows the Brillouin zone with the large Fermi surface.
}
\label{PseudogapVsK}
\end{figure}

In Fig.~\ref{PseudogapVsK} \cite{DING_NATURE} we plot the leading edge shifts
for three samples at 14K: the slightly overdoped 87K and UD 83K samples are
in their SC states while the UD 10K sample is in the pseudogap regime.
The gap estimate for each sample was made on the ``minimum gap locus''
(explained earlier in the context of the SC gap; see further below). 
The large error bars on the UD 10K sample come from the difficulty of
accurately locating the midpoint of a broad spectrum. 
Also there is a flattening of the gap near the node, a feature
that we discussed earlier for the SC gap in UD samples.

The remarkable conclusion is that the normal state pseudogap
has a very similar $\bk$-dependence and magnitude as the SC gap below $T_c$. 

\subsection{Fermi Arcs}

The $T$-dependence and anisotropy of the pseudogap was investigated 
in more detail in Ref.~\cite{Nature98} motivated by the following question.
Normal metallic systems are characterized by a Fermi surface,
and optimally doped cuprates are no different despite the absence
of sharp quasiparticles (see Section \ref{NormalState}).
On the underdoped side of the phase diagram, however,
how does the opening of pseudogap affect the locus of low lying 
excitations in $\bk$-space?

\begin{figure}
\includegraphics[width=.8\textwidth]{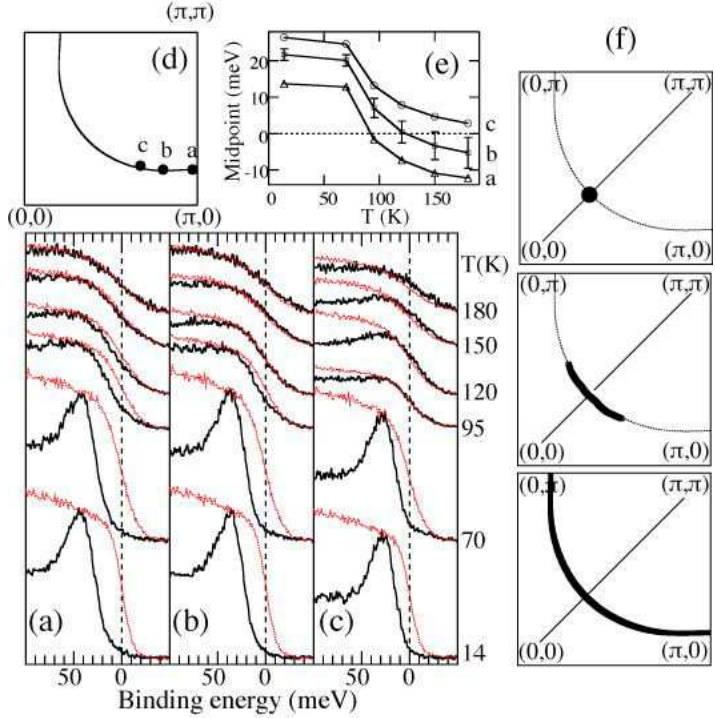}
\caption{
(a,b,c): Spectra taken at three ${\bf k}$ points in the
$Y$ quadrant of the zone (shown in (d)) for an 83K underdoped
Bi2212 sample at various temperatures (solid curves). The
dotted curves are reference spectra from polycrystalline
Pt (in electrical contact with the sample) used to determine
the chemical potential (zero binding energy).  Note the
closing of the spectral gap at different $T$ for different
${\bf k}$'s,  which is also apparent in the plot (e)
of the midpoint of the leading edge of the spectra as a
function of $T$. Panels (f) show a schematic illustration
of the temperature evolution of the Fermi surface in
underdoped cuprates. The d-wave node below $T_c$ (top panel)
becomes a gapless arc above $T_c$ (middle panel) which
expands with increasing $T$ to form the
full Fermi surface at $T^*$ (bottom panel).
}
\label{GapClosingVsK}
\end{figure}

In Fig.~\ref{GapClosingVsK} we show ARPES spectra for an
UD 83K sample at three {\bf k} points on the Fermi
surface for various temperatures. The superconducting gap,
as estimated by the position of the sample leading edge
midpoint at low $T$, is seen to decrease as one moves from point
{\it a} near $(\pi,0)$ to {\it b} to {\it c}, closer to
the diagonal $(0,0) \to (\pi,\pi)$ direction, consistent
with a $d_{x^2-y^2}$ order parameter. At each {\bf k} point
the quasiparticle peak disappears above $T_c$ as $T$
increases, with the pseudogap persisting well above $T_c$,
as noted earlier.

The striking feature which is apparent from Fig.~\ref{GapClosingVsK} is that
the pseudogap at different {\bf k} points closes at different
temperatures, with larger gaps persisting to higher $T$'s.
At point {\it a}, near $(\pi,0)$, there is a pseudogap at
all $T$'s below 180K, at which the Bi2212 leading edge
matches that of Pt. As discussed above, this defines $T^*$
above which the the largest pseudogap has
vanished within the resolution of our experiment, and a
closed contour of gapless excitations -- a Fermi surface --
is obtained. The surprise is that if we
move along this Fermi surface to point {\it b} the sample
leading edge matches Pt at 120K, which is smaller than
$T^*$. Continuing to point {\it c}, about halfway to the
diagonal direction, we find that the Bi2212 and Pt leading
edges match at an even lower temperature of 95K. In addition,
spectra measured on the same sample along the Fermi
contour near the $(0,0) \to (\pi,\pi)$ line shows no gap
at any $T$ (even below $T_c$) consistent with $d_{x^2-y^2}$
anisotropy. 

One simple way to quantify the behavior of the gap is to
plot the midpoint of the leading edge of the spectrum;
see Fig.~\ref{GapClosingVsK}(e). 
We note that a leading edge midpoint at a negative binding energy,
particularly for {\bf k} point {\it c}, indicates the formation of a
peak in the spectral function at $\omega = 0$ at high $T$.
Further, we will say that the pseudogap 
has closed at a {\bf k} point when the midpoint equals zero energy, in
accordance with the discussion above. A clearer way of determining
this will be presented below when we discuss the symmetrization method,
but the results will be the same.

\begin{figure}
\sidecaption
\includegraphics[width=.4\textwidth]{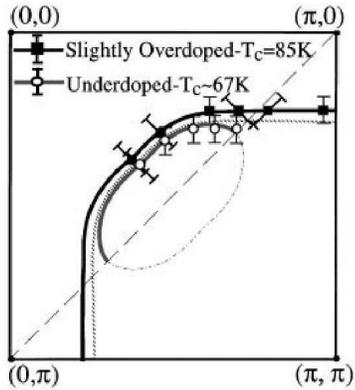}
\caption{
Fermi level crossings from two Bi2212 samples of
differing oxygen content. The entire zone can be reconstructed
by fourfold rotation about (0,0) (from Ref.~\cite{MARSHALL}).
}
\label{marshf4}
\end{figure}

From Fig.~\ref{GapClosingVsK}, we
find that the pseudogap closes at point {\it a} at a $T$
above 180K, at point {\it b} at 120 K, and at point
{\it c} just below 95 K. If we now view these data as a
function of decreasing $T$, the picture of Fig.~\ref{GapClosingVsK}(f)
clearly emerges. With decreasing $T$, the pseudogap first opens
up near $(\pi,0)$ and progressively gaps out larger
portions of the Fermi contour. Thus one obtains gapless arcs
which shrink as $T$ is lowered, eventually leading
to the four point nodes of the $d$-wave SC gap.
The existence of such arcs is apparent from the first ARPES work on
the pseudogap \cite{MARSHALL}, where it was noted that the Fermi contours
in the pseudogap phase did not extend all the way to the zone boundary
(see Fig.~\ref{marshf4}).

Whether the arcs shrink to a point precisely at $T_c$ or
below $T_c$ is not clear from the existing data. As discussed
in the preceding Section, we do believe that arcs do not survive
deep into the SC state where there is point node at $T \ll T_c$
in clean samples, as also evidenced by the linear $T$ drop in the
superfluid density at low $T$.

We next turn to a powerful visualization aid that makes 
these results very transparent. This is the symmetrization method
introduced in Ref.~\cite{Nature98}, which effectively eliminates
the Fermi function $f$ from ARPES data and permits us to focus directly
on the spectral function $A$. Given ARPES data described by\cite{NK}
$I(\omega)= \sum_{\bf k} I_0 f(\omega)A({\bf k},\omega)$ with
the sum over a small momentum window about the Fermi momentum ${\bf k}_F$,
we can generate the symmetrized spectrum $I(\omega) + I(-\omega)$.
Making the reasonable assumption of particle-hole (p-h) symmetry
for a small range of $\omega$ and $\epsilon_{\bf k}$, we have
$A(\epsilon_{\bf k},\omega)=A(-\epsilon_{\bf k},-\omega)$
for $|\omega|,|\epsilon|$ less than few tens of meV.
It then follows, using the identity $f(-\omega) = 1-f(\omega)$,
that $I(\omega) + I(-\omega) =  \sum_{\bf k} I_0 A({\bf k},\omega)$
which is true even after convolution with a (symmetric) energy resolution
function; for details see the appendix of Ref.~\cite{mesot01}.
The symmetrized spectrum coincides with the
raw data for $\omega \le -2.2T_{eff}$, where $4.4T_{eff}$ is
the 10\%-90\% width of the Pt leading edge, which
includes the effects of both temperature and resolution.
Non-trivial information is obtained for the range
$|\omega| \le 2.2T_{eff}$, which is then the scale on which p-h
symmetry has to be valid.  We have extensively checked this method,
and studied in detail the errors introduced by
incorrect determination of the chemical potential or of ${\bf k}_F$
(which lead to spurious narrow features in the symmetrized spectra),
and the effect of the small ($1^\circ$ radius) {\bf k}-window of the
experiment (which was found to be small).

\begin{figure}
\sidecaption
\includegraphics[width=.6\textwidth]{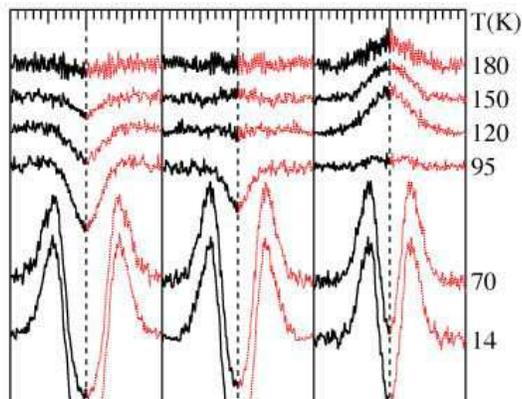}
\caption{
Symmetrized spectra corresponding to the raw spectra
(a,b,c) of Fig.~\ref{GapClosingVsK}.
The gap closing in the raw spectrum of Fig.~\ref{GapClosingVsK}
corresponds to when the pseudogap depression disappears in the 
symmetrized spectrum. Note the appearance of a spectral peak at 
higher temperatures in c.}
\label{symmetrized}
\end{figure}

In Fig.~\ref{symmetrized} we show symmetrized data for the 
UD 83K underdoped sample corresponding to the raw data of 
Fig.~\ref{GapClosingVsK}.
To emphasize that the symmetry is put in by hand, we show
the $\omega > 0$ curve as a dotted line. At {\bf k} point a near
$(\pi,0)$ the sharp quasiparticle peak disappears above
$T_c$ but a strong pseudogap suppression, on the same scale as
the superconducting gap, persists all the way up to 180K ($T^*$).
Moving to panels b and c in Fig.~\ref{symmetrized} we again see pseudogap
depressions on the scale of the superconducting gaps at those 
points, however the pseudogap fills up at lower temperatures: 
120K at b and 95K at c. In panel c, moreover, a spectral peak 
at zero energy emerges as $T$ is raised. All of the conclusions 
drawn from the raw data in Fig.~\ref{GapClosingVsK} are 
immediately obvious from the simple symmetrization analysis of 
Fig.~\ref{symmetrized}.

There are many important issues related to these results that will
be taken up in Section 7.3 where we describe 
modeling the electron self-energy in the pseudogap state. We will 
discuss there the remarkable $T$-dependent lineshape changes and 
the $T$-dependence of the gap itself. Here we simply 
note that, without any detailed modeling, the data\cite{Nature98}
clearly show qualitative differences in the $T$-dependence at 
different $\bk$-points. Near the $(\pi,0)$ point the gap goes away
with increasing temperature with the spectral weight filling-in, but no
perceptible change in the gap scale with $T$. On the other hand,
at $\bk_F$ points halfway to the node, one sees a 
suppression of the gap scale with increasing temperature.

We conclude this discussion with a brief mention of the implications of
our results.  We believe that the unusual
$T$-dependence of the pseudogap anisotropy will be a very important
input in reconciling the different crossovers seen in the
pseudogap regime by different probes. The point here
is that each experiment is measuring a {\bf k}-sum weighted with
a different set of {\bf k}-dependent matrix elements or kinematical
factors (e.g., Fermi velocity).  For instance, quantities which involve
the Fermi velocity, like dc resistivity above $T_c$ and the penetration depth
below $T_c$ (superfluid density),
should be sensitive to the region near the zone diagonal,
and would thus be affected by the behavior we see at {\bf k} point c.
Other types of measurements (e.g. specific heat and tunneling) are 
more ``zone-averaged" and will have
significant contributions from {\bf k} points a and b as well,
thus they should see a more pronounced pseudogap effect. 
Interestingly, other data we have indicate that the region in the
Brillouin zone where 
behavior like {\bf k} point c is seen shrinks as the doping is reduced,
and thus appears to be correlated with the loss of
superfluid density\cite{UEMURA}.
Further, we speculate that the disconnected Fermi arcs
should have a profound influence on magnetotransport given the lack
of a continuous Fermi contour in momentum space.

\subsection{Evolution of the Fermi Surface with Doping}

We now discuss the doping dependence of the normal state Fermi surface
on the underdoped side of the phase diagram. The first issue to face up to
is: can the Fermi arcs described above be a manifestation of 
a Fermi surface with small closed contours centered about $(\pi/2,\pi/2)$?
Such hole-pockets enclosing $x$ holes (per planar Cu) are
suggested by some theories of lightly doped Mott insulators \cite{LEE}
as alternatives to the large Fermi surfaces containing $(1+x)$ holes
which would be consistent with the Luttinger counting.

The $T$-dependence of the arcs is by itself evidence
against their being part of a pocket Fermi surface. Nevertheless,
if there were such small hole pockets then one should observe
two features in the ARPES data: a closure of the Fermi arc on
the other side of $(\pi/2,\pi/2)$, which would be clear evidence for a 
``shadow band''-like dispersion ($(\pi,\pi)$-foldback of the
main band) in the UD samples.
In a variety of UD samples we have carefully searched for both these
features and found no evidence for either \cite{DING_97}. 
However this is a tricky issue, given the very broad spectra
and possible materials problems in the highly UD samples.
Nevertheless, given the available evidence,
the gapless arcs that we observe \cite{Nature98} are simply an
intermediate state in the smooth evolution of $d$-wave nodes
into a full Fermi surface. This smooth evolution was carefully checked
on an UD 83K sample where a dense mapping was done in {\bf k} space
at $T=90$K, revealing only a small Fermi arc just above $T_c$.

\begin{figure}
\sidecaption
\includegraphics[width=.8\textwidth]{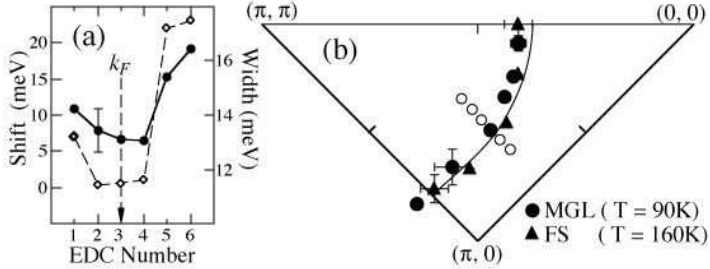}
\caption{
(a) Midpoint shifts (dots) and widths (diamonds) for an UD 83K
sample at a photon energy of 22 eV at 90K for a cut shown by
the open dots in (b). (b)  Fermi surface at 160K (solid triangles)
and minimum gap locus at 90K (solid dots). Notice that the two
surfaces coincide within error bars. The error bars represent
uncertainties of Fermi crossings as well as possible sample
misalignment. The solid curve is a rigid band estimate of the
Fermi surface.
}
\label{MinGapLocus}
\end{figure}

The other issue related to the Fermi surface is: what is its
doping dependence above $T^*$ where the pseudogap
effects are absent. While one can easily compare the near optimal
and lightly UD Fermi surfaces, the rapid rise of $T^*$ with underdoping
does not permit us to address this question.  However, one can
study the ``minimum gap locus'' in any gapped state, in close analogy
with the manner in which this was defined in the SC state;
see Section 5.

There is also a more fundamental reason to study the 
``minimum gap locus'' in the pseudogap regime. One wants to know
whether the pseudogap is ``tied'' to the Fermi surface, or if it has
some other characteristic momentum {\bf Q} (unrelated to $\bk_f$).
In Fig.~\ref{MinGapLocus}(a) \cite{DING_97} we follow the dispersion 
of an UD 83K sample in the pseudogap regime. Moving
perpendicular to the (expected) Fermi surface from occupied to
unoccupied states, one finds that that the dispersion first approaches
the chemical potential and then recedes away from it. This locates a
$\bk$-point on the minimum gap locus. For a lightly UD sample
we find in Fig.~\ref{MinGapLocus}(b) that this locus in the
pseudogap regime coincides, within experimental error bars,
with the Fermi surface determined above $T^*$ where there is no pseudogap.
The pseudogap is thus tied to the Fermi surface in the same
way the SC gap is, and is in contrast with, say, 
charge or spin density waves, which are tied to other characteristic
{\bf Q} vectors. 

\begin{figure}
\includegraphics[width=.5\textwidth]{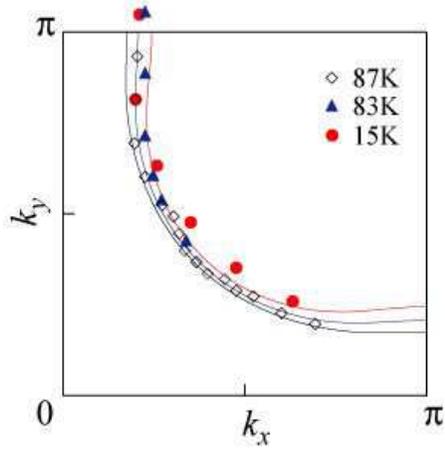}
\caption{
Fermi surfaces of the 87K, 83K, and 15K samples.
All surfaces enclose a large area consistent with the Luttinger
count (see text).
The solid lines are tight binding estimates of the Fermi surface
at 18\%, 13\%, and 6\% doping assuming rigid band behavior.
}
\label{fs_doping}
\end{figure}

In the more heavily underdoped samples, it is not possible
to compare the minimum gap locus in the pseudogap state
with the Fermi surface above $T^*$, or the minimum gap locus
below $T_c$, since the latter two are not measurable
with $T^*$ too high and $T_c$ too low.
Nevertheless, if one {\it assumes}, by
continuity, that the minimum gap locus in the pseudogap
state gives information about the Fermi surface 
that got gapped out, then even for an highly UD sample
one finds a large underlying Fermi surface, satisfying the Luttinger
count of $(1+x)$ holes per planar Cu \cite{DING_97} as shown in
Fig.~\ref{fs_doping}.

\begin{figure}
\includegraphics[width=1\textwidth]{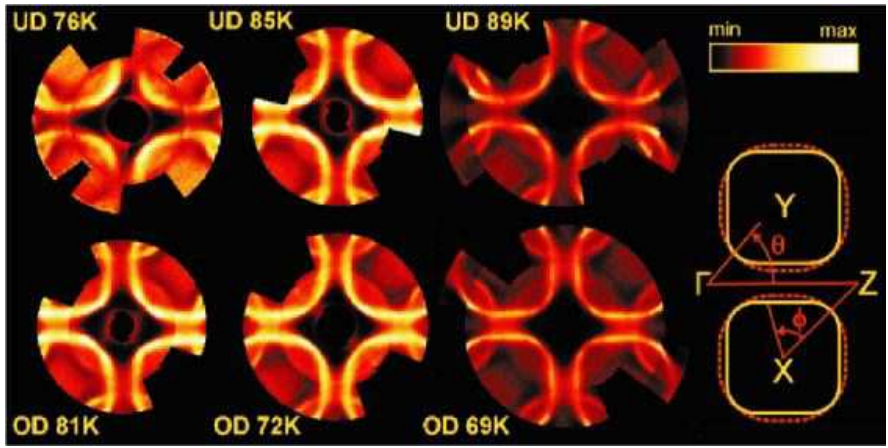}
\caption{
Basal plane projection of the normal-state (300 K) Fermi surface of 
Bi(Pb)-2212 from high-resolution ARPES. The $E_{F}$ intensity (normalized 
to the signal at $\omega$ = 0.3 eV) is shown in color. The $T_{c}$ of each 
sample is indicated. The raw data cover half of the colored area of each 
map and have been rotated by 180¡ around the $\Gamma$ point to give a better 
k-space overview. The line dividing raw and rotated data runs almost 
vertically for the UD76K map and from top left to bottom right in all 
other maps. The sketch shows the Fermi surface for the OD69K data set as yellow 
barrel-like shapes defined by joining the maxima of fits to the normalized 
$E_{F}$ MDCs. (from Ref.~\cite{KORDYUK})}
\label{kordf1}
\end{figure}

The same conclusion has been recently reached by the Dresden 
group\cite{KORDYUK}.
Fig.~\ref{kordf1} reproduces their Fermi energy intensity maps as a
function of doping, where a large Fermi surface (plus its shadow band
image) is always visible.  They argue, though, that the volume is not
quite 1+x, and they attribute this difference to the presence of bilayer 
splitting.

\subsection{Low Energy vs High Energy Pseudogaps}

In all of the preceding discussion we have focussed on the
``low energy'' or leading-edge pseudogap. It is important to
point out that the phrase pseudogap is (somewhat confusingly)
also used to describe a higher energy feature, which we call
the ``high energy pseudogap''.

\begin{figure}
\includegraphics[width=.5\textwidth]{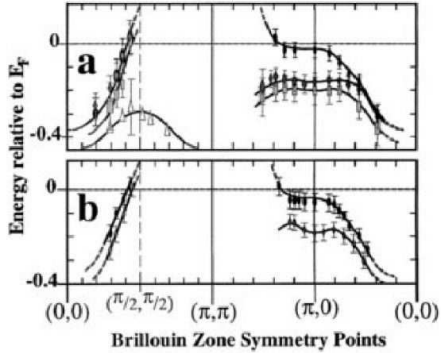}
\caption{
(a),(b) Map peak centroids vs k for
Bi$_{2}$Sr$_{2}$Ca$_{1-x}$Dy$_{x}$Cu$_{2}$O$_{8+\delta}$ thin films and deoxygenated
Bi$_{2}$Sr$_{2}$CaCu$_{2}$O$_{8+\delta}$ bulk samples, respectively, with various
hole doping levels. (a) Filled oval, 1\% Dy near optimal
doping with T$_c$ = 85 K; gray diamond, 10\% Dy underdoped
with T$_c$ = 65 K; gray rectangular, 17.5\% Dy underdoped
with T$_c$ = 25 K; triangle, 50\% Dy insulator. (b) Filled oval,
600 air annealed slightly overdoped with T$_c$ = 85 K; gray diamond,
550 argon annealed underdoped with T$_c$ = 67 K (from Ref.\cite{MARSHALL}).
}
\label{marshf3}
\end{figure}

The presence of a high energy pseudogap was evident in the first ARPES
work on the pseudogap, reproduced in Fig.~\ref{marshf3} \cite{MARSHALL}.
As the doping is reduced from optimal doped, a gap opens up in a region around
the $(\pi,0)$ points of the zone.  The energy of this gap is significantly
higher than the leading edge gap emphasized in later
work \cite{LOESSER,DING_NATURE}.  The resulting dispersion of this high
energy feature looks
reminiscent of what is expected for a spin density wave gap.
As the doping is further reduced, an energy gap then opens up along
the $(\pi,\pi)$ direction, and the material becomes truly insulating.

\begin{figure}
\sidecaption
\includegraphics[width=.5\textwidth]{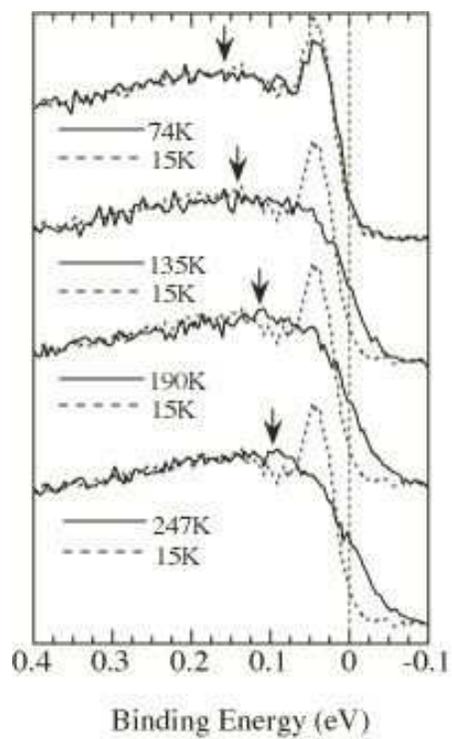}\caption{
Spectra at the $(\pi,0)$ point for an UD 89K sample at various
temperatures compared with the low temperature (15K) spectrum.
The position of the high energy feature is marked by an arrow.
}
\label{HighEPseudo}
\end{figure}

In Fig.~\ref{HighEPseudo} \cite{JC99} we show the temperature
dependence of the $(\pi,0)$ spectrum for an UD 89K sample.
Note that there is no coherent quasiparticle peak until the system
is cooled below $T_c$, with only a broad incoherent spectrum
observed for all $T > T_c$. The leading edge pseudogap
which develops below $T^{*}$ is difficult to see on the energy
scale of this figure; the midpoint shift at 135 K is 3 meV.
However, a higher energy feature can easily be identified by 
a change in slope of the spectra as function of binding energy;
this is also very clear in the data of Fig.~\ref{fig8.13} of 
Section 7.6.
At the highest temperature $T=247$K this feature is just due to
the Fermi function cutoff,
but in the pseudogap regime, this feature actually represents 
the onset of loss of spectral weight on a high energy scale,
and hence may be called the ``high energy pseudogap''.
It can also be seen from  Fig.~\ref{HighEPseudo}
that the energy scale of this feature is very similar to that
of the well-known $(\pi,0)$-hump of the peak-dip-hump structure
seen in the SC state. This connection will be discussed in detail 
in Section 7.6 where we also argue that the 
high energy pseudogap and the hump have similar 
dispersions \cite{JC99}.

\subsection{Origin of the Pseudogap?}

We conclude with a summary of ARPES results on the
pseudogap and a brief discussion of its theoretical understanding.

As described above, the low-energy (leading edge) pseudogap
has the following characteristics.
\begin{enumerate}
\item[{$\bullet$}] 
The magnitude of the pseudogap near $(\pi,0)$, i.e., the scale of
which there is suppression of low energy spectral weight above $T_c$,
is the same as the maximum SC gap at low temperatures. Further
both have the same doping dependence.

\item[{$\bullet$}] 
There is a crossover temperature scale $T^*$ above which the full
Fermi surface of gapless excitations is recovered. The pseudogap 
near $(\pi,0)$ appears below $T^*$.

\item[{$\bullet$}] 
The normal state pseudogap evolves smoothly through $T_c$ into
the SC gap as a function of decreasing temperature.

\item[{$\bullet$}] 
The pseudogap is strongly anisotropic with $\bk$-dependence 
which resembles that of the $d$-wave SC gap. The anisotropy of
the pseudogap seems to be $T$-dependent leading to the formation
of disconnected Fermi arcs below $T^*$.

\item[{$\bullet$}] 
The pseudogap is ``tied'' to the Fermi surface, i.e., the minimum
gap locus in the pseudogap regime coincides with the Fermi surface above
$T^*$ and the minimum gap locus deep in the SC state, at least in those
samples where all three loci can be measured.
\end{enumerate}

The simplest theoretical explanation of the pseudogap, 
qualitatively consistent with the ARPES observations, is that
it arises to due pairing fluctuations above $T_c$ \cite{RANDERIA,MR_VARENNA}.
The SC gap increases with underdoping while $T_c$ decreases. 
Thus in the underdoped regime $T_c$ is not controlled by the destruction 
of the pairing amplitude, as in conventional BCS theory, 
but rather by fluctuations of the phase
\cite{EMERY} of the order parameter leading to the Uemura scaling
$T_c \sim \rho_s$ \cite{UEMURA}. Even though SC order is destroyed at $T_c$,
the local pairing amplitude survives above $T_c$ giving rise
to the pseudogap features. A natural mechanism for such a 
pseudogap coming from spin pairing in a doped Mott insulator
exists within the RVB framework \cite{RVB}, with possibility of additional
chiral current fluctuations \cite{RVB2}.

More recently the pairing origin of the pseudogap has been challenged. 
Some experiments \cite{TALLON} have been argued to suggest a non-pairing 
explanation with a competition between the pseudogap and the SC gap.
A specific realization of this scenario is the staggered flux or d-density
wave mechanism \cite{DDW} in which $T^*$ is actually a phase transition
below which both time-reversal and translational invariance are broken.
A more subtle phase transition with only broken time-reversal has
also been proposed \cite{VARMA} as the origin of the pseudogap.

Although a qualitative understanding of some of the characteristics of
the pseudogap within the non-pairing scenarios is not clear at this time,
these theories make sharp predictions about broken symmetries below $T^*$
which can be tested. A very recent ARPES study \cite{KAMINSKY} of
circular dichroism finds evidence in favor of broken time reversal,
thus casting some doubt on the pairing fluctuation ideas. The last word
has clearly not been said on this subject, either theoretically or
experimentally, and the origin of the pseudogap remains one of the
most important open questions in the field of high $T_c$ superconductors.

\section{Photoemission Lineshapes and the Electron Self-Energy}

Under certain conditions, which were discussed in Section 2, ARPES measures
the occupied part of the single particle spectral function,
$A({\bf k},\omega)f(\omega)$, with 
$A=ImG/\pi$ where $G$ is the Greens function.  The latter can be 
expressed as $G^{-1}=\omega-\epsilon_{\bf k}-\Sigma({\bf k},\omega)$ 
where $\epsilon_{\bf k}$ is the single-particle energy (defined by the kinetic 
energy and single-particle potential energy terms of the Hamiltonian) and 
$\Sigma$ is the Dyson self-energy (i.e., everything else).  Often, this 
form is associated with a perturbative expansion used to estimate $\Sigma$, but 
of course the expression is itself tautological.  The purpose 
of writing $G$ in this form is that it isolates all many-body effects in the 
function $\Sigma$.  An advatange 
of ARPES is that one has the possibility of extracting $\Sigma$ directly 
from the data, allowing comparison to various microscopic predictions for 
$\Sigma$.

One of the more trivial examples of this is when one fits ARPES data to 
determine the superconducting gap, $\Delta$.  For instance, the work 
described in Section 5 \cite{DING95} used a broadened form of BCS theory 
to fit the leading edge of the spectra.  This is equivalent to $\Sigma = 
-i\Gamma + \Delta_{\bf k}^2/(\omega+\epsilon_{\bf k}+i\Gamma)$, $\Gamma=0$ 
describing standard BCS theory.  The advantage of this procedure is the 
actual gap function, $\Delta$, is extracted from the data, rather than ill
defined quantities, such as the often utilized leading edge shift 
(midpoint of the leading edge) which is not the same as $\Delta$ because 
of lifetime and resolution effects.  When this is done, a $\Delta_k$ is 
obtained which has rather spectacular agreement with that expected for a 
d-wave order parameter.  Although ARPES contains no phase information of 
the order parameter, the linear behavior of $\Delta_k$ along the Fermi 
surface near the gap zero (node) implies a sign change.  Morevoer, ARPES 
has the additional advantage of determining the shape of $\Delta_k$ 
in the Brillouin zone, which gives important information on the 
spatial range of the pairing interaction \cite{MESOT}, as also discussed 
in Section 5.

Even when fitting data at low temperatures including energy and momentum
resolution, a non-zero 
$\Gamma$ is always needed.  The origin of this residual $\Gamma$ is still 
debated.  It is larger than what is expected based on impurity 
scattering, and certainly larger than that implied by various 
conductivity probes (thermal, microwave, and infrared).  Although the 
transport scattering rate is different from $Im\Sigma$ (and in 
particular, only Umklapp processes contribute to electrical 
conductivity), the discrepancy is still large enough to be noticable, 
even when taking into account the fact that in the simple approximation 
being employed here, $\Gamma$ represents some average of 
$Im\Sigma$ over a frequency range of order $\Delta$.

Although it has been suggested that the residual $\Gamma$ is due to 
surface inhomogeneity effects (in particular, a distribution of $\Delta$ 
due to local oxygen inhomogeneities \cite{Pan}), a more likely possibility is 
that it is the same effect which is seen in normal metals like $TiTe_2$.
In the latter case, it was convincingly argued that this was the expected 
final state lifetime contamination effect when attempting to extract 
$\Sigma$ from ARPES spectra \cite{TiTe2}.  Although the latter is expected to 
vanish in the pure 2D limit, 
even small 3D effects can lead to a noticable effect, since final state 
lifetimes are large.  For instance, in simple models, its contribution to 
$\Gamma$ is of order $(v_c^i/v_c^f)\Gamma_f$, where $v_c^i$ is the c-axis 
velocity of the initial state, $v_c^f$ that of the final state, and 
$\Gamma_f$ is the final state lifetime \cite{SMITH}.  Since $\Gamma_f$ is
typically of order 1 eV, then a velocity ratio of only 0.01 is sufficient to 
cause a residual $\Gamma$ of 10 meV.

With this as an introduction,
in this section, we desire to take a more serious look at the issue of 
extracting $\Sigma$ from the data.  The most commonly employed strategy 
is to come up with some model for $\Sigma$, and then see how well it 
fits the data, as illustrated by the simple example above.  We will discuss
this approach in more detail later.  We start, though, 
with discussing an alternate approach which we have recently advocated.

\subsection{Self-Energy Extraction}

Let us first assume we know $A$.  Given that,
we can easily obtain $\Sigma$.  A Kramers-Kronig transform of $A$ will
give us the real part of $G$
\begin{equation}
ReG(\omega) = P\int_{-\infty}^{+\infty} d\omega'
\frac{A(\omega')}{\omega'-\omega}
\label{eq8.1}
\end{equation}
where $P$ denotes the principal part of the integral.
Knowing now both $ImG$ and $ReG$, then
$\Sigma$ can be directly read off from the definition of $G$.
\begin{eqnarray}
Im\Sigma = \frac{ImG}{(ReG)^2 + (ImG)^2} \nonumber \\
Re\Sigma = \omega-\epsilon-\frac{ReG}{(ReG)^2 + (ImG)^2}
\label{eq8.2}
\end{eqnarray}

To obtain $ReG$ using Eq.~\ref{eq8.1}, we need to know $A$
for {\it all} energies.  From ARPES, though, we only know the product of $A$
and $f$.   (While unoccupied states can be studied by inverse photoemission, 
its resolution at present is too poor to be useful for our purposes).
This is not a limitation if an occupied ${\bf k}$-state is being analyzed and
one can either ignore the unoccupied weight or use a simple 
extrapolation for it (except that only $Re\Sigma+\epsilon$ is
determined).  On the other hand, one is usually interested in $k$ vectors near
the Fermi surface.  Therefore a key assumption will have to be made.
We can implement our procedure if we make the assumption 
of particle-hole symmetry, 
$A(\epsilon_{\bf k},\omega)=A(-\epsilon_{\bf k},-\omega)$, 
within the small ${\bf k}$-window centered at ${\bf k}_F$.
Then, $A$ is obtained by exploiting the identity
$A(\epsilon_{\bf k},\omega)f(\omega)+
A(-\epsilon_{\bf k},-\omega)f(-\omega)=A(\epsilon_{\bf k},\omega)$,
which holds even in the presence of the energy resolution convolution.
Note, this can only be invoked at ${\bf k}_F$, and was used
previously to remove the Fermi function from ARPES
data \cite{Nature98}, where it was denoted as the symmetrization 
procedure (note that the ``symmetrized'' data will correspond to the raw 
data for $\omega < \sim -2.2kT$).
Although the particle-hole symmetry assumption is reasonable for small
$|\omega|$ where it can be tested in the normal 
state by seeing whether the ``symmetrized'' spectrum has a maximum at the
Fermi energy ($E_F$), it will almost certainly
fail for sufficiently large $\omega > 0$.
Nevertheless, since we only expect to derive $\Sigma$ for $\omega < 0$, 
then the unoccupied spectral weight will affect the result only in two
ways.  The first is through the sum rule $\int d\omega A(\omega) = 1$
which must be used to eliminate the intensity prefactor of the ARPES 
photocurrent.  From Eq.~\ref{eq8.2}, we see that violation of the sum rule will
simply rescale $Im\Sigma$, but not $Re\Sigma$ due to the $\omega-\epsilon$
factor.  Our normalization, though, is equivalent to assuming $n_{k_F}$=0.5,
and thus does not involve ``symmetrized'' data.
The second influence comes from the Kramers-Kronig
transformation in Eq.~\ref{eq8.1}, which is a bigger problem.
Fortunately, the contribution from large
$\omega' > 0$, for which our assumption is least valid, is suppressed
by $1/(\omega'-\omega)$.
Further, for ${\bf k}_F$, $\epsilon_{\bf k}$=0 and
thus $Re\Sigma$ is not plagued by an unknown constant.

When using real data, it is sometimes desirable to 
filter the noise out of the data, as well as to deconvolve the energy 
resolution, before employing the above procedure.  These details can be found
in Ref.~\cite{Mike99}.  Moreover, it is assumed that any ``background'' 
contribution (see Section 2) has been subtracted from the data as well.

\begin{figure}
\sidecaption
\includegraphics[width=.5\textwidth]{fig1.epsf}
\caption{
\label{fig8.1}
(a) Symmetrized spectrum for overdoped Bi2212
($T_c$=87K) at $T$=14K at $(\pi,0)$ with (dotted line) and without
(solid line) energy resolution deconvolution.
The resulting $Im\Sigma$ and $Re\Sigma$ are shown in (b) and (c).
The dashed line in (c) determines the condition $Re\Sigma=\omega$.}
\end{figure}

In Fig.~\ref{fig8.1}a, we show $T$=14K symmetrized data for a
$T_c$=87K Bi2212 overdoped sample at the $(\pi,0)$ point \cite{Mike99}. 
We note the important differences in this superconducting state
spectrum, compared with the normal state one (which can be fit by
a simple Lorentzian),
due to the opening of the superconducting gap, with
the appearance of a sharp quasiparticle peak displaced from $E_F$ 
by the superconducting gap, followed by a spectral dip, then by a 
``hump'' at higher binding energies.
The resulting $\Sigma$ is shown in Fig.~\ref{fig8.1}b and c.
At high binding energies, one obtains a constant $Im\Sigma$ 
with a very large value ($\sim$ 300 meV).
Near the spectral dip, $Im\Sigma$ has a small peak followed by a sharp 
drop.

Despite this sharp drop below 70 meV,
$Im\Sigma$ remains quite large at low frequencies.
Then, below 20 meV, there is a narrow spike in $Im\Sigma$.  
This is the imaginary part of the 
BCS self-energy, $\Delta^2/(\omega+i0^+)$, which
kills the normal state pole at $\omega$=0.  The resulting 1/$\omega$
divergence of the real part $Re\Sigma$, which creates new poles at 
$\pm\Delta$=32meV, is easily seen in Fig.~\ref{fig8.1}c.
This is followed by a strong peak in $Re\Sigma$ near the spectral dip 
energy, which follows from the Kramers-Kronig transformation of the 
sharp drop in $Im\Sigma$. The strong peak in $Re\Sigma$ explains why the low
energy peak in $A$ is so narrow despite the large value of $Im\Sigma$.
The halfwidth of the spectral peak is given by $\Gamma=zIm\Sigma$ where
$z^{-1}=1-\partial Re\Sigma/\partial \omega$ ($z$ is the quasiparticle
residue).  In the vicinity of the
spectral peak, $z^{-1}$ is large ($\sim$9), giving a $\Gamma$ of $\sim$14 meV.
We note, though, that $\Gamma$ is still quite sizeable, and thus the peak 
is not resolution limited, as discussed above.

\begin{figure}
\sidecaption
\includegraphics[width=.5\textwidth]{fig2.epsf}
\caption{
\label{fig8.2}
(a) Symmetrized spectrum for underdoped Bi2212 ($T_c$=85K)
at $T$=95K (pseudogap phase) at the $(\pi,0)-(\pi,\pi)$ Fermi crossing with
(dotted line) and without (solid line) energy resolution deconvolution.
The resulting $Im\Sigma$ and $Re\Sigma$ are shown in (b) and (c).  The dashed
line in (c) determines the condition $Re\Sigma=\omega$.}
\end{figure}

We can contrast this result with that obtained in the pseudogap phase.
In Fig.~\ref{fig8.2}a, we show $T$=95K symmetrized data from a
$T_c$=85K underdoped Bi2212 sample at the 
$(\pi,0)-(\pi,\pi)$ Fermi crossing.  One again sees (Fig.~\ref{fig8.2}b) a
peak in $Im\Sigma$ at $\omega$=0, but it is broadened relative to
that of the superconducting state, and the corresponding divergence of
$Re\Sigma$ (Fig.~\ref{fig8.2}c) is smeared out.  Such behavior would be 
consistent with 
replacing the BCS self-energy $\Delta^2/(\omega+i0^+)$ by 
$\Delta^2/(\omega+i\Gamma_0)$, 
and can be motivated by considering the presence
of pair fluctuations above $T_c$, as will be discussed further below.
Note from Fig.~\ref{fig8.2} that although the equation
$\omega-Re\Sigma(\omega)=0$ is still satisfied at $|\omega| \sim \Delta$, 
$zIm\Sigma$ is so large that the spectral peak is strongly broadened
in contrast to the sharp peak seen below $T_c$.  Actually, to a good
approximation, the spectral function is essentially the inverse of $Im\Sigma$
in the range $|\omega| < \sim 2\Delta$.
We can also contrast this case with data taken
above $T^*$, the temperature at which the pseudogap ``disappears".  In
that case, the spectrum is featureless, and the peak in $Im\Sigma$
is strongly broadened.  As the doping increases, this peak in $Im\Sigma$
disappears.
Further doping causes a depression in $Im\Sigma$ to develop around
$\omega=0$, indicating a crossover to more Fermi liquid like behavior.

\subsection{Temperature Dependence of $\Sigma$}

We now turn to the rather controversial issue of how the spectrum 
at $(\pi,0)$ varies as a function of temperature, that is, how one 
interpolates between Figs.~\ref{fig8.1} and \ref{fig8.2}.

\begin{figure}
\sidecaption
\includegraphics[width=.5\textwidth]{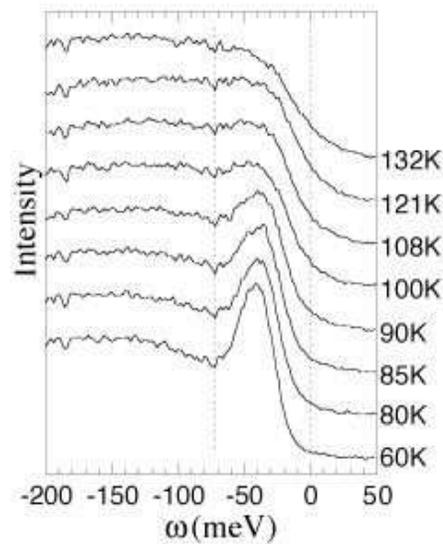}
\caption{Temperature dependence of ARPES data at $(\pi,0)$
for a $T_c$=90K Bi2212 sample.  The vertical dotted lines
mark the spectral dip energy and the chemical potential.}
\label{fig8.3}
\end{figure}

In Fig.~\ref{fig8.3}, we show data taken for an optimal doped ($T_c$=90K) Bi2212
sample \cite{Mike01}.  The leading edge of the
spectral peak is determined by the superconducting gap, whose energy stays
fairly fixed in temperature, and persists above $T_c$ (the pseudogap).
On the trailing edge, one sees a spectral dip, whose energy also remains fixed
in temperature, and becomes filled in above $T_c$ due to broadening of 
the trailing edge of the peak.

\begin{figure}
\sidecaption
\includegraphics[width=.7\textwidth]{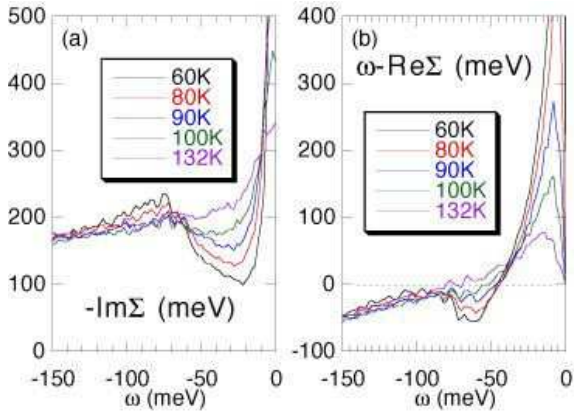}
\caption{Temperature dependence of (a) $Im\Sigma$ and (b) $Re\Sigma$
derived from the data of Fig.~\protect\ref{fig8.3}.}
\label{fig8.4}
\end{figure}

In Fig.~\ref{fig8.4}a, $Im\Sigma$ is plotted for various temperatures.
At low temperatures and energies, it is again characterized by a peak
centered at zero energy due to the
superconducting energy gap, and a ``normal" part which can be treated as
a constant plus an $\omega^2$ term.
A maximum in $Im\Sigma$ occurs near the energy
of the spectral dip.  Beyond this, $Im\Sigma$ has a large, nearly frequency
independent, value.  As the temperature
is raised, the zero energy peak broadens, the constant term increases, and
the $\omega^2$ term goes away.

In Fig.~\ref{fig8.4}b, the quantity $\omega-Re\Sigma$ is plotted.  At low 
temperatures
and energies, there is a $1/\omega$ term due to the energy gap, and
a ``normal" part which is linear in $\omega$.  As expected, the zero crossing
is near the location of the spectral peak.  Beyond this, there is a minimum
near the specral dip energy, then the data are approximately linear again,
but with a smaller slope than near the zero crossing.  As the temperature is
raised, the gap (1/$\omega$) term broadens out and the low energy linear
in $\omega$ term decreases,
paralleling the behavior discussed above for $Im\Sigma$.

From Figs.~\ref{fig8.3} and \ref{fig8.4}, we see that rather than the spectral
peak decreasing in weight with increasing temperature,
it disappears by broadening strongly in energy.  This
can be seen directly by inspecting Fig.~\ref{fig8.4}, in that as the temperature
increases, $Im\Sigma$ (Fig.~\ref{fig8.4}a) in the vicinity of the peak increases
in magnitude with $T$, and $z^{-1}$ (roughly the slope in Fig.~\ref{fig8.4}b 
near the zero crossing) decreases with $T$.  In fact, it is the strong $T$
variation of $Im\Sigma$
and $z^{-1}$, and the fact that they operate in concert, which is responsible
for the rapid variation in the effective width of the spectral peak with $T$.

The above analysis is important in that it shows how coherence is lost in
the system.  It is apparent from Figs.~\ref{fig8.3} and \ref{fig8.4} that once 
a temperature is reached where the spectral
peak is no longer discernable in the data, the difference in behavior
of the self-energy between low energies and high energies is lost.  That is,
once the spectral dip is filled in, the low and high energy behaviors have
merged, and the sharp peak and broad hump at low temperature is simply
replaced by a single broad peak (with a leading edge gap due to the
pseudogap).  This is
consistent with the spectral peak simply losing its integrity as the
temperature is raised.  The analysis does not support a
picture of a well defined quasiparticle peak whose weight simply disappears
upon heating, as has been suggested by other authors \cite{FENG_00}.

\subsection{Modeling $\Sigma$}

This behavior can be further quantified by fitting the self-energy for 
binding energies smaller than the dip energy to that expected for a 
superfluid Fermi liquid, and 
exploring the temperature dependence of the resulting parameters.  The 
reader can find this analysis in Ref.~\cite{Mike01}.  Rather, we will 
discuss 
here a simpler analysis we performed where we contrasted the temperature
dependence of overdoped and underdoped samples \cite{Phen98}.
In this case, we chose to look at data at the $(\pi,0)-(\pi,\pi)$ Fermi 
crossing (antinode).  The dip/hump structure is considerably weaker here
than at the $(\pi,0)$ point, allowing us to concentrate on more general
aspects of the spectra.

We begin with the overdoped sample, where there is no
strong pseudogap effect. The simplest
self-energy which can describe the low energy data at all $T$ is
\begin{equation}
\Sigma({\bf k},\omega) = -i\Gamma_1
+ \Delta^2/[(\omega +i0^+) + \epsilon({\bf k})].
\label{eq8.3}
\end{equation}
Here $\Gamma_1$ is a single-particle scattering rate
taken, for simplicity, to be an $\omega$-independent constant.
It is effectively an average of the (actual $\omega$-dependent)
$\Sigma''$ over the frequency range of the fit.
(In Ref.~\cite{Mike01}, we generalized this to a constant plus an 
$\omega^2$ 
term, and included the resulting linear $\omega$ contribution to $\Sigma'$).
The second term is the BCS self-energy discussed previously.

\begin{figure}
\sidecaption
\includegraphics[width=.7\textwidth]{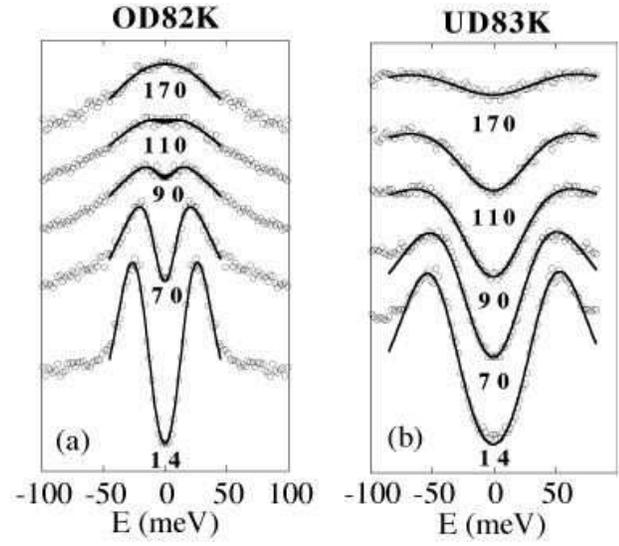}
\caption{
Symmetrized data for (a) a $T_c$=82K overdoped sample and (b) a $T_c$=83K 
underdoped sample at the $(\pi,0)-(\pi,\pi)$ Fermi
crossing at five temperatures, compared to
the model fits described in the text.}
\label{fig8.5}
\end{figure}

In Fig.~\ref{fig8.5}a, we show symmetrized data for an overdoped $T_c$=82K 
sample at the antinode together with
the fits obtained as follows \cite{Phen98}. 
Using Eq.~\ref{eq8.3}, we calculate the spectral function
\begin{equation}
\pi A({\bf k},\omega) = \Sigma''({\bf k},\omega) / 
\left[(\omega - \epsilon_{\bf k} - \Sigma'({\bf k},\omega))^2
+ \Sigma''({\bf k},\omega)^2\right],
\label{eq8.4}
\end{equation}
convolve it with the experimental resolution, and fit to 
symmetrized data.  The fit is restricted to a range of $\pm$45 meV 
given the small gap in the overdoped case and the sharpness of the 
quasiparticle peaks below $T_c$. We find that Eq.~\ref{eq8.3}
describes the low energy data quite well. 

\begin{figure}
\sidecaption
\includegraphics[width=.7\textwidth]{fig6.epsf}
\caption{
$\Delta$ (open circles), $\Gamma_1$ (solid circles), and $\Gamma_0$ (solid
squares) versus $T$ at the $(\pi,0)-(\pi,\pi)$ Fermi crossing for (a) a
$T_c$=82K overdoped sample and (b) a $T_c$=83K underdoped sample.
 The dashed line marks $T_c$.  The error bars for
$\Delta$ are based on a 10\% increase in the RMS error of the fits.}
\label{fig8.6}
\end{figure}

The $T$-variation of the fit parameters
$\Delta$ and $\Gamma_1$ are shown in Fig.~\ref{fig8.6}a.
$\Delta (T)$ decreases with $T$, and although small at $T_c$, it only vanishes
above $T_c$, indicating the possibility of a weak pseudogap.  This effect is
sample dependent, in that several overdoped samples we have looked at, the gap
vanishes closer to $T_c$.  We caution that the error bars shown in 
Fig.~\ref{fig8.6}a
are based on the RMS error of the fits, but do not take into account
experimental errors in $\mu$ and ${\bf k}_F$.

$\Gamma_1(T)$ is found to be
relatively $T$-independent in the normal state.
Below $T_c$,
we see that $\Gamma_1$ decreases very rapidly, and 
can be perfectly fit to the form $a + bT^6$.  
This rapid drop in linewidth leading to sharp quasiparticle peaks
at low $T$, which can be seen directly in the ARPES data, is
consistent with microwave and thermal conductivity 
measurements, and implies that electron-electron 
interactions are responsible for $\Gamma_1$.
Note the clear break in $\Gamma_1$ at $T_c$, despite the fact $\Delta$ has not
quite vanished.
We have seen similar behavior to that described above
for a variety of overdoped samples at several {\bf k} points.

We next turn to the more interesting underdoped case.  We find 
that near $(\pi,0)$ the self-energy (\ref{eq8.3}) cannot give an adequate
description of the data, in that it does not
properly describe the pseudogap and its unusual ``filling in" above $T_c$. 
Theoretically, we cannot have a divergence in $\Sigma({\bf k}_F,\omega=0)$
in a state without broken symmetry.
A simple modification of the BCS self-energy rectifies both these problems:
\begin{equation}
\Sigma({\bf k},\omega) = -i\Gamma_1 + \Delta^2 /
[\omega + \epsilon({\bf k}) + i\Gamma_0].
\label{eq8.5}
\end{equation}
The new term $\Gamma_0 (T)$ should be viewed as the inverse pair lifetime.
The theoretical motivation for Eq.~\ref{eq8.5} is given in Ref.~\cite{Phen98}.
We stress that this three parameter form is again a minimal representation 
of the pseudogap self-energy. Since it is not obviously a unique 
representation, it is very important to see what one learns from the fits.

In Fig.~\ref{fig8.5}b, we show symmetrized data at the antinode for a $T_c$=83K
underdoped sample. 
Below $T_c$ we see quasiparticle peaks. Above $T_c$ these 
peaks disappear but there is still a large suppression of spectral
weight around $\omega$=0.  As $T$ is raised further, the pseudogap fills in
(rather than 
closing) leading to a flat spectrum at a temperature of $T^*$ (200K). 
The self-energy, Eq.~\ref{eq8.5}, gives a good fit to the data.  These
fits were done below $T_c$ over a larger energy range ($\pm$75 meV) than in
the overdoped case because of the larger gap.
The range above $T_c$ was increased to $\pm$85 meV so as to
properly describe the pseudogap depression.

In Fig.~\ref{fig8.6}b, we show the $T$-dependence of the fit parameters.
We find a number of surprises.  First, $\Delta$
is independent of $T$ within error bars.  
Similar behavior has been inferred from specific heat
and tunneling data.
This $T$-independence is in contrast to the behavior of 
the overdoped 82K sample with almost identical $T_c$ at the same {\bf k} 
point. In addition, for the underdoped sample,
the gap evolves smoothly through $T_c$.  

The single-particle scattering rate $\Gamma_1(T)$ for the
underdoped sample is found to be qualitatively similar to the overdoped 
case. It is consistent with being
$T$-independent above $T_c$, but with a value over twice as large as the
overdoped case (allowing $\Gamma_1$ to vary above $T_c$ does not improve the
RMS error of the fits). 
Second, we see the same rapid decrease in $\Gamma_1$
below $T_c$ as in the overdoped case.
Note again the clear break at $T_c$.

The most interesting result is $\Gamma_0(T)$.
We find $\Gamma_0 = 0$ below $T_c$ and proportional to $T-T_c$ above.  
This behavior is robust, and is seen in all the fits that we have tried.
Moreover, a non-zero $\Gamma_0$ is needed above $T_c$ to obtain a proper fit
to the data (its effect cannot be reproduced by varying the other
parameters).
The fact that this $T$-dependence is exactly
what one expects of an inverse pair lifetime
is a non-trivial check on the validity of the
physics underlying Eq.~\ref{eq8.5}.
Further, we observe from Fig.~\ref{fig8.6} that $T^*$ corresponds to
where $\Delta(T) \sim \Gamma_0(T)$.  This condition can be understood from
the small $\omega$ expansion of Eq.~\ref{eq8.5}.

\begin{figure}
\sidecaption
\includegraphics[width=.7\textwidth]{fig7.epsf}
\caption{
(a) Symmetrized data for a $T_c$=77K underdoped sample for three temperatures
at (open circles) ${\bf k}_F$ point 1 in the zone inset, and at (open
triangles) ${\bf k}_F$ point 2, compared to the model fits.
(b) $\Delta(T)$ for these two k points (filled
and open circles), with $T_c$ marked by the dashed line.}
\label{fig8.7}
\end{figure}

The next important question is whether the $T$-dependence at the antinode
described above exists at other ${\bf k}_F$-points. 
To answer this, we have looked at 
$T$-dependent data for a number of underdoped samples at two different 
{\bf k} vectors.  All data at the antinode give results similar to those 
for the 83K sample.  
However at the second {\bf k}-point, about
halfway between the antinode and the node along the $(0,0)-(\pi,\pi)$ 
direction, we see quite different behavior.  
We demonstrate this in Fig.~\ref{fig8.7}a where symmetrized data for a 77K 
underdoped sample are shown. For the antinode, 
one clearly sees the gap fill in above $T_c$, with little evidence 
for any $T$-dependence of the position of the spectral feature 
defining the gap edge, just as for the 83K sample.  
In contrast, at the second {\bf k} point, the gap is clearly closing, 
indicating a strong $T$-dependence of $\Delta$. 
Similar behavior is seen in other underdoped samples with
$T_c$ between 75 and 85K.  

In Fig.~\ref{fig8.7}b, we show the $T$-dependence of
$\Delta$ obtained from fits (over a range of $\pm$66 meV)
at the second {\bf k} point for the 77K sample.
$\Delta$ is found to be strongly 
$T$-dependent, being roughly constant below $T_c$, then dropping 
smoothly to zero above. The strong $T$-dependence of $\Delta$ makes
it difficult to unambiguously determine $\Gamma_0$ from the fits at this
{\bf k}-point. On theoretical grounds, we expect that, here too, 
there is a non-zero $\Gamma_0$, and the closing of the pseudogap
is again determined by $\Delta(T) \sim \Gamma_0(T)$, however this condition
is satified by the rapid drop in $\Delta(T)$, rather than the rise in 
$\Gamma_0(T)$.  For completeness, we also show $\Delta(T)$ for this sample
at the antinode, which has a similar behavior to
that of the 83K sample.

We see that these results give further evidence for the unusual
{\bf k}-dependences first noted in Ref.~\cite{Nature98}.
Strong pairing correlations are seen over a very wide $T$-range
near $(\pi,0)$, but these effects are less pronounced
and persist over a smaller $T$-range as one moves closer to the 
zone diagonal. This is clearly tied to the strong
{\bf k}-dependence of the effective interaction and the unusual
(anomalously broad and non-dispersive) nature of electronic states 
near $(\pi,0)$.

\subsection{Peak/Dip/Hump - Experiment}

We now turn to a detailed discussion of the peak/dip/hump lineshape.
As mentioned above, 
a very broad normal state spectrum near the $(\pi,0)$
point of the zone evolves quite rapidly for $T < T_c$ into a narrow
quasiparticle peak,
followed at higher binding energies by a dip
then a hump, the latter corresponding to where
the spectrum recovers to its normal state value \cite{PDH}.
Similar effects are observed in tunneling spectra \cite{PDHT}.

\begin{figure}
\sidecaption
\includegraphics[width=.7\textwidth]{fig8.epsf}
\caption{
Spectra in (a) the normal state (105 K) and (b) the superconducting state (13 K)
along
the line $\Gamma-\bar{M}-Z$, and (c) the superconducting state (13 K) along
the line $\bar{M}-Y$ for an overdoped ($T_c$=87K) Bi2212 sample.
The zone is shown as an inset in (c) with
the curved line representing the observed Fermi surface.}
\label{fig8.8}
\end{figure}

In Fig.~\ref{fig8.8}, we show spectra for a $T_c=$87K Bi2212 sample
along $\Gamma-\bar{M}-Z$, 
i.e., $(0,0)-(\pi,0)-(2\pi,0)$, in (a) the normal state (105 K) and 
(b) the superconducting state (13 K), from which we note two striking
features \cite{Mike97}.
First, we see that the low energy peak in the superconducting state
persists over a large range in ${\bf k}$-space,
even when the normal state spectra have dispersed away from the Fermi energy. 
Second, when the hump in the superconducting state disperses, it essentially
follows that of the normal state spectrum.  This is accompanied by
a transfer of weight to the hump from the low frequency peak, which is fairly
fixed in energy.  The same phenomena are also seen along $\bar{M}$ to $Y$
(Fig.~\ref{fig8.8}c).  We will argue that the unusual dispersion seen in the 
superconducting state of Fig.~\ref{fig8.8} is closely tied
to the lineshape change discussed earlier (Fig.~\ref{fig8.3}).

The simplest explanation of the superconducting state spectra would be
the presence of two bands (e.g., due to
bilayer splitting), one responsible for the peak and the other 
for the hump.  However, this explanation is untenable \cite{Bilayer}.  First, 
if the sharp peak were associated with a second
band, then this band should also appear above $T_c$.  But there is no
evidence for it in the normal state data.  Second, if the peak and
hump were from two different bands, then their intensities must be governed
by different matrix elements.  However, we found \cite{DING_96} that the
intensities of both features scaled together as the photon polarization was
varied from in to out of plane, as if they were governed by a common matrix
element (Section 4.5).  These arguments suggest that the unusual lineshape 
and dispersion
represent a single electronic state governed by non-trivial many-body 
effects, as assumed in the previous discussion 
(Figs.~\ref{fig8.1}-\ref{fig8.4}).
For more overdoped materials, though, bilayer splitting should be taken into 
account, as discussed in Section 4.5.

Under this assumption,
the data are consistent with a strong reduction of the imaginary part of the
self-energy ($Im\Sigma$) at low energies in the superconducting state 
(Fig.~\ref{fig8.1}).  If
the scattering is electron-electron like in nature, then $Im\Sigma$ at
frequencies smaller than $\sim 3\Delta$ will be suppressed due to the opening
of the superconducting gap \cite{3Delta}.  On closer inspection, though,
a more interesting story emerges.
First, from Figs.~\ref{fig8.1} and \ref{fig8.8}, we see that the superconducting 
and normal state 
data match beyond 90 meV.  From 90 meV, the dip is quickly
reached at 70 meV, then one rises to the sharp peak.  
Notice that since the width of the peak is around 20 meV, then
the change in behavior of the spectra (from hump, to dip, to the trailing
edge of the peak) is occuring on the scale of the energy resolution.  That
means that the intrinsic dip must be quite sharp.  This implies that the large
$Im\Sigma$ at high energies must drop to a small value over a
narrow energy interval to be consistent with the data, i.e., there is
essentially a step in $Im\Sigma$.
In fact, the data are
not only consistent with a step in $Im\Sigma$, but the depth of the dip is
such that it is best fit by a peak in $Im\Sigma$ at the dip energy, followed
by a rapid drop to a small value.  This behavior can again be seen from 
the independent analysis shown in Figs.~\ref{fig8.1} and \ref{fig8.4}.

What are the consequences of this behavior in $Im\Sigma$?  If $Im\Sigma$ has
a sharp drop at $\tilde{\omega}$, then by Kramers-Kronig transformation,
$Re\Sigma$
will have a sharp peak at $\tilde{\omega}$ (Fig.~\ref{fig8.1}).  This peak
can very simply explain the unusual dispersion shown in
Fig.~\ref{fig8.8}, as it will cause a low energy quasiparticle
pole to appear even if the normal state binding energy is large.
The most transparent way to appreciate this result
is to note that a sharp step in $Im\Sigma$ is equivalent to
the problem of an electron interacting with a sharp (dispersionless) mode,
since in that case,
the mode makes no contribution to $Im\Sigma$ for energies below the mode
energy, and then makes a constant contribution for energies above.  This
problem has been
treated by Engelsberg and Schrieffer \cite{EngSch}, and extended to
the superconducting state by Scalapino and coworkers \cite{Parks}.
The difference in our case is that since the effect only occurs 
{\it below} $T_c$, it is
a consequence of the opening of the superconducting gap in the electronic
energy spectrum, and thus of a collective origin, rather than a phonon.

\begin{figure}
\sidecaption
\includegraphics[width=.4\textwidth]{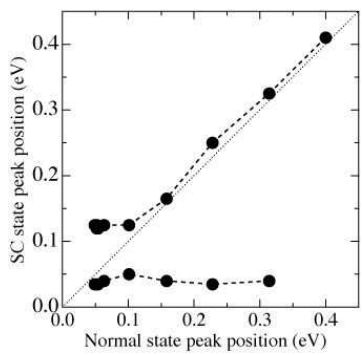}
\caption{
Positions of the sharp peak and the broad hump in the 
superconducting state versus normal state peak position obtained from
Fig.~\protect\ref{fig8.8}a and \protect\ref{fig8.8}b.  
Solid points connected by a dashed line are the data, 
the dotted line represents the normal state dispersion.}
\label{fig8.9}
\end{figure}

To facilitate comparison to this classic work, in Fig.~\ref{fig8.9} we plot the 
position
of the low energy peak and higher binding energy hump as a function of the
energy of the single broad peak in the normal state.  This plot has a striking
resemblance to that predicted for electrons interacting with a sharp mode
in the superconducting state, and one clearly sees the low
energy pole which we associate with the peak in $Re\Sigma$.
On general grounds, the flat dispersion of the
low energy peak seen in Fig.~\ref{fig8.9} is a combination of two effects:
(1) the
peak in $Re\Sigma$, which provides an additional mass renormalization of the
superconducting state relative to the normal state, and thus pushes spectral 
weight towards
the Fermi energy, and (2) the superconducting gap, which pushes spectral
weight away.  This also explains the strong drop in intensity of the low energy
peak as the higher binding energy hump disperses.

An important feature of the data is the dispersionless nature of the sharp
peak.  The mode picture discussed above would imply a dispersion of
the peak from $\Delta_k$ to $\tilde{\omega}=\omega_0+\Delta_k$ as the normal
state binding energy increases (where $\omega_0$ is the mode energy). However,
this dispersion turns out to be weak.
From the data of Fig.~\ref{fig8.1}, we infer an $\omega_0=1.3\Delta_{max}$,
$\omega_0$ being essentially the energy separation of the peak and dip.
Since $\Delta_k$ is known to be of the $d_{x^2-y^2}$ form, then $\Delta_k$
should go to zero as we disperse towards the
$\Gamma$ point.  Therefore, the predicted dispersion is only from 
$\Delta_{max}$ to $1.3\Delta_{max}$ (32 to 42 meV).

Since the dip/hump
structure is most apparent at the $(\pi,0)$ points, it is natural
to assume that it has something to do
with $Q=(\pi,\pi)$ scattering, as discussed by Shen and
Schrieffer \cite{Shen97}. But here, we find a new effect. 
If one compares the
data of Figs.~\ref{fig8.8}b and \ref{fig8.8}c, one sees that a low energy peak 
also exists along
$(\pi,0)-(\pi,\pi)$ for approximately the same momentum range as the one
from $(\pi,0)-(0,0)$.  That is, if there is a peak for momentum $p$,
one also exists for momentum $p+Q$.  This can be understood, since
the self-energy equations for $p$ and $p+Q$ will be strongly coupled if
$Q$ scattering is dominant.

\subsection{Mode Model}

\begin{figure}
\sidecaption
\includegraphics[width=.4\textwidth]{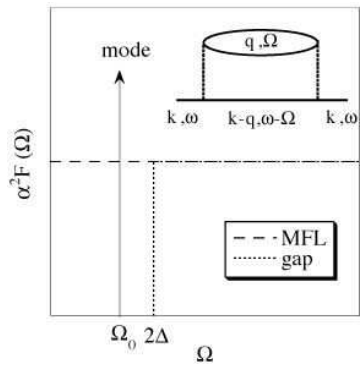}
\caption{
$\alpha^2F$ for three models, MFL 
(dashed line), gapped MFL (dotted line), and gapped MFL plus mode (dotted
line plus $\delta$ function).  Inset:
Feynman diagram for the lowest order contribution to $\Sigma$ from 
electron-electron scattering.}
\label{fig8.10}
\end{figure}

For now, though, we ignore the complication of momentum dependence.  
The lowest order contribution to electron-electron scattering is 
represented by the Feynman diagram shown in the inset of Fig.~\ref{fig8.10}.
In the 
superconducting state, each internal line will be gapped by $\Delta$.  
This implies that the scattering will be suppressed for $|\omega| < 
3\Delta$.  This explains the presence of a sharp 
quasiparticle peak at low temperatures.  What is not 
so obvious is whether this in addition explains the strong spectral dip.  
Explicit calculations show only a weak dip-like feature \cite{Little92}.  To 
understand this in detail, we equate the bubble plus interaction lines
(Fig.~\ref{fig8.10} inset) to an ``$\alpha^2F$'' as 
in standard strong-coupling literature.  In a marginal Fermi liquid 
(MFL) at T=0, $\alpha^2F(\Omega)$ is simply a constant in $\Omega$.  The 
effect of the gap is to force $\alpha^2F$ to zero for $\Omega < 
2\Delta$.  The question then arises where the gapped weight goes.  It 
could be distributed to higher energies, but in light of the above 
discussion, we might expect it to appear as a collective mode inside of 
the $2\Delta$ gap.  For instance, if the bubble represents spin
fluctuations, a sharp 
mode will appear if the condition $1-U\chi_0({\bf q},\Omega)=0$ is 
satisfied for $\Omega < 2\Delta$.  These three cases (MFL, gapped MFL, 
gapped MFL plus mode) are illustrated in Fig.~\ref{fig8.10}.

\begin{figure}
\includegraphics[width=.7\textwidth]{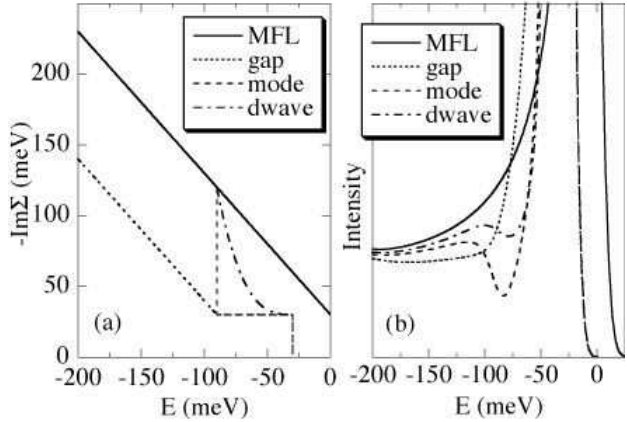}
\caption{
(a) $Im\Sigma$ for MFL (solid line), gapped MFL (dotted line), gapped 
MFL plus mode (dashed line), and simple d-wave model (dashed-dotted line).
Parameters are $\alpha$=1, $\omega_c$=200meV, $\Delta$=30meV (0 for MFL),
$\Omega_0$=2$\Delta$, and $\Gamma_0$=30meV.
(b) Spectral functions (times a Fermi function with $T$=14K) convolved with a 
resolution gaussian of $\sigma$=7.5 meV for these four cases
($\epsilon$=-34meV).}
\label{fig8.11}
\end{figure}

$\Sigma$ is easy to obtain analytically if we ignore the 
complication of the superconducting density of states from the 
$\bf k-q$ line of Fig.~\ref{fig8.10} and just replace this by a step function at 
$\Delta$.  The resulting $Im\Sigma$ for the gapped MFL and gapped MFL plus
mode models are shown in Fig.~\ref{fig8.11}a \cite{ND98} in comparison to the 
normal state MFL. 
Note that structure in $\alpha^2F$ at $\Omega$ appears in $\Sigma$ at 
$|\omega|=\Omega+\Delta$ due to the gap in the $\bf k-q$ line.  Moreover, the
MFL plus mode is simply the normal state MFL cut-off at $3\Delta$ (this is
obtained under the assumption that all the gapped weight in $\alpha^2F$ 
shows up in the mode).  In contrast, the gapped MFL decays linearly to zero
at $3\Delta$.

The Nambu spectral function is given by
\begin{equation}
A(\omega) = \frac{1}{\pi}Im \frac{Z\omega + \epsilon}
{Z^2(\omega^2-\Delta^2)-\epsilon^2}
\label{eq8.6}
\end{equation}
with (a complex) $Z(\omega) = 1 - \Sigma(\omega)/\omega$.
These are shown in Fig.~\ref{fig8.11}b and were convolved with a gaussian of
$\sigma$=7.5 meV, typical of high resolution ARPES, with a
constant $Im\Sigma$ ($\Gamma_0$) added
for $|\omega| > \Delta$ to reduce the size of the quasiparticle peak.
We note that there is no dip as such for the gapped MFL model, whereas the
addition of the mode causes a significant dip.  The latter behavior is
consistent with 
experiment.  Moreover, the mode model has the additional advantage that  
$Im\Sigma$ recovers back to the normal state value by $3\Delta$, which is 
also in agreement with experiment in that the normal and superconducting 
state spectra agree beyond 90 meV (Figs.~\ref{fig8.1} and \ref{fig8.8}).

We contrast this behavior with that expected for a simple d-wave 
model.  To a first approximation, this can be obtained by replacing the 
step drop in $Im\Sigma$ in the MFL plus mode model with 
$(|\omega|-\Delta)^3$ for $|\omega| < 3\Delta$ \cite{Quinlan96}.  This is 
shown in Fig.~\ref{fig8.11}a as well, with the resulting spectrum in 
Fig.~\ref{fig8.11}b.  Only a
weak dip appears.  Moreover, we have analyzed models with 
the exponent 3 replaced by some n and have found that n must be large
to obtain a dip as strong as seen in experiment.  
Therefore, the upshot is that at the least, something similar to a step 
is required in $Im\Sigma$ to be consistent with experiment.

In principle, we could take the above MFL plus mode model and fit 
experiment with it.  We consider a simpler model.  
There are several reasons for this.  First, the MFL model has a number 
of adjustable parameters associated with it.  There is the coupling
constant ($\alpha$), the cut-off frequency ($\omega_c$), and
the mode energy (which is not in general $2\Delta$).  Moreover, 
the spectrum for $\bf k$ points near the $(\pi,0)$ point does not appear to be
MFL-like in nature.  We have found that the 
normal state Bi2212 spectrum is fit very well by a Lorentzian plus a 
constant background in an energy range less than $0.5$eV.  This is also true for
Bi2201 spectrum where 
the normal state can be accessed to much lower temperatures.

In the resulting Lorentzian model, the normal state $\Sigma$ is purely an 
imaginary constant, and $\alpha^2F$ is a mode at zero 
energy.  In the superconducting state, this mode gets pushed back to some 
energy within $2\Delta$.  This model is artificial in the sense that 
all the self-energy is being generated by the mode.  That is why we went 
through the above discussion motivating the mode more properly as a 
rearrangement of $\alpha^2F$ due to the superconducting gap.  In 
practice, though, the results are very similar to the MFL plus mode model, 
and has the further advantage of having the several parameters of that 
model collapse to just the mode strength ($\Gamma_1$) and mode position 
($\Omega_0$) of the Lorentzian model.  Moreover, analytic results can 
still be obtained for $\Sigma$ when the superconducting density of states 
for the $\bf k-q$ line of Fig.~\ref{fig8.10} is taken into account.  The 
result is \cite{ND98}
\begin{eqnarray}
-Im\Sigma(\omega)&=&\Gamma_0N(|\omega|) + \Gamma_1N(|\omega|-\Omega_0), \;
 |\omega| > \Omega_0+\Delta \nonumber \\
                &=&\Gamma_0N(|\omega|), \;
 \Delta < |\omega| < \Omega_0+\Delta \nonumber \\
                &=&0, \;  |\omega| < \Delta
\label{eq8.7}
\end{eqnarray}
where $N(\omega)= \omega/\sqrt{\omega^2-\Delta^2}$ 
is the BCS density of states, and
\begin{eqnarray}
\pi Re\Sigma(\omega) = \Gamma_0N(-\omega) \ln\left[{|-\omega+
\sqrt{\omega^2-\Delta^2}|}/{\Delta}\right] \nonumber \\
 + \Gamma_1N(\Omega_0-\omega) \ln\left[{|\Omega_0-\omega
+\sqrt{(\omega-\Omega_0)^2-\Delta^2}|}/{\Delta}\right] \nonumber \\
- \{\omega \rightarrow -\omega\}
\label{eq8.8}
\end{eqnarray}
where it has again been assumed that $\Delta$ is a real constant in 
frequency.  An s-wave density of states has been used to obtain an 
analytic result.  A d-wave density of states will not be that different.
The advantage of 
an analytic result is that it is useful when having to take spectra and 
convolve with resolution to compare to experiment. 
Our results are not very sensitive to $\Gamma_0$, included 
again to damp the quasiparticle peak.

\begin{figure}
\sidecaption
\includegraphics[width=.7\textwidth]{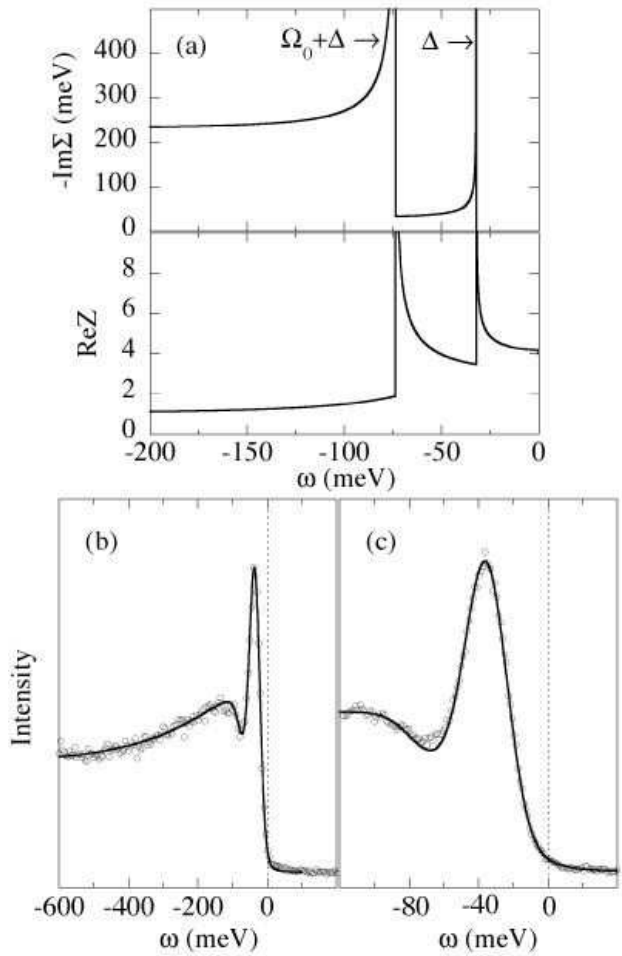}
\caption{
(a) $Im\Sigma$ and $ReZ$ at $(\pi,0)$ from Eqs.~\protect\ref{eq8.7} and 
\protect\ref{eq8.8}
($\Gamma_1$=200meV, $\Gamma_0$=30meV, $\Delta$=32meV, $\Omega_0$=1.3$\Delta$).
Comparison of the data at $(\pi,0)$ for (b) wide and (c) narrow energy scans
with calculations based on Eqs.~\protect\ref{eq8.6}-\protect\ref{eq8.8}, with 
an added step edge background contribution.}
\label{fig8.12}
\end{figure}

The resulting real (Eq.~\ref{eq8.8}) and imaginary (Eq.~\ref{eq8.7}) parts 
of $\Sigma$ at
$(\pi,0)$ are shown in Fig.~\ref{fig8.12}a. Note the singular behaviors 
at $\Delta$ (peak energy) due to the $\Gamma_0$
term and at $\Omega_0+\Delta$ (dip energy) due to the $\Gamma_1$ term.
In both cases, 
step drops in $Im\Sigma$ would also give singularities in $Re\Sigma$.  
The advantage of peaks in $Im\Sigma$ (due to the SC density of states)
is that it makes the dip deeper in better agreement with experiment.
In Fig.~\ref{fig8.12}b and \ref{fig8.12}c, we show a comparison of the 
resulting spectral
function (convolved with the experimental energy and momentum resolution)
to experimental data at $(\pi,0)$ for both wide and narrow energy
scans, where a step edge background with a gap of $\Delta$ is added
to the calculated spectrum.  The resulting agreement 
is excellent.

It is interesting to note that the mode energy we infer from 
the data is 41 meV, equivalent to a magnetic resonant mode 
energy observed in YBCO \cite{YBCO} and Bi2212 \cite{BSCO} by neutron 
scattering data at ${\bf Q}=(\pi,\pi)$.  
The $Q$ dependence of this mode correlates well with the
observations of Fig.~\ref{fig8.8}.  To explore this in greater detail, we now 
consider the doping dependence of the peak/dip/hump structure.

\subsection{Doping Dependence}

\begin{figure}
\sidecaption
\includegraphics[width=.7\textwidth]{fig13.epsf}
\caption{
Spectra along $(\pi,0) \to (\pi,\pi)$ in (a)
the superconducting state (T=60K), and (b) the pseudogap state
(T=100K) for an underdoped 75K sample (curves are labeled in units of
$\pi/a$). The thick vertical bar indicates the position of the higher
energy feature, at which the spectrum changes slope as highlighted by 
the intersecting straight lines.}
\label{fig8.13}
\end{figure}

We show 
data along $(\pi,0) \to (\pi,\pi)$ for an underdoped 75K sample in the 
superconducting state (Fig.~\ref{fig8.13}a) and in the pseudogap state
(Fig.~\ref{fig8.13}b) \cite{JC99}. 
Below $T_c$, the sharp peak at low energy is essentially dispersionless, 
while the higher energy hump rapidly disperses from
the $(\pi,0)$ point towards the $(\pi,0) \to (\pi,\pi)$ Fermi crossing
seen above $T^*$.  Beyond this, the intensity drops dramatically,
but there is clear evidence that the hump disperses back to higher 
energy.  In the pseudogap state, the high energy feature also shows 
strong dispersion, much like the
hump below $T_{c}$, even though the leading edge is non-dispersive
like the sharp peak in the superconducting state.

\begin{figure}
\sidecaption
\includegraphics[width=.7\textwidth]{fig14.epsf}
\caption{
Doping dependence of the dispersion from (a) $(\pi,0) \to (\pi\pm\pi,0)$, 
(b) $(\pi,0) \to (\pi,\pm\pi)$, and (c) both directions, for the
peak and hump in the superconducting state. U is underdoped and O is
overdoped. Points were obtained by polynomial fits to the data, and 
are consistent with the simpler criterion used in Fig.~\protect\ref{fig8.13}.}
\label{fig8.14}
\end{figure}

In Fig.~\ref{fig8.14} we show the dispersion of the sharp peak and hump
(below $T_c$), for a variety of doping levels,
in the vicinity of the $(\pi,0)$ point along the two principal axes.
The sharp peak at low energies is seen to be essentially
non-dispersive along both directions for all doping levels, while
the hump shows very interesting dispersion.
Along $(\pi,0) \to (0,0)$ (Fig.~\ref{fig8.14}a), the hump exhibits
a maximum, with an eventual dispersion away from the Fermi energy,
becoming rapidly equivalent to the binding energy of the broad peak in
the normal state as one moves away from the region near $(\pi,0)$.
In the orthogonal direction (Fig.~\ref{fig8.14}b), since the hump
initially disperses towards the $(\pi,0) \to (\pi,\pi)$
Fermi crossing, which is known to be a weak function of
doping, one obtains the rather dramatic effect that the
dispersion becomes stronger with underdoping.
We also note that there is an energy separation between
the peak and the hump due to the spectral dip. In essence, the hump
disperses towards the spectral dip, but cannot cross it, with its
weight dropping strongly as the dip energy is approached.  Beyond this point,
one sees evidence of the dispersion bending back to higher binding energy for
more underdoped samples.

\begin{figure}
\includegraphics[width=.7\textwidth]{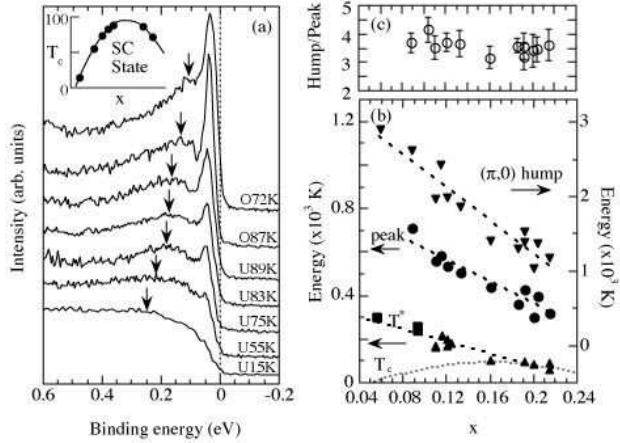}
\caption{Dependence of energy scale on carrier density:
(a) Doping dependence of the spectra (T=15K) at
the $(\pi,0)$ point. The inset shows $T_c$ vs. doping.
(b) Doping dependence of $T^{*}$, and the peak and hump binding energies
in the superconducting state along with their
ratio (c), as a function of doping, $x$. The empirical relation between $T_c$
and $x$ is given by $T_c/T_c^{max}=1-82.6(x-0.16)^2$ with
$T_c^{max}$=95K. For $T^{*}$, solid squares represent lower bounds.}
\label{fig8.15}
\end{figure}

Fig.~\ref{fig8.15}a shows the evolution of the low temperature spectra at the 
$(\pi,0)$ 
point as a function of doping. The sharp quasiparticle peak moves to higher 
energy, indicating that the gap increases with underdoping 
(although this is difficult to see on the scale of Fig.~\ref{fig8.15}a).
We see that 
the hump moves rapidly to higher energy with underdoping. 
These trends can be seen very clearly in Fig.~\ref{fig8.15}b, where the energy 
of the 
peak and hump are shown as a function of doping for a large number of samples.
Finally, we observe that the quasiparticle peak loses spectral weight with 
increasing underdoping, as expected for a doped Mott 
insulator; in addition 
the hump also loses spectral weight though less rapidly.

\begin{figure}
\includegraphics[width=.7\textwidth]{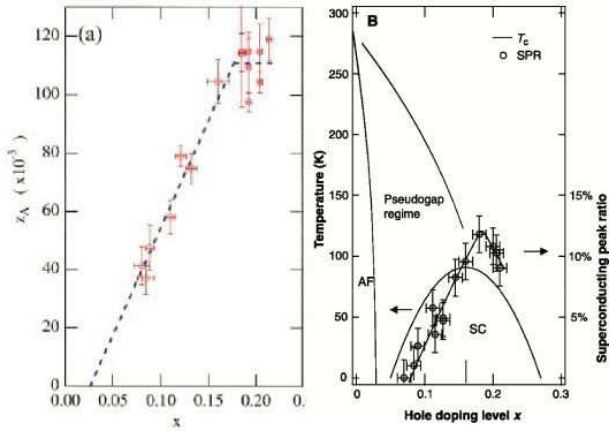}
\caption{(a) Doping dependence of the low-T (14 K)
coherent weight (z$_A$). The dashed line is a guideline showing
that z$_A$ increases linearly on the underdoped side, and tapers off
on the overdoped side. (from Ref.~\cite{DING_00})
(b) The doping dependence of the superconducting peak
ratio (SPR) is plotted over a 
typical Bi2212 phase diagram. The solid line is a 
guide to the eye. Horizontal error bars denote uncertainty in determining 
the doping level ($\pm$0.01); vertical error bars denote uncertainty in 
determining the SPR ($\pm$1.5\%). AF, antiferromagnetic regime; SC, 
superconducting regime. (from Ref.~\cite{FENG_00}) }
\label{specwt}
\end{figure}

This effect has recently been quantified in greater detail, where it was
found that the spectral weight of the peak varies linearly with
doping, as reproduced in Fig.~\ref{specwt}. \cite{FENG_00,DING_00}.  We
remark that Ding \etal \cite{DING_00} also found the unusual relation that the
product of the peak weight times the peak energy is constant with doping.

The hump below $T_c$ is clearly related to the
superconducting gap, given the weak doping
dependence of the ratio between the hump and quasiparticle peak positions at
$(\pi,0)$, shown in Fig.~\ref{fig8.15}c.
Tunneling data find this same correlation on a wide variety of
high-$T_{c}$ materials whose energy gaps vary by a factor of 
30 \cite{JohnZ96}.

To motivate the analysis below that firmly establishes
the mode interpretation of of the peak/dip/hump spectra and its connection
with neutron data, we note
that the spectral dip represents a pairing induced gap in the incoherent
part of the spectral function at $(\pi,0)$ occurring at an energy
$\Delta + \Omega_0$,  where $\Delta$ is the superconducting gap and
$\Omega_0$ is the mode energy. We can estimate the mode energy from 
ARPES data from the energy difference between the dip ($\Delta + \Omega_0$)
and the quasiparticle peak ($\Delta$).

\begin{figure}
\includegraphics[width=.7\textwidth]{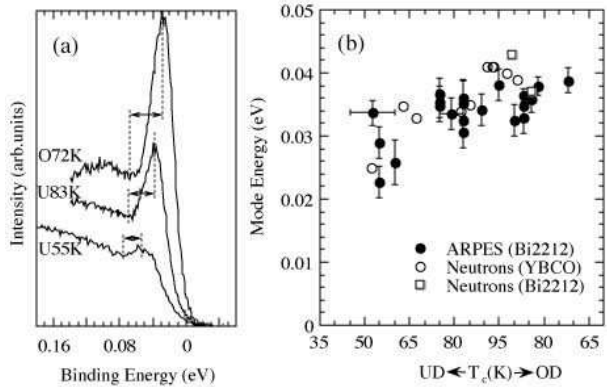}
\caption{Doping dependence of the mode energy:
(a) Spectra at $(\pi,0)$ showing the decrease in the energy separation
of the peak and dip with underdoping. Peak and dip locations were 
obtained by independent polynomial fits and carefully checked for the 
effects of energy resolution.
(b) Doping dependence of the collective mode energy inferred from ARPES
together with that inferred from neutron data \cite{Bourges98}.}
\label{fig8.16}
\end{figure}

\begin{figure}
\sidecaption
\includegraphics[width=.5\textwidth]{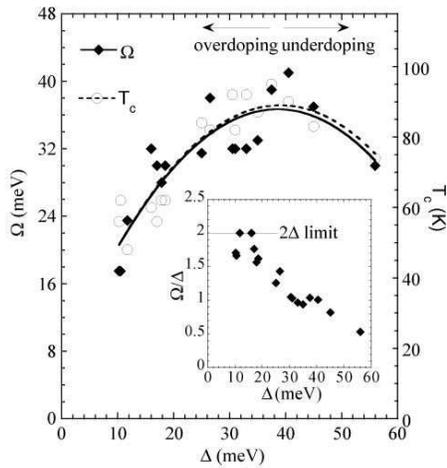}
\caption{
Measured mode energy $\Omega$ and bulk $T_c$ value
vs.\ measured gap value $\Delta$ for 17 junctions over a wide doping
range from UD74K to OD48K.  Solid and dashed lines are quadratic fits of
$\Omega$ and $T_c$ vs.\ $\Delta$.  Inset shows $\Omega /\Delta$ vs.\ $\Delta$
(from Ref.~\cite{JohnZ01}).}
\label{johnz}
\end{figure}

In Fig.~\ref{fig8.16}b we plot the mode energy as estimated from ARPES for 
various
doping levels as a function of $T_c$ and compare it with neutron measurements.
We find striking agreement both in terms of the energy scale
and its doping dependence \cite{Bourges98}. The same agreement, in greater
detail, has been recently found using tunneling data \cite{JohnZ01}, as shown
in Fig.~\ref{johnz}.
We note that the mode energy inferred
from ARPES decreases with underdoping, just like the
neutron data, unlike the gap energy (Fig.~\ref{fig8.15}b), which increases.
This can be seen directly in the raw data, shown in Fig.~\ref{fig8.16}a.
This is also seen from the tunneling data, where they have found that the
mode energy scales with doping as 5$T_c$, just like the neutron resonance.
An interesting point from the tunneling is that the ratio of the mode energy
to the gap energy saturates to 2 in the overdoped limit, as would be expected
for a collective mode sitting below a continuum with a gap of 2$\Delta$.
Moreover, there is strong correlation between the temperature
dependences in the ARPES and neutron data.  While neutrons see a sharp mode
only below $T_c$, a smeared out remnant persists up to $T^*$ \cite{Dai96}.
As the sharpness of the mode is responsible for the sharp spectral dip, one
then sees the
correlation with ARPES where the dip disappears above $T_c$, but with
a remnant of the hump persisting to $T^*$.

An important feature of the neutron data is that the mode only exists
in a narrow momentum range about $(\pi,\pi)$, and is magnetic in
origin.  To see a further connection
with ARPES, we return to the results of Fig.~\ref{fig8.14}.  Note the dispersion
along the two orthogonal directions are similar (Fig.~\ref{fig8.14}c),
unlike the dispersion inferred in the normal state.
As these two directions are related by a $(\pi,\pi)$ translation
($(x,0)\equiv (0,-x); (0,-x)+(\pi,\pi) = (\pi,\pi-x)$), we see that the
hump dispersion is clearly reflecting the $(\pi,\pi)$ nature of the
collective mode.  This dispersion is also consistent
with a number of models
in the literature which identify the high energy
feature in the pseudogap regime as a remnant of the insulating 
magnet \cite{InsTh}.
We note, though, that the mode is due to quasiparticle pair creation and thus
not just a continuation of the spin wave mode from the
antiferromagnet \cite{ModeTh}.

This brings up a question that is at the heart of the high $T_c$ problem:
how can a feature which can be understood as a strong coupling effect
of superconductivity, as discussed above, turn out to have a dispersion that
resembles
that of a magnetic insulator?  The reason is that the collective mode
has the same wavevector, $(\pi,\pi)$, which characterizes the magnetic order
of the insulator.
It is easy to demonstrate that in the limit that the mode energy goes
to zero (long range order), one actually reproduces a
symmetric dispersion similar to that in Fig.~\ref{fig8.14}c, with the spectral 
gap determined by the strength of the mode \cite{KAMPF}.  This is in 
accord with the
increase in the hump energy with underdoping (Fig.~\ref{fig8.15}b) tracking the 
rise in the neutron mode intensity.
Since the hump scales with the superconducting gap,
the obvious implication is that the mode is intimately connected with
pairing, a conclusion which can also be made by relating the mode to the
superconducting condensation energy \cite{Dai99}.  That is, high $T_c$
superconductivity is likely due to the same magnetic correlations which
characterize the insulator and give rise to the mode.

\subsection{Dispersion Kink of Nodal Quasiparticles}

So far, our discussion has largely been centered on behavior near the 
$(\pi,0)$ point of the zone.  We now turn to consideration of other ${\bf 
k}$ vectors.

Remarkably, we find that the effects discussed above are manifest
even on the zone diagonal where the gap vanishes,
with significant changes in both the
spectral lineshape and dispersion below $T_c$,
relative to the normal state where the nodal points exhibit
quantum critical scaling \cite{Valla99}.
Specifically, below $T_c$ a kink in the dispersion develops along 
the diagonal at a finite energy ($\sim$70 meV) \cite{Kink,Adam01}. This 
is accompanied, as required by Kramers-Kr\"onig relations, by a 
reduction in the linewidth leading
to well-defined quasiparticles \cite{Adam00}.  As one moves
away from the node, the renormalization
increases, and the kink in dispersion along the diagonal smoothly
evolves into the spectral dip, with the same characteristic
energy scale throughout the zone.

\begin{figure}
\sidecaption
\includegraphics[width=.7\textwidth]{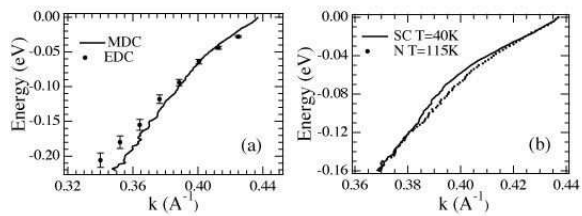}
\caption{
ARPES data along the $(\pi,\pi)$ direction at $h\nu$=28eV.
(a) EDC dispersion in the normal state compared to the MDC dispersion.
The EDCs are shown in Fig.~\protect\ref{fig8.18}d.
(b) MDC dispersions in the superconducting state (T=40K) and
normal state (T=115K). (c) change in MDC dispersion from (b).
(d) Kramers-Kr\"onig transform of (c).
}
\label{fig8.17}
\end{figure}

In Fig.~\ref{fig8.17}a, we plot the dispersion of the spectral peak above $T_c$
obtained from constant ${\bf k}$ scans (energy distribution curves or
EDCs),
and the peak in momentum obtained from constant $\omega$ scans
(momentum distribution curves or MDCs) \cite{Valla99}
from data for a $T_c$=90K sample along the $(\pi,\pi)$ direction \cite{Adam01}.
We find that the EDC and MDC peak dispersions
are very different, a consequence of the $\omega$ dependence of
$\Sigma$.

To understand this, we start by
noting that since $\epsilon_{{\bf k}} \simeq v_F^0(k-k_F)$,
then from Eq.~8 the
MDC at fixed $\omega$ is a Lorentzian centered at
$k = k_F + \left[\omega-\Sigma^{\prime}(\omega)\right]/v_F^0$,
with a width (HWHM) $W_M = |\Sigma^{\prime\prime}(\omega)|/v_F^0$,
{\it provided} (i) $\Sigma$ is essentially independent
of $k$ normal to the Fermi surface, {\it and} (ii) the dipole matrix 
elements do not vary significantly with $k$
over the range of interest.  That these two conditions are
fulfilled can be seen by the nearly Lorentzian
MDC lineshape observed in ARPES \cite{Valla99}.

On the other hand, in general, the EDC at fixed ${\bf k}$ 
has a non-Lorentzian lineshape reflecting the non-trivial
$\omega$-dependence of $\Sigma$, in addition to the Fermi cutoff at 
low energies. Thus the EDC peak is {\it not} given by
$\omega - v_F^0(k-k_F) - \Sigma^{\prime}(\omega) = 0$
but also involves $\Sigma^{\prime\prime}$, unlike the MDC peak.
Further, if the EDC peak is sharp enough, making a Taylor expansion 
we find that its width (HWHM) is given by $W_E \simeq |\Sigma^{\prime\prime}
(E_k)|/[1 - \partial\Sigma^{\prime}/\partial\omega|_{E_k}]$, 
where $E_k$ is the peak position.

We see that it is much simpler to interpret
the MDC peak positions, and thus focus on the
change in the MDC dispersion going from the normal (N) to the
superconducting (SC) state shown in Fig.~\ref{fig8.17}b.
The striking feature of Fig.~\ref{fig8.17}b is the development of a kink
in the dispersion below $T_c$.
At fixed $\omega$ let the dispersion change from $k_N$ to $k_{SC}$.
Using $v_F^0(k_N - k_{SC}) = \Sigma_{SC}^{\prime}(\omega)
- \Sigma_N^{\prime}(\omega)$, we directly obtain the change in
real part of $\Sigma$ plotted in Fig.~\ref{fig8.17}c. The
Kramers-Kr\"onig transformation of $\Sigma_{SC}^{\prime} -
\Sigma_N^{\prime}$
then yields $\Sigma_{N}^{\prime\prime} - \Sigma_{SC}^{\prime\prime}$,
plotted in Fig.~\ref{fig8.17}d, which shows that
$|\Sigma_{SC}^{\prime\prime}|$ is smaller than
$|\Sigma_N^{\prime\prime}|$
at low energies.

\begin{figure}
\includegraphics[width=.7\textwidth]{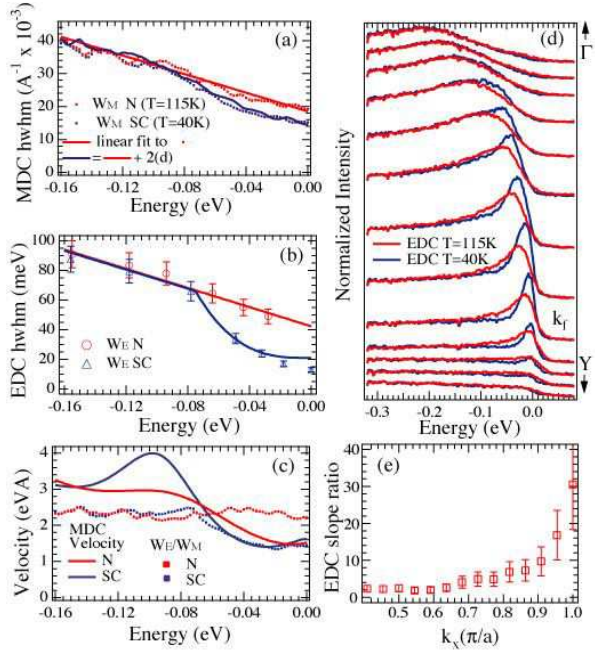}
\caption{
(a) Comparison of change in $\Sigma^{\prime\prime}$ obtained directly
from the MDC widths (HWHM) to the one obtained from the dispersion in 
Fig.~\protect\ref{fig8.17}d by using the Kramers-Kr\"onig transform.  
(b) HWHM width obtained 
from EDCs shown in (d). Lines marked by fit are linear in normal state 
and linear/cubic in superconducting state.  The data in (b) fall below 
the fits at low energies because of the Fermi cut-off of the EDCs.
(c) Renormalized MDC velocity obtained from differentiating 
Fig.~\protect\ref{fig8.17}b 
(solid lines), compared to the ratio $W_{E}/W_{M}$ from (a) and (b). 
(e) Ratio of EDC dispersion slopes above and below the kink energy at various
points along the Fermi surface (from middle panels of 
Fig.~\protect\ref{fig8.19}).}
\label{fig8.18}
\end{figure}

We compare these results in Fig.~\ref{fig8.18}a with the
$W_M = |\Sigma^{\prime\prime}|/v_F^0$ estimated directly from the
MDC Lorentzian linewidths.
The normal state curve was obtained
from a linear fit to the corresponding MDC width data points in 
Fig.~\ref{fig8.18}a,
and then the data from Fig.~\ref{fig8.17}d was added to it to generate the
low temperature curve.
We are thus able to make a quantitative connection between the
appearance of a kink in the (MDC) dispersion below $T_c$ and
a drop in the low energy scattering
rate in the superconducting state relative to the normal state,
which leads to the appearance of quasiparticles below $T_c$ \cite{Adam00}.
We emphasize that we have estimated these $T$-dependent changes in the
complex self-energy without making fits to the EDC
lineshape, thus avoiding the problem of modeling the $\omega$ dependence
of $\Sigma$ and the extrinsic background.

In Fig.~\ref{fig8.18}b, we plot the EDC width obtained as explained in 
Ref.~\cite{Adam00} 
from Fig.~\ref{fig8.18}d.  As an interesting exercise, we present in 
Fig.~\ref{fig8.18}c the 
ratio of this EDC width to the MDC width of Fig.~\ref{fig8.18}a (dotted lines), 
and compare it to the renormalized MDC velocities, 
$1/v\equiv dk/d\omega$, obtained directly by numerical differentiation 
of Fig.~\ref{fig8.17}b (solid lines). We note that only for a sufficiently 
narrow EDC lineshape is the ratio $W_E/W_M \simeq v_F^0 / 
[1 - \partial\Sigma^{\prime}/\partial\omega] = v_F$. Interestingly, 
only in the superconducting state below the kink energy do these two
quantities agree, 
which implies that only in this case does one have a Fermi liquid.

Similar kinks in the dispersion have been seen by ARPES in normal metals 
due to the electron-phonon interaction \cite{Baer99}. Phonons cannot be 
the cause here, since the kink disappears above $T_c$. Rather, this effect 
is suggestive of coupling to an electronic collective
excitation which only appears below $T_c$.  Recently, this view has been
challenged by Lanzara \etal \cite{Shen01}, and we discuss this work at the
end of this subsection.

We now study how the lineshape and dispersion evolve as we
move along the Fermi surface.  An analysis similar to the above is 
possible, but more complicated due to the presence of an energy
gap \cite{MDC01}.  We will thus confine ourselves here to a general 
description of the data.  In Fig.~\ref{fig8.19},
we plot raw intensities for a series of
cuts parallel to the $MY$ direction (normal state in left panels,
superconducting state in middle panels). We start from
the bottom row that corresponds to a cut close to the node and reveals
the same kink described above. As we move
towards $(\pi,0)$, the dispersion kink (middle panels) becomes more
pronounced and
at around $k_{x}$=0.55 develops into a break separating the
faster dispersing high energy part of the spectrum from the slower
dispersing low energy part. This break leads to the
appearance of two features in the EDCs, shown in the right panels of
Fig.~\ref{fig8.19}. Further towards $(\pi,0)$, the low energy feature,
the quasiparticle peak, becomes almost dispersionless.
At the $(\pi,0)$ point, this break effect becomes the most pronounced,
giving rise to the peak/dip/hump in the EDC.
We note that there is a continuous evolution in the zone from kink to
break, and these features all occur at exactly the same energy.

\begin{figure}
\sidecaption
\includegraphics[width=.7\textwidth]{fig19.epsf}
\caption{
Left panels: Log of normal state ($h\nu$=22eV, T=140K) ARPES intensity
along selected cuts parallel to $MY$. EDC peak positions are indicated by
crosses. Middle panels: Log of superconducting state (T=40K) intensity
at the
same cuts as for left panels. Crosses indicate positions of broad high
energy
peaks, dots sharp low energy peaks.  Right panels: EDCs at locations
marked
by the vertical lines in the middle panels.}
\label{fig8.19}
\end{figure}

\begin{figure}
\includegraphics[width=1\textwidth]{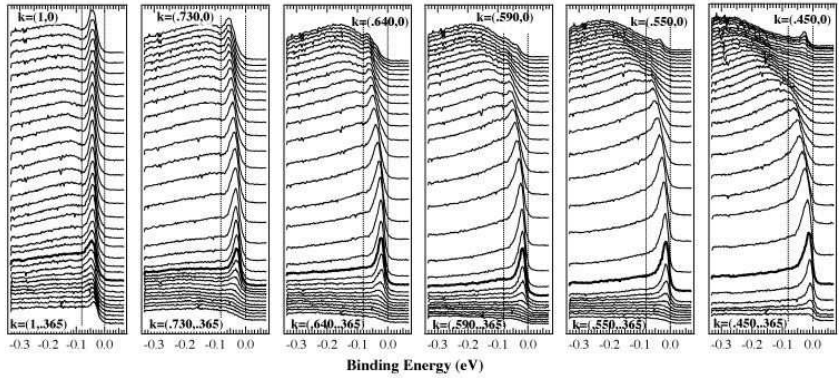}
\caption{
ARPES intensity (T=40K) along selected cuts from Fig.~\protect\ref{fig8.19}.
The thick lined curves
correspond approximately to ${\bf k}_{F}$.  Vertical lines are at 0 and
-80 meV.
}
\label{fig8.20}
\end{figure}

The above evolution is suggestive of the self-energy becoming stronger 
as the $(\pi,0)$ point is approached. This can be quantified from the 
observed change in the dispersion. In Fig.~\ref{fig8.18}e we plot the ratio of 
the EDC dispersion slope above and below the kink energy at various
points along the Fermi surface obtained from middle panels of 
Fig.~\ref{fig8.19}.
Near the node, this ratio is around 2, but becomes large near the 
$(\pi,0)$ point because of the nearly dispersionless quasiparticle 
peak.  A different behavior was inferred in Ref.~\cite{Shen01},
but in their 
case, the cuts near $(\pi,0)$ were perpendicular to ours, and thus not 
normal to the Fermi surface.

The lineshape also indicates that the self-energy is larger near 
$(\pi,0)$, as is evident in Fig.~\ref{fig8.20}. Along the diagonal, there is a 
gentle reduction in $\Sigma^{\prime\prime}$ at low energies, as shown 
in Fig.~\ref{fig8.18}a and b, with an onset at the dispersion kink energy scale.
In contrast, near the $(\pi,0)$ point there must be a very rapid change
in $\Sigma^{\prime\prime}$ in order to produce a spectral dip, as discussed
above. Despite these differences, it is important 
to note that these changes take place throughout the zone at the
same characteristic energy scale (vertical line in Fig.~\ref{fig8.20}).

As also discussed above, the $(\pi,0)$ ARPES spectra can be 
naturally explained in terms of the interaction of the electron with a 
collective mode of electronic origin which only exists below $T_c$. It 
was further speculated that this mode was the neutron resonance.
Here we have shown that dispersion and lineshape anomalies 
have a continuous evolution throughout the zone and are characterized 
by a single energy scale. This leads us to suggest that the same 
electron-mode interaction determines the superconducting lineshape 
and dispersion at all points in the zone, including the nodal direction.
In essence, there is a suppression of the low
energy scattering rate below the finite energy of the mode.
Of course, since the neutron mode is characterized by a
$(\pi,\pi)$ wavevector, one would expect its effect on the
lineshape to be much stronger at points in the zone
which are spanned by $(\pi,\pi)$, as observed here.

\begin{figure}
\includegraphics[width=1\textwidth]{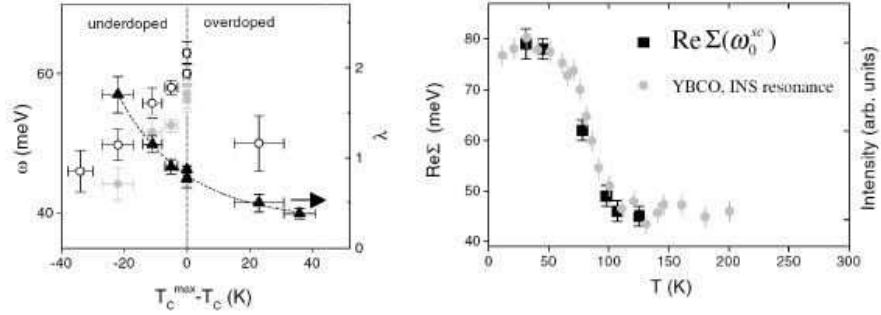}
\caption{
Left panel: Plot of $\omega_{0}$, the energy of the maximum value of 
Re$\Sigma$ in the superconducting state (open circles), and 
$\omega_{0}^{SC}$ (gray circles), the energy of the maximum in 
difference between the superconducting and normal state values plotted 
as a function of T$_c$ referenced to the maximum $T^{max}_{c}$ 
($\equiv$ 91 K). The coupling constant $\lambda$ (black triangles) is 
referenced to the right-hand scale. Right panel: Temperature dependence 
of Re$\Sigma(\omega^{SC}_{0})$ from the nodal line for the UD69K sample 
(black squares) compared with the temperature dependence of the intensity 
of the resonance mode observed in inelastic neutron scattering studies of 
underdoped 
YBa$_{2}$Cu$_{3}$O$_{6+x}$, $T_{c}$ = 74 K (gray circles) (adapted 
from Ref.\cite{PJ01}).
}
\label{johnf3}
\end{figure}

A similar conclusion has been reached by Johnson {\it et al.} \cite{PJ01},
where they find that the kink energy scales with doping like the neutron
resonance (Fig.~\ref{johnf3}a), and that the temperature dependence of 
$\Sigma^{\prime}$ tracks that of the resonance intensity 
(Fig.~\ref{johnf3}b). Moreover, they find that $\Sigma^{\prime}$
increases with underdoping (Fig.~\ref{johnf3}a), much like that extracted 
from the peak/dip/hump lineshape at $(\pi,0)$.

Detailed calculations which take into account the momentum 
dependence of the neutron resonance give an extremely good description 
of the experimental data \cite{EN00}.  These 
results can be understood by studying the Feynman diagram of 
Fig.~\ref{fig8.10}.  The key point is that the neutron resonance has a finite 
width in momentum space, corresponding to a short correlation length of 
order 2 lattice constants.  Because of this, there is now an 
internal sum over momentum in the diagram, which will be dominated by the 
flat regions of the fermionic dispersion around the $(\pi,0)$ points.  
This means that structure in the electron self-energy will occur at an 
energy of $\Delta_{max}+\Omega_{res}$, independent of external momentum.  
This explains why the energy scale is invariant throughout the zone.  On 
the other hand, as the external momentum is swept, the momentum 
dependence of the neutron form factor is probed.  Since the latter peaks 
at $Q=(\pi,\pi)$, then the magnitude of the self-energy will be maximal at 
$(\pi,0)$, since these points are connected by $Q$, and minimum at the 
node.  This explains why the peak/dip/hump effect first weakens into a 
``break'' effect and then into a dispersion kink as the node is approached.
These calculations have been recently extended to incorporate bilayer splitting
effects \cite{EN02}, and are able to explain a
number of unusual lineshape and dispersion features present in data on heavily 
overdoped Bi2212 \cite{BILAY1,BILAY2,BILAY3,BILAY4}

\begin{figure}
\includegraphics[width=1\textwidth]{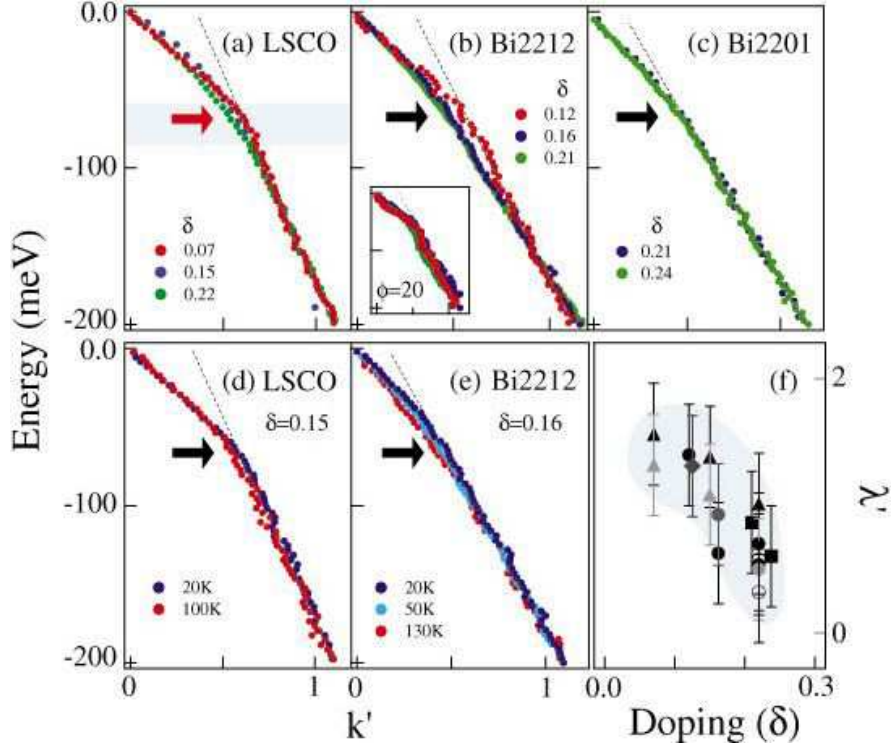}
\caption{
Ubiquity of a sudden change (`kink') in the dispersion. Top 
panels are plots of the dispersion (derived from the momentum distribution 
curves) along $(0, 0) - (\pi,\pi)$ (except panel {\bf b} inset, which is off 
this line) versus the rescaled momentum k' for different samples and at 
different doping levels. a-c, Doping ($\delta$) dependence of LSCO 
(at 20 K; a), Bi2212 (superconducting state, 20 K; b), and Bi2201 (normal 
state, 30 K; c). Dotted lines are guides to the eye. The kink position 
in {\bf a} is compared with the phonon energy at $q = (\pi, 0)$ (thick red 
arrow) and the phonon width and dispersion (shaded area) from neutron data. 
The doping was determined from the $T_{c}$ versus doping universal curve. 
Inset in b, dispersions off the (0, 0) - $(\pi,\pi)$ direction, showing 
also a sharpening of the kink on moving away from the nodal direction. 
The black arrows indicate the position of the kink in the dispersions. 
{\bf d,e}, Temperature dependence of the dispersions for LSCO (d, optimally 
doped) and Bi2212 (e, optimally doped). f, Doping dependence of $\lambda$ 
along the (0, 0) - $(\pi,\pi)$ direction as a function 
of doping. Data are shown for LSCO (filled triangles) and NdLSCO (1/8 doping; 
filled diamonds), Bi2201 (filled squares) and Bi2212 (filled circles in 
the first Brillouin zone, and unfilled circles in the second zone). The 
different shadings represent data obtained in different experimental runs. 
Blue area is a guide to the eye. (adapted from Ref.\cite{Shen01}).
}
\label{lanzf1}
\end{figure}

As mentioned above, this picture has been challenged by 
Lanzara \etal \cite{Shen01}.  These authors claim that the kink is still
present above $T_{c}$, except it is smeared in energy.
Moreover, they find that a kink is present in a large variety of cuprates,
including Bi2201 and LSCO, with an energy which is material and doping
independent, as shown in Fig.~\ref{lanzf1}.  They argue that all of these
observations are in support 
of a phonon interpretation of the kink.

Although initially attractive, there are some problems with this scenario.
First, in regards to the kink above $T_{c}$, it has been claimed by
Johnson \etal \cite{PJ01} that the ``kink'' above $T_{c}$ is simply the
curvature in the dispersion one expects based on marginal Fermi liquid
theory.  In support of this, they argue that the maximum in the real part
of $\Sigma$ is at a different energy in the normal state than in
the superconducting state.  Though this appears to be the case, there is
indeed residual structure in the normal state self-energy at the kink energy
in the optimal doped sample we have looked at.  On the other hand, our
normal state data actually corresponds to the pseudogap phase, and as a
residual of the neutron resonance is present in the pseudogap phase, the
residual ``kink'' above $T_{c}$ (if really there) does not rule out a
magnetic intepretation.  Moreover, as Johnson \etal convincingly showed
(Fig.~\ref{johnf3}b), there is definitely a large component to the self-energy
which follows the same order paramter like temperature variation that the
neutron resonance intensity does.  This observation is supported by recent
work of Gromko \etal \cite{BILAY4} concerning a dispersion kink in the
bonding band of heavily overdoped Bi2212 near $(\pi,0)$.

Second, in regards to constancy of the energy scale, this is indeed an
interesting observation, though we note this statement contradicts that of 
Johnson \etal concerning the doping dependence of the kink energy mentioned 
above (Fig.~\ref{johnf3}).  Also, even
in a phonon model, the kink energy should occur at the sum of the maximum
gap energy plus the phonon energy.  Why the sum of these two numbers should
be doping and material independent is a real puzzle (as it would be for
a magnetic interpretation as well).  And, why only one phonon would be
relevant, despite the large number of phonons present in the cuprates, is
another puzzle.  Still, a phonon model for the kink has certain attractions,
as discussed by Lanzara \etal \cite{Shen01}.  Certainly, more work is
needed to definitively resolve the controversies surrounding the origin of
the dispersion kink.

\subsection{Condensation Energy}

We conclude this section by discussing the relation of ARPES data to the 
superconducting condensation energy.

We begin with the assumption that the condensation energy does not
have a component due to phonons.  
To proceed, we assume an effective single-band Hamiltonian
which involves only two particle interactions.
Then, simply exploiting standard formulas
for the internal energy $U = \langle H - \mu N \rangle$ 
($\mu$ is the chemical potential, and $N$ the number of particles)
in terms of the one-particle Green's function, we obtain \cite{Mike00}
\begin{eqnarray}
\lefteqn{U_{N} - U_{S} =} \nonumber \\
& & \sum_{\bf k} \int_{-\infty}^{+\infty}d\omega 
(\omega + \epsilon_k) f(\omega)
\left[A_{N}({\bf k},\omega) - A_{S}({\bf k},\omega)\right]
\label{eq8.9}
\end{eqnarray}
where the spin variable has been summed over.
Here and below the subscript $N$ stands for the normal state,
$S$ for the superconducting state. 
$A({\bf k},\omega)$ is the single-particle spectral function, 
$f(\omega)$ the Fermi function, and $\epsilon_k$
the bare energy dispersion which defines the kinetic energy part of 
the Hamiltonian.
Note that the $\mu N$ term has been absorbed into $\omega$
and $\epsilon_k$, that is, these quantities are defined relative to the
appropriate chemical potential, $\mu_N$ or $\mu_S$.  In general, $\mu_N$ and
$\mu_S$ will be different.  This difference has to be taken into
account, since the condensation energy is small.

The condensation energy is defined by the
zero temperature limit of $U_{N} - U_{S}$ in the above expression.
Note that this involves defining (or somehow extrapolating to) the
normal state spectral function at $T=0$. Such an
extrapolation, which we return
to below, is not specific to our approach, but required in all
estimates of the condensation energy.
We remark that Eq.~\ref{eq8.9} yields the correct condensation
energy, $N(0) \Delta^2/2$, for the BCS theory of 
superconductivity.

We also note that Eq.~\ref{eq8.9} can also be broken up into two pieces to
individually yield the thermal expectation value of the
kinetic energy (using $2\epsilon_k$ in the parentheses in front
of $f(\omega)$), and that of the potential energy
(using $\omega-\epsilon_k$ instead).

The great advantage of Eq.~\ref{eq8.9} is that it involves just the occupied 
part of the single particle spectral function, which is measured
by ARPES.
Therefore, in principle, one should be able to derive the condensation energy
from such data, if an appropriate extrapolation of the normal state spectral
function to T=0 can be made.
On the other hand, a disadvantage is that the bare energies, $\epsilon_k$, are
{\it a priori} unknown.  Note that these are not directly obtained from
the measured ARPES dispersion, which already includes many-body
renormalizations. 
Rather, they could be determined by projecting the kinetic
energy operator onto the single-band subspace.

Some of the problems
associated with an analysis based on experimental data
can be appreciated.  First, the condensation energy is obtained by 
subtracting two large numbers.  Therefore, normalization
of the data becomes a central concern.  Problems in this regard when 
considering $n({\bf k})$, which is the zeroth moment of the ARPES data,
were discussed previously \cite{JC98}.
For the first moment, these problems are further amplified
due to the $\omega$ weighting in the integrand.  When analyzing
real data, we have found that the high energy tail contribution to the 
first moment is very sensitive to how the data are normalized.  Different
choices
of normalization can even lead to changes in sign of the first moment.

Another concern concerns the ${\bf k}$ sum in Eq.~\ref{eq8.9}. 
ARPES has ${\bf k}$-dependent 
matrix elements, which lead to weighting
factors not present in Eq.~\ref{eq8.9}.  These effects can in principle be
factored out by either theoretical estimates of the matrix 
elements \cite{Bansil99},
or by comparing data at different photon energies to obtain information on
them \cite{mesot01}.

Another issue in connection with experimental data is an appropriate
extrapolation of the normal state to zero temperature.  Information on this
can be obtained by analyzing the temperature dependence of the normal state
data, remembering that the Fermi function will cause a temperature dependence
of the data which should be factored out before attempting the $T=0$
extrapolation.  We finally note that the temperature dependence issue is
strongly coupled to the normalization problem mentioned above.  In ARPES, the
absolute intensity can change due to temperature dependent changes in absorbed
gasses, surface doping level, and sample location.  Changes of background
emission with temperature is another potential problem.

Despite these concerns,
we believe that with careful experimentation, many of these difficulties
can be overcome, and even if an exact determination of Eq.~\ref{eq8.9} is not 
possible,
insights into the origin of the condensation energy will certainly be
forthcoming from the data.  This is particularly true for ARPES, which has
the advantage of being ${\bf k}$ resolved and thus giving one information on the
relative contribution of different ${\bf k}$ vectors to the condensation energy.

Insights into what real data might indicate have been offered by
us \cite{Mike00} in the context of the ``mode'' model illustrated in 
Fig.~\ref{fig8.12}.  What we found was that for parameters characteristic 
of optimal doped ARPES data, the superconducting condensation was driven 
by kinetic energy lowering, as opposed to the potential energy lowering 
found in BCS theory.  This occurs because $n_k$ becomes sharper in the 
superconducting state than in the normal state.  In essence, the normal 
state is a non Fermi liquid and the superconducting state is a Fermi 
liquid, so what occurs is that the effect of quasiparticle formation on 
sharpening $n_k$ is greater than the effect of particle-hole mixing on 
smearing it.  The net result is a sharpening, leading to a lowering in 
kinetic energy.  In BCS theory, the normal state is a Fermi liquid, and 
thus only the particle-hole mixing effect is present, leading to a net 
smearing of $n_k$ and thus an increase in the kinetic energy.  The same 
model can be used to evaulate the optical sum rule \cite{NP02},
and what is found is a violation of the sum rule with a sign and 
magnitude consistent with recent optics experiments \cite{OPT}.  
It will be of great 
interest to see whether these results can be confirmed directly from 
ARPES data, as speculated early on by Anderson \cite{And90}.

\section{Acknowledgments}

Much of the experimental work described in this article was done in
collaboration with Hong Ding, Adam Kaminski, Helen Fretwell, Kazimierz
Gofron, Joel Mesot, Stephan Rosenkranz, Tsunehiro Takeuchi, and the group
of Takashi Takahashi, including Takafumi Sato and Takayoshi Yokoya.  We
were very fortunate to have available to us the samples from Kazuo
Kadowaki, T. Mochiku, David Hinks, Prasenjit Guptasarma, Boyd Veal, Z. Z. Li,
and Helene Raffy.  We have also benefited from many interactions over the
years with Phil Anderson, Alex Abrikosov, Jim Allen, Cliff Olson, Ole Andersen,
Al Arko, Bertram Batlogg, Arun Bansil, Matthias Eschrig,
Atsushi Fujimori, Peter Johnson, Bob
Laughlin, Bob Schrieffer, Z. X. Shen, and Chandra Varma.

This work was supported by the National Science Foundation, Grant
No. DMR 9974401 (JCC)
and the U.S. Department of Energy, Office of Science, under Contract
No. W-31-109-ENG-38 (JCC and MRN).  MR is grateful for partial 
support from the Indian DST through the Swarnajayanti scheme.


\begin{thebibliography}{8.}
\bibitem{BIO_SL} 
This would allow us to understand (1) why polarization
selection rules are obeyed for the $\Gamma\bar{M}$ mirror plane,
and (2) qualitatively, why the intensities for odd and even 
polarizations along $\Gamma X$ are comparable.
\bibitem{caveat}
The data in Ref.~\cite{fretwell00} is taken in the superconducting
state ($T=40$K) of the $T_c = 90$K sample.
We note that the integration range ($\pm$100 meV) employed for the 
intensity patterns is much larger than the gap energy scale, and the
minimum gap locus (to be discussed below in Section 6) in the
superconducting state is essentially identical to the normal
state Fermi surface.
\bibitem{Bilayer}
Bilayer splitting is discussed in Section 4.5.
\end{thebibliography}
\end{document}